\documentclass[11pt]{article}
\usepackage{cite}
\usepackage{jheppubmod}
\usepackage{enumitem}
\usepackage{amsmath}
\usepackage{physics}
\usepackage{graphicx}
\usepackage{amsfonts} 
\usepackage{amssymb}
\usepackage[dvipsnames, svgnames, x11names ]{xcolor}
\usepackage{color}
\usepackage{braket}
\graphicspath{ {./1 /} }
\usepackage{subcaption}

\def\qq{{\cal Q}}

\def\pb[#1,#2]{\{#1, #2\}}
\def\deb[#1,#2]{[#1,#2]_{\text{D.B.}}}

\def\tr{{\rm tr}}

\def\Or[#1]{{\text{O}}\left({#1}\right)}
\def\dotl[#1,#2]{\left\langle #1,\, #2 \right\rangle}
\def\dotlb[#1,#2]{\left\langle #1,\, #2 \right\rangle}
\def\dotlm[#1,#2]{\left[ #1,\, #2 \right]}
\def\dotp[#1,#2]{(\vect{#1} \cdot\vect{#2})}
\def\aff[#1,#2]{\hat{#1}(#2)}

\def\n4sym{{\cal N}=4 SYM}
\def\>{\rangle}
\def\<{\langle}
\def\weight[#1,#2,#3]{\{(#1),#2,#3\}}
\def\ads[#1]{$\text{AdS}_{#1}$}

\def\tarelr[#1]{\widetilde{a}^{\text{rel}}_{R#1}}
\def\Oright[#1]{{\cal O}_{R#1}}
\def\Oleft[#1]{{\cal O}_{L#1}}
\def\aleft[#1]{a_{L#1}}
\def\arelr[#1]{a^{\text{rel}}_{R#1}}

\def\Tr{{\rm {Tr}}}
\def\hkll{{\Phi}}

\newcommand{\be}{\begin{equation}}
\newcommand{\ee}{\end{equation}}
\newcommand{\beq}{\begin{equation}}
\newcommand{\eeq}{\end{equation}}
\newcommand{\ben}{\begin{displaymath}}
\newcommand{\een}{\end{displaymath}}
\newcommand{\beqa}{\begin{eqnarray}}
\newcommand{\eeqa}{\end{eqnarray}}
\newcommand{\bea}{\begin{eqnarray}}
\newcommand{\eea}{\end{eqnarray}}
\newcommand{\bean}{\begin{eqnarray*}}
\newcommand{\eean}{\end{eqnarray*}}
\newcommand{\ba}{\begin{array}}
\newcommand{\ea}{\end{array}}
\newcommand{\bi}{\begin{itemize}}
\newcommand{\ei}{\end{itemize}}

\newcommand{\bs}{\begin{split}}
\def\sess\end{split}
\def\D{\cal D}

\newcommand{\vect}[1]{{#1}}

\def\rs{|\Psi\rangle}
\def\ls{\langle \Psi|}

\def\rsz{|\Psi_{0}\rangle}
\def\lsz{\langle \Psi_0|}

\def\Tr{{\rm Tr}}

\def\rs{|\Psi\rangle}
\def\ls{\langle \Psi|}

\def\drop{\widehat{\Phi}}

\preprint{CERN-TH-2023-003}
\title{Holography and Localization of Information in Quantum Gravity}
\author[a,b,c]{Eyoab Bahiru}
\author[d,e,f]{Alexandre Belin}
\author[g]{Kyriakos Papadodimas}
\author[g]{Gabor Sarosi}
\author[a,b]{Niloofar Vardian}

\affiliation[a]{SISSA, International School for Advanced Studies, via Bonomea 265, 34136 Trieste, Italy}
\affiliation[b]{INFN, Sezione di Trieste, via Valerio 2, 34127 Trieste, Italy}
\affiliation[c]{International Centre for Theoretical Physics, Strada Costiera 11, Trieste 34151 Italy}
\affiliation[d]{Dipartimento di Fisica, Universit\`a di Milano - Bicocca, I-20126 Milano, Italy}
\affiliation[e]{INFN, Sezione di Milano-Bicocca, Piazza della Scienza 3, 20126 Milano, Italy}
\affiliation[f]{Institute of Physics, Ecole Polytechnique F\'ed\'erale de Lausanne, CH-1015 Lausanne, Switzerland}
\affiliation[g]{Theoretical Physics Department, CERN, CH-1211 Geneva 23, Switzerland}
\emailAdd{ebahiru@sissa.it}
\emailAdd{alexandre.belin@unimib.it}
\emailAdd{kyriakos.papadodimas@cern.ch}
\emailAdd{gabor.sarosi@cern.ch}
\emailAdd{nvardian@sissa.it}

\abstract{Within the AdS/CFT correspondence, we identify a class of CFT operators which represent diff-invariant and approximately local observables in the gravitational dual. Provided that the bulk state breaks all asymptotic symmetries, we show that these operators commute to all orders in $1/N$ with asymptotic charges, thus resolving an apparent tension between locality in perturbative quantum gravity and the gravitational Gauss law. The interpretation of these observables is that they are not gravitationally dressed with respect to the boundary, but instead to features of the state. We also provide evidence that there are bulk observables  whose commutator vanishes to all orders in $1/N$ with the entire algebra of single-trace operators defined in a space-like separated time-band. This implies that in a large $N$ holographic CFT, the algebra generated by single-trace operators in a short-enough time-band has a non-trivial commutant when acting on states which break the symmetries. It also implies that information deep in the interior of the bulk is invisible to single-trace correlators in the time-band and hence that it is possible to localize information in perturbative quantum gravity.
}

\begin{document} 
\maketitle

\section{Introduction}

It is generally believed that in quantum gravity, space-time locality is an emergent notion which becomes accurate and useful in certain limits of the underlying theory. This perspective is realized in the AdS/CFT correspondence \cite{Maldacena:1997re}: bulk locality becomes precise in the large $N$, strong coupling limit and when probing the theory with simple enough operators. Moreover, a large number of proposals aiming to resolve the black hole information paradox rely on a certain amount of non-locality \cite{tHooft:1984kcu, Susskind:1993if,Giddings:2012gc,Bousso:2012as,Papadodimas:2012aq, Verlinde:2012cy, Maldacena:2013xja,penington2020entanglement, Almheiri:2019hni, Almheiri:2019psf,Penington:2019kki,Laddha:2020kvp}. A natural question is to understand whether non-local features of quantum gravity are visible only in the non-perturbative regime, or whether remnants of non-locality are also visible at the perturbative level.

Even in classical general relativity it is not entirely straightforward to formulate the concept of locality, as it is non-trivial to define local observables. Physical observables need to be diff-invariant and,  in order for them to also be local, they have to be associated to points in space-time which have to be specified in a diff-invariant way. If the space-time has a boundary, a standard approach is to define points relationally with respect to the boundary or by completely fixing the gauge. We say that these observables are {\it gravitationally dressed} with respect to the boundary. However, the resulting observables, while diff-invariant, are not strictly localized and have non-vanishing Poisson brackets at space-like separation. A particular aspect of this difficulty is related to the gravitational Gauss law: in gravitational theories defined with asymptotically flat or AdS boundary conditions, the Hamiltonian, and other asymptotic symmetry charges, are boundary terms. Acting with a candidate local, diff-invariant observable in the interior of space will generally change the energy of the state, which is immediately measurable at space-like separation due to Gauss's law.

Despite these difficulties, at the classical level, there are ways of defining local and diff-invariant observables in the neighborhood of a state, provided that the state is sufficiently complicated. A class of such observables introduced a long time ago \cite{PhysRev.111.1182,Bergmann:1960wb,DeWitt:1962cg} will be reviewed in sub-section \ref{sdob}, see also \cite{Giddings:2005id, Marolf:2015jha,Khavkine:2015fwa} for more recent discussions. These observables respect the causal structure of the underlying space-time, in the sense that their Poisson brackets at space-like separation vanish. In particular, provided that the state we are considering is complicated enough, the action of these observables is not visible by the boundary Hamiltonian, as these observables only rearrange energy in the interior of space. The price we have to pay is that these observables are not defined globally on the phase space of solutions. They have desired properties only for certain states.

A natural question is to what extent can such local diff-invariant observables be defined at the quantum level. As mentioned above, we do not expect to be able to find exactly local diff-invariant observables at the non-perturbative level, however it may be possible to do so in perturbation theory. This question is important in order to be able to quantify  departures from locality in quantum gravity and to understand if there is a way to generalize the structure of algebras of observables of quantum field theory to situations where gravity is included perturbatively. 

It is useful to formulate these questions in the context of the AdS/CFT correspondence. We consider a CFT state $\rsz$ that is dual to a semi-classical asymptotically AdS$_{d+1}$ geometry in global coordinates and a short time-band near the boundary as shown in Fig. \ref{fig1}. We consider the algebra ${\cal A}$ of observables in semi-classical gravity which are localized in this time band. This algebra includes the Hamiltonian and other asymptotic charges. From the point of view of the dual CFT, it is natural to identify the algebra ${\cal A}$ with the algebra generated by single-trace operators localized in this time-band, we will call it the "single-trace algebra". The expectation is that the single-trace algebra ${\cal A}$ corresponds to the causal wedge of the time-band \cite{Banerjee:2016mhh}\footnote{A different approach for studying time-bands based on gravitational entropy and minimal surfaces was initiated in \cite{Balasubramanian:2013rqa,Balasubramanian:2013lsa,Myers:2014jia,Headrick:2014eia} and also \cite{Hubeny:2014qwa}. It would be interesting to understand possible connections between those ideas and the results presented in this paper.}. Notice that here we have causal-wedge reconstruction and not entanglement wedge reconstruction, as we are looking only at the single-trace subalgebra. In the CFT the notion of a time-band algebra only makes sense at large $N$, since large $N$ generates a natural hierarchy between operators that are small combinations of single-trace operators and arbitrarily complicated operators. For finite $N$ there is no such hierarchy and the time-slice axiom would imply that ${\cal A}$ is the full CFT algebra\footnote{We do not include in the single-trace algebra elements like $e^{i Ht}{\cal O}(t=0,x)e^{-i H t}$ with $t=O(N^0)$ and large enough to exit the time-band. Such "precursor" operators are complicated from the point of view of operators in the time-band and go beyond the semi-classical description.}. Algebras of single-trace operators in holographic CFTs have been discussed in \cite{Papadodimas:2012aq,Papadodimas:2013jku,Papadodimas:2013wnh} and more recently in \cite{Leutheusser:2021qhd, Leutheusser:2021frk, Witten:2021unn,Chandrasekaran:2022eqq,Leutheusser:2022bgi}.

If the time-band is short enough, then there is a region in the bulk which is space-like with respect to the time-band. We will refer to this region as the "diamond"\footnote{ For now we assume that the state has simple topology and there are no black hole horizons in the interior.}. If we were able to define diff-invariant observables localized in the diamond, they should commute with the algebra ${\cal A}$. As already mentioned, the question is non-trivial as these observables must be gravitationally dressed and if we use the boundary to dress them, then they will not commute with ${\cal A}$. For example, it appears that since the Hamiltonian $H$ is an element of ${\cal A}$ it would be able to detect any excitation added in the interior of the diamond using the gravitational Gauss law. To summarize, the question we want to examine: 
\begin{figure}
    \centering
    \includegraphics[width=0.15\textwidth]{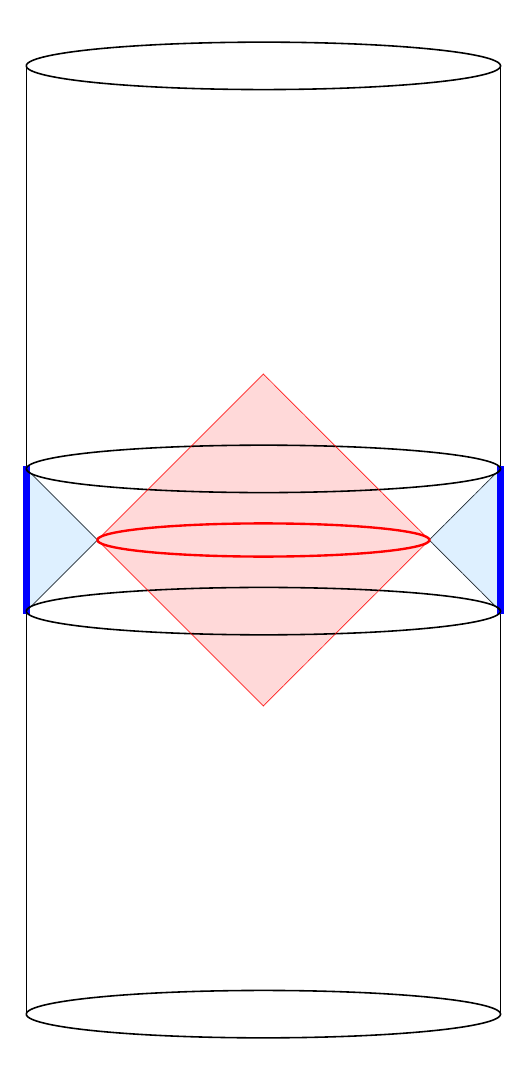}
  \caption{The single-trace operators localized in the time band $t\in(-\epsilon,\epsilon)\times {\mathbb S}^{d-1}$ (dark blue region on boundary) form an algebra ${\cal A}$ which is conjectured to be dual to the causal wedge of the region (light blue). If the state $\rsz$ of the system breaks all symmetries, then the causal diamond in the middle (light red), which is spacelike separated from the time-band, corresponds to the commutant ${\cal A}'$ of the algebra ${\cal A}$ when acting on the code subspace of the state $\rsz$.}
    \label{fig1}
\end{figure}
 \begin{quote}
 {\it Does the algebra ${\cal A}$, when acting on the state $\rsz$ and small perturbations around it, have a non-trivial commutant in the $1/N$ expansion?}
 \end{quote}
As we will discuss later, we need to refine the question by demanding that the commutant acts non-trivially within the code-subspace of the state, in order to avoid obvious but uninteresting constructions\footnote{\label{trivialun} For example, for a complicated state with energy of $O(N^2)$, a unitary which rotates the phase of a single energy eigenstate will have commutators of $O(e^{-N^2})$ with all elements of ${\cal A}$. However, this would not be an interesting example, as this operator is generally "invisible" from the bulk point of view and does not create excitations inside the diamond.}. We emphasize that we do not expect the algebra to have a commutant at finite $N$ \cite{Banerjee:2016mhh}.

A closely related question is that of localization of information. According to AdS/CFT the quantum state of the CFT at any moment in time contains the full information of the bulk. In particular, if we had considered the {\it full} algebra of all operators in the time-band, as opposed to the algebra generated by few (relative to $N$) single-trace operators, then we would be able to reconstruct the interior of the diamond. Suppose however, that we only have access to the algebra ${\cal A}$ of single-trace operators in the time band. Can we then reconstruct the information of whatever is hidden inside the diamond? This can also be rephrased as follows: 
\begin{quote}
{\it Given a state $\rsz$, can we find another state $\rsz'$ such that the correlators of the single-trace algebra ${\cal A}$ in the time-band, evaluated on these two states agree to all orders in $1/N$, but correlators of single-trace operators differ at $O(N^0)$ outside the time-band?}
\end{quote}
The intuition here is that we want to find a state $\rsz'$ which contains an additional excitation relative to $\rsz$ in the interior of the diamond which becomes visible by single-trace operators only after a light-ray has reached the boundary i.e. in the future or past of the time-band. If the algebra ${\cal A}$ had a commutant then we could take $\rsz' = U(A') \rsz$ for some unitary $U$ built out of operators $A'$ in the commutant. 

We will provide evidence that the answer to the two aforementioned questions is positive, provided that the state $\rsz$ is complicated enough. The reasoning was first outlined in \cite{Bahiru:2022oas}. In this paper we extend the construction in a few ways and provide additional arguments and examples.

Standard approaches to bulk reconstruction lead to observables which are relationally defined with respect to the boundary. This is the case for approaches based on the HKLL reconstruction \cite{Banks:1998dd, Bena:1999jv,Hamilton:2005ju,Hamilton:2006az, Hamilton:2006fh, Hamilton:2007wj,Heemskerk:2012mn, Almheiri:2017fbd}, as well as approaches based on the Petz map \cite{Cotler:2017erl,Chen:2019gbt} or modular reconstruction \cite{Jafferis:2015del,Faulkner:2017vdd}, as they all require some sort of boundary dressing. For concreteness we start with a standard HKLL operator given by
\be
\label{hkll0}
\hkll(t,r,\Omega) = \int_{\rm bdry} dt' \,d\Omega'_{d-1} K(t,r,\Omega; t',\Omega') {\cal O} (t',\Omega') \,.
\ee
Here $K$ is a particular Green's function which depends on the background metric. Implicit in this expression is a gauge-fixing scheme in a particular coordinate system, which is uniquely determined by making use of the boundary. If we pick the point $(t,r,\Omega)$ to be in the diamond, the operator \eqref{hkll0} commutes with all single-trace operators in the time band at large $N$. At subleading orders multi-trace corrections need to be added to \eqref{hkll0} to ensure vanishing commutators. However the commutator with the Hamiltonian and other asymptotic charges, which is nonzero at order $1/N$, cannot generally be corrected by multi-trace corrections. The physical reason is that the operator \eqref{hkll0} is gravitationally dressed with respect to the boundary. The non-vanishing commutator with $H$ appears to be an obstacle in identifying \eqref{hkll0} as an element of the commutant of ${\cal A}$ \cite{Giddings:2018umg,Donnelly:2016rvo}.

In this paper we present a way to find operators which commute with the asymptotic charges to all orders in $1/N$, while at the same time create excitations in the interior of the diamond similar to those of the HKLL operator. These operators can be defined provided the state $\rsz$ that we are considering breaks all asymptotic symmetries. These operators correspond to observables gravitationally dressed with respect to features of the state. 

A crucial starting observation is that, if a state $\rsz$ is dual to a bulk geometry which breaks the asymptotic symmetries, then the overlap
\be
\label{decayintro}
\lsz U(g) \rsz  \qquad g \in SO(2,d) \,,
\ee
is generally exponentially small, of order $ O(e^{-a N^2})$ with  $\Re(a)>0$ provided that the element $g$ is sufficiently far from the identity\footnote{But not too far. The state may return to itself in compact directions of the conformal group or approximately back to itself due to Poincare recurrences.}. Here $SO(2,d)$ represents the asymptotic symmetry group of AdS$_{d+1}$. We will quantify this statement more precisely in the later sections. In fact, we will provide evidence that if we introduce the code subspace around the state $\rsz$, defined as
\be
\label{code0intro}
{\cal H}_{ 0} = {\rm span}\{|\Psi_0\rangle, {\cal O}(t,\Omega) |\Psi_0\rangle,...,{\cal O}_1(t_1,\Omega_1)...{\cal O}_n(t_n,\Omega_n)\rsz\} \,,
\ee
and similarly ${\cal H}_g$ for the state $U(g)\rsz$ then any inner product between unit normalized states of ${\cal H}_0, {\cal H}_g$  will also be of order $ O(e^{-a N^2})$.

Starting with a standard HKLL operator $\Phi$ we consider the operator
\be
\label{defopintro}
\drop = c\int_{B} d\mu(g) U(g)P_0 \Phi P_0 U(g)^{-1} \,,
\ee
where $P_0$ denotes the projector on \eqref{code0intro} and $d\mu(g)$ is the Haar measure on $SO(2,d)$ and $B$ is a reasonably sized neighborhood of $SO(2,d)$ around the identity. The overall normalization constant $c$ will be specified later. The main claim, which will be discussed in section \ref{sec:operators}, is that operators \eqref{defopintro} have the desired properties: their commutators with the asymptotic symmetry charges $Q$ of $SO(2,d)$ are exponentially small
\be
\label{zerocom}
[Q,\hat{\Phi}] = O(e^{-N^2}) \,,
\ee
when acting on the code subspace, while at the same time, the {\it leading} large-$N$ action  of $\hat{\Phi}$ on the code subspace \eqref{code0intro} is the same as that of the corresponding HKLL operator $\Phi$, that is
\be
\label{acthkll}
\langle \Psi_1| \hat{\Phi} |\Psi_2\rangle = \langle \Psi_1| \Phi |\Psi_2\rangle + O(1/N) \qquad \qquad \forall \,\,|\Psi_1\rangle,|\Psi_2\rangle \in {\cal H}_0 \,.
\ee
The interpretation is that by performing the integral \eqref{defopintro} we have removed the gravitational dressing of the operators from the boundary and moved it over to the state. This is only possible on states where \eqref{decayintro} decays sufficiently fast.

The operators \eqref{defopintro}  have vanishing commutators with the asymptotic charges to all orders in $1/N$. This demonstrates that the apparent obstacle to identifying a commutant due to Gauss's law can be overcome. In order to find a true commutant we need to ensure vanishing commutators to all orders in $1/N$ with all single-trace operators in the time-band algebra. It would be interesting to explore whether a formula achieving this goal and similar to \eqref{defopintro} can be derived, possibly by integrating over the unitary orbits generated by ${\cal A}$. 

We provide an alternative formal argument supporting the idea that the algebra ${\cal A}$ has a nontrivial commutant when acting on the code subspace ${\cal H}_{code}$ of a complicated state $\rsz$. To see that we consider an operator $\hat{\Phi}$ {\it defined} by
\be
\label{lineqs}
\hat{\Phi} A \rsz = A \Phi \rsz \qquad \forall A \in {\cal A} \,,
\ee
where again $\Phi$ is a standard HKLL operator. 
This represents a set of linear equations, one for each $A\in {\cal A}$,  which define the action of $\hat{\Phi}$ on ${\cal H}_0$. A sufficient condition for the consistency of these  equations is that for all non-vanishing operators $A\in {\cal A}$ we have $A\rsz \neq 0$.\footnote{An intuitive way to think about this condition is that since the states we will consider have a large energy, they cannot be annihilated by single-trace operators. There is however an important subtlety with symmetries, which we will discuss in detail.} In section \ref{otconop} we provide evidence that this is true in the $1/N$ expansion. Given that these equations are consistent, we will show in section \ref{otconop} that the operators $\hat{\Phi}$ defined by \eqref{lineqs} obey the following properties: i) by construction they commute with operators in ${\cal A}$ and ii) to leading order at large $N$ they act like HKLL operators. This provides  evidence that the algebra ${\cal A}$ has a commutant in the $1/N$ expansion. As mentioned earlier, a commutant is not expected at finite $N$. Indeed, at finite $N$ it is possible to find complicated operators in the time-band which annihilate the state $\rsz$ and equations \eqref{lineqs} do not have a consistent solution.

If we take the state $\rsz$ to be the vacuum, i.e. empty AdS, then the previous construction fails: since the vacuum is invariant under the asymptotic symmetries we no longer have the decay of \eqref{decayintro} and \eqref{acthkll} fails. Also \eqref{lineqs} fails because there are operators in the time-band, in particular $H$, which annihilate the state. We emphasize that this failure is not a limitation of our particular construction. Instead the interpretation of this failure is that  since empty AdS has no bulk features, the only way to specify a point in the bulk is by dressing it to the boundary. Hence any bulk diff-invariant operators acting around the vacuum will not commute with the asymptotic charges \cite{Giddings:2018umg,Donnelly:2016rvo}. This can also be seen from the fact that even classically, local diff-invariant observables cannot be defined properly in the vacuum.

We clarify that the results of this paper do not contradict the claim of \cite{Chowdhury:2021nxw} that {\it specifically} for perturbative states around empty AdS, it is possible to reconstruct the state from correlators in the time-band. However we notice that  {\it interesting} states, that is, states which have bulk observers capable of performing physical experiments, are generally expected to be of the form where the symmetries are broken and the construction presented in this paper can be applied.

If the state $\rsz$ corresponds to a black hole state, and if the variance of the asymptotic charges scales like $N^2$\,\,\footnote{For example, this is true for black hole states with energy spread similar to the canonical ensemble.} we find that using the operators \eqref{defopintro} we can create excitations behind the horizon which cannot be detected by correlators of single-trace operators in the $1/N$ expansion. Understanding how to diagnose these excitations from a CFT calculation remains an outstanding open problem. We emphasize that this does not contradict the fact that, generally, excitations created by unitaries on top of typical states with small energy spread can be detected by single-trace correlators \cite{Papadodimas:2013jku,Papadodimas:2015jra,Papadodimas:2017qit}. Such states with small energy spread are those for which our construction cannot be applied. 

The operators we identify provide evidence supporting the idea that locality is respected in perturbative quantum gravity and that information can be localized in subregions at the level of perturbation theory, provided that the underlying state is sufficiently complicated. It also suggests that it should be possible to associate algebras of observables to subregions. However these observables have certain features of state-dependence, since both \eqref{defopintro} and \eqref{lineqs} give operators which are defined only on the code-subspace of the original state $\rsz$. It is certainly possible to extend the domain of definition of our operators by combining together code subspaces of sufficiently different states, each one of which must break the asymptotic symmetries, thus partly eliminating the state-dependence of the operators. However the number of these states must not be too large, otherwise the small overlaps between the code subspaces start to accumulate and modify the correlators. This becomes particularly relevant for black hole states, where we do not expect to have operators with the desired properties defined globally for most microstates and some genuine state-dependence is expected.

The plan of the paper is as follows: in section \ref{sec:aspectsloc} we review background material about various aspects of locality in field theory and gravity. In section \ref{sectionAdSCFT} we describe the setup in AdS/CFT and study the decay of the inner product \eqref{decayintro}. In section \ref{sec:operators} we introduce the operators \eqref{defopintro} and discuss their basic properties. In section \ref{otconop} we provide an alternative argument for the existence of a commutant based on equations \eqref{lineqs}. In section \ref{sec:examples} we consider various examples. In section \ref{sec:bhs} we consider aspects of our operators in the presence of black holes. Finally we close with a discussion of open problems in \ref{sec:discuss}.

\section{Aspects of locality in field theory and gravity}\label{sec:aspectsloc}

In this section, mostly addressed to non-experts, we review some background necessary to explore the question of localizing information in different regions of space. A closely related question is the association of algebras of observables to subregions and the factorization of the Hilbert space. We start with non-gravitational field theories, where a non-dynamical background space-time can be used in order to define sub-regions and their causal relations, and then we consider the additional complications when gravity is taken into account.

In relativistic theories we expect that signals and information cannot travel faster than light. Consider an initial space-like slice $\Sigma$ and divide it into a compact subregion $D$ and its complement $D'$. We denote by $J(D')$ the domain of dependence of $D'$. We then want to address the following question: is it possible to modify the state\footnote{Either classical state, or quantum density matrix.} in region $D$  without affecting the state in $J(D')$. If the answer is positive then an observer initially in  $D'$, and confined to move in $J(D')$, cannot reconstruct information about the interior of $D$. Then we say that information can be localized.

\subsection{Classical field theories}\label{subsecclass}

At the classical level this question can be addressed by studying  the initial value problem: we specify initial data ${\cal C}$ on a spacelike slice $\Sigma$ and then look for a solution in the entire space-time, or at least a neighborhood of the slice $\Sigma$, compatible with the initial data. The initial data will typically include the values and time-derivatives of various fields of the theory. The theories we will be considering have gauge invariance. One of the implications is that the existence of a solution is guaranteed only if the initial data satisfy certain {\it constraints}. In relativistic field theories the dynamical equations are hyperbolic, which ensures that signals propagate forward from $\Sigma$ at most at the speed of light. On the other hand the constraint equations for initial data are of elliptic nature. This makes the question of being able to specify the initial data independently in region $D$ and its complement $D'$  non-trivial. It is thus convenient to divide the question formulated above in two steps:

\begin{itemize}[label={}]

\item {\bf A. Localized preparation of states:} for given initial data ${\cal C}_1$  on $\Sigma $ satisfying the constraints, to what extent can we deform to other initial data ${\cal C}_2$, also satisfying the constraints, such that ${\cal C}_1,{\cal C}_2$ agree on $D'$, possibly up to a gauge transformation, but differ essentially\footnote{ i.e. cannot be matched by a gauge transformation on $D$.  } on $D$?

{\bf B. No super-luminal propagation:} suppose we are given two initial data ${\cal C}_1,{\cal C}_2$ which satisfy the constraints, which agree on $D'$ and differ on $D$. We then want to show that the two corresponding solutions agree on $J(D')$, possibly up to a gauge transformation.
\end{itemize}
We will return to the classical problem in theories with gauge invariance in the following subsections. For now we briefly consider the simplest example of a free Klein-Gordon field in flat space obeying
$
\Box \phi =m^2\phi
$.
We consider initial data on the slice $\Sigma$ corresponding to $t=0$. The initial data on this slice are parametrized by ${\cal C} = \{\phi(t=0,x), \partial_0 \phi (t=0,x)\}$. 
In this case condition $A$ mentioned earlier is clearly satisfied: the initial data do not need to obey any constraint, so we can simply select the functions $\phi,\partial_0\phi$ to have any smooth profile with features strictly localized in $D$. Notice that this requires the use of non-analytic initial data. Condition $B$ is also satisfied, see \cite{Wald:1984rg} for a basic review.\footnote{In the case of non-relativistic theories, for example the heat equation, which is first order in time and hence not hyperbolic, we are able to specify the initial data in subregions independently but the speed of propagation is unbounded. Hence the heat equation obeys condition $ A$ but not $ B$.}

\subsection{Localization of information  in QFT}

In non-gravitational QFT we can associate algebras of observables to space-time regions \cite{Streater:1989vi,Haag:1963dh,Haag:1992hx}. Locality is exact, and is expressed by the condition that algebras corresponding to  space-like separated regions commute. An analogue of the initial value problem in QFT is expressed by the condition of {\it primitive causality}  or relatedly the {\it time-slice axiom} which postulates that the only operators commuting with the algebra generated by operators in a time-band are proportional to the identity. Moreover a local version of these statements postulates that the algebra of operators in a subregion coincides with the algebra of operators in the causal domain of dependence of the subregion \cite{haagschroer}.

An intuitive way to see that information can be localized in QFT is as follows: suppose $\rsz$ is a state in the Hilbert space of the QFT. Consider a unitary operator $U_D$ constructed out of observables localized in $D$ and the new state $\rs = U_D \rsz$. The unitary $U_D$ modifies the state by creating an excitation in region $D$ which encodes the desired information in that region. For any observation ${\cal O}_{D'}$ in region $D'$, and more generally in $J(D')$, we have
\be
\label{unitaryd}
\ls {\cal O}_{D'} \rs = \lsz U_D^\dagger {\cal O}_{D'}U_D \rsz =\lsz {\cal O}_{D'} \rsz \,,
\ee
where we used $[U_D,{\cal O}_{D'}]=0$. Hence states $|\Psi\rangle, \rsz$ are indistinguishable by measurements in $J(D')$ and the excitation created by $U_D$ in $D$ is invisible in $J(D')$.

\subsubsection{Comments on the split property}\label{sec2.2.2}

More generally we would like to know whether it is possible to independently specify the quantum state in space-like separated regions. The question is non-trivial since in most quantum states these regions will be entangled. It is believed that, as long as the regions in question are separated by a finite buffer region, then the answer should be positive. This is related to the {\it split property} of quantum field theory\cite{Haag:1992hx,Roos:1970fm,Buchholz:1973bk,doplicher1984standard}.

The split property can be defined as follows: consider the causal diamond whose base is a ball $D_1$ and the corresponding operator algebra ${\cal A}_{D_1}$. Consider a slightly larger ball $D_2$, containing $D_1$, with corresponding operator algebra ${\cal A}_{D_2}$ in its causal diamond. The split property is satisfied if we can find a type I von Neumann algebra of operators ${\cal N}$ such that ${\cal A}_{D_1} \subset {\cal N} \subset {\cal A}_{D_2}$.  It has been shown that quantum field theories with a reasonable thermodynamic behavior, as expressed in terms of nuclearity conditions (see \cite{Haag:1992hx} for an introduction), satisfy the split property. Using the algebra ${\cal N}$ we can have strict localization of quantum information which is completely inaccessible from $J(D_2')$.

Equivalently, the split property can be defined by the existence of state $|{\cal \phi}\rangle$ which is cyclic and separating for the algebra ${\cal A}_{D_1\cup D_2'}$ and such that
\be
\label{splitvac}
\langle {\cal \phi}| a \,b |{\cal \phi}\rangle  = \langle 0| a |0\rangle \langle 0| b |0 \rangle\qquad \forall \,\,a\in {\cal A}_{D_1}, b\in {\cal A}_{D_2'} \,,
\ee
where $|0\rangle$ is the Minkowski vacuum and $D_2'$ denotes the complement of $D_2$. In the state $|\phi\rangle$ the mutual information between regions $D_1$ and $D_2'$ is vanishing. Such a state is not uniquely defined, since for any unitary $U\in {\cal A}_{(D_1'\cap D_2)}$ a state of the form $U |\phi\rangle$ will also satisfy \eqref{splitvac}. 

Starting with a split state $|\phi\rangle$ we can construct more general states by exciting the two regions $D_1$ and $D_2'$ acting with localized operators in the corresponding algebras. Since there is no entanglement between $D_1$ and $D_2'$ in the split state $|\phi\rangle$ the two algebras act independently and we can arbitrarily approximate an excited state in $D_1$ and another state in $D_2'$.

An interesting question is to estimate the energy of a split state\footnote{Since the split state is not unique, a reasonable question might be finding the lowest possible expectation value for the energy of a split state.}. We do not expect a split state to be an energy eigenstate, so in general it will have non-vanishing energy variance. Here we provide some very heuristic arguments about the expectation value of the energy. As a starting point, let us consider a CFT on ${\mathbb R}^{1,d-1}$ with coordinates $x^0,x^1,...,x^{d-1}$. We define two regions to be the causal domains of two slightly displaced Rindler wedges with bases $x^0=0,x^1 < -\epsilon$ and $x^0=0,x^1 > \epsilon$ respectively. At $x^0=0$ the two wedges are separated by the buffer region $-\epsilon<x^1<\epsilon$. In this case the total energy of the split state will be infinite due to the infinite planar extension of the regions in the transverse directions. However, we expect to have a finite energy per unit area ${\cal E}$. Since we are dealing with a CFT then the only scale in the problem is the size $\epsilon$ of the buffer region. Hence by dimensional analysis the energy per unit area will scale like
$  
{\cal E} = \frac{s}{\epsilon^{d-1}}
$
where $s$ is a constant depending on the CFT. If we now consider a more general compact region $D_1$ of typical size $R$, which is separated by a small buffer region of typical size $\epsilon$ from $D_2'$ then we would expect that a split state with respect to $D_1,D_2'$ will have energy which in the $\epsilon\rightarrow 0 $ limit will scale like
\be
\label{energysplit}
E = s\frac{ \,\,\,\,A_{(\partial D_1)} }{\epsilon^{d-1}} + O(\frac{\epsilon}{ R}) \,,
\ee
where $A_{(\partial D_1)}$ is the area of the boundary of $D_1$. This is a heuristic estimate and it would be interesting to investigate it more carefully.
 
As mentioned above, this is the expectation value of the energy and it would be interesting to understand the spectral decomposition of a split state in the energy basis. Notice that a split state does not respect the Reeh-Schlieder property with respect to the algebra ${\cal A}_{D_1}$\footnote{Since there is no entanglement between $D_1$ and $D_2'$ we cannot create excitations in region $D_2'$ by acting with operators in $D_1$.}. This implies in particular that the split state should have non-compact support in energy, since otherwise the Reeh-Schlieder property would have to hold for $D_1$, see for example \cite{Witten:2018zxz}.

\subsubsection{Subtleties with gauge invariance}

Consider U(1) gauge theory minimally coupled to a charged scalar with Lagrangian 
${\cal L} = -\frac{1}{ 4}F_{\mu\nu}F^{\mu\nu}-(D_\mu\phi)^* D^\mu \phi\,\,,\,\,D_\mu \phi = \partial_\mu \phi -i g A_\mu \phi$. The system has $U(1)$ gauge invariance 
$A_\mu \rightarrow A_\mu + \partial_\mu \Lambda$, $\phi \rightarrow e^{i g\Lambda} \phi$.
The dynamical equations are
\be
\begin{split}
\label{eqsqed}
& \partial^\nu F_{\mu\nu} = i g (\phi \partial_\mu \phi^* - \phi^* \partial_\mu \phi) -2 g^2 A_\mu \phi^* \phi \cr
&\Box \phi = ig(\partial_\mu A^\mu)\phi + 2i g A^\mu \partial_\mu \phi + g^2 A_\mu A^\mu \phi \,.
\end{split}
\ee
In this case the initial data are ${\cal C} = \{A_\mu(t=0,x),\partial_0 A_\mu(t=0,x),\phi(t=0,x),\partial_0 \phi(t=0,x)\}$. Here we encounter the subtleties mentioned for gauge systems: initial data related by a gauge transformation are physically equivalent and initial data are admissible (i.e. lead to a solution) only if the obey a constraint, the Gauss law, which is the $\mu=0$ component of the first equation in \eqref{eqsqed}
\be
\label{qedconstraint}
\partial^i(\partial_0 A_i - \partial_i A_0) = i g(\phi \partial_0 \phi^* -\phi^* \partial_0 \phi) -2g^2
A_0 \phi^* \phi \,.
\ee
We now revisit the two properties mentioned in subsection \ref{subsecclass}. The fact that the dynamical part of \eqref{eqsqed} obey condition $B$ follows from general properties of hyperbolic equations of this type. Let us now examine question $A$ in this theory. From \eqref{qedconstraint} we see that if we try to deform the initial data in region $D$, then we may be forced to change them in $D'$ too. For example if we turn on a profile for the scalar in region $D$ with total non-zero charge, then the gauge field has to be turned on in region $D'$. 
The Gauss law constraint \eqref{qedconstraint} is of the familiar form $\nabla \cdot \vec{E} = \rho$. This imposes the constraint that $\oint_{\partial D} \vec{E}\cdot d\vec{S} = Q_{D}$. 

However it is clear that once we make sure that the initial data in $\D'$ are compatible with the Gauss constraint from the total charge $Q_{D}$ enclosed in $D$, there are many ways of rearranging the initial data in region $D$ keeping those in $D'$ fixed. In other words there are deformations of the constraint equation \eqref{qedconstraint}, which are not gauge-equivalent, and  which have compact support localized in $D$. This means that theory under consideration obeys condition $A$.

Moving on to the quantum theory, we can consider $U(1)$ gauge theory weakly coupled to matter. As in the classical theory the total charge $Q$ enclosed in a region can be measured on its boundary and the total charge of the entire state can be measured at space-like infinity. At the quantum level we can get information not only about the expectation value of the charge but all the higher moments 
\be
\label{chargespectral}
\langle \Psi| Q^n |\Psi\rangle\qquad,\qquad n=1,2,....
\ee

To proceed it is useful to consider observables in this theory. Physical observables must be gauge invariant. In a $U(1)$ gauge theory there are several examples of such observables which are also local, for example local operators constructed out of $F_{\mu\nu}(x)$ or $\phi^*(x)\phi(x)$. Other interesting gauge invariant operators which are not completely local, but can be contained in compact regions are closed Wilson loops $e^{ig\oint_C A_\mu dx^\mu}$ or bilocals of the form $\phi^*(x)e^{ig\int_{C,x}^y A_\mu dx^\mu} \phi(y)$. All these operators are neutral and do not change the electric charge of the region $D$, if they are entirely contained in $D$. We can use such operators localized in region $D$ to construct unitaries $U_D$ which can be used to modify the state inside $D$ leaving all correlators outside invariant, as in \eqref{unitaryd}. So information can be localized in this theory if we work with neutral operators.

But what if we want to create an excitation in region $D$ which has non-zero charge? We already know from the classical problem that it will not be possible to add a charge in $D$ without affecting the exterior due to Gauss law \eqref{qedconstraint}. The same is true at the quantum level. A charged operator $\phi$ in $D$ is not gauge invariant. It can be made gauge invariant by dressing it with a Wilson line extending all the way to infinity. We can think of the Wilson line as a localized tube of electric flux ensuring that Gauss law is satisfied. It may be energetically better to smear the Wilson line in a spherically symmetric configuration. The important point is that the dressed operator $\Phi(x) = e^{i g\int_{\infty}^x A_\mu dx^\mu}\phi(x)$ is no longer a local operator, though it is gauge invariant. If we act with a unitary made out of this operator, we will modify correlators outside $D$ and \eqref{unitaryd} will fail. This means that the addition of the charge in $D$ can be detected immediately outside. This is not surprising, as the same thing is already visible at the classical level.

However, looking a bit more carefully, we run into certain surprising features of the quantum theory. Suppose we have several charged fields $\phi_i$, labeled by a flavor index $i$, with the same electric charge. We construct the corresponding dressed operators $\Phi_i(x) = e^{ig \int_{\infty}^x A_\mu dx^\mu}\phi_i(x)$, using some particular prescription for the Wilson line. These obey
\be
[Q,\Phi_i(x)] = g\, \Phi_i(x) \,,
\ee
where $Q=\int_{\mathbb S^{2}_{\infty}} *F$ is the charge operator which can be measured at space-like infinity. 
Suppose the point $x=0$ is inside $D$. We create a charged excitation of type $i$ in region $D$ by acting on $|0\rangle$ with a unitary $U_i = e^{i \epsilon \Phi_i(0)}$. Then we study correlators in region $D'$ in the state $U_i|0\rangle$ in perturbation theory. Consider a correlator of $Q$ and $\Phi_j(x)$ in region $D'$.
\be
\langle 0| U_i^\dagger \Phi_j(x) Q U_i |0 \rangle  = \langle 0| \Phi_j(x) |0\rangle + i \epsilon \langle 0| [\Phi_j(x) Q,\Phi_i(0)] |0\rangle + {\cal O}(\epsilon^2) \,,
\ee
where to leading order in the perturbative expansion the second term is
\be
\label{chargedetect}
\langle 0| [\Phi_j(x) Q,\Phi_i(0)] |0\rangle =g \langle 0|\Phi_j(x) \Phi_i(0) |0 \rangle \propto \delta_{ij} \,.
\ee
Hence by measuring correlators of all $\phi_j(x)$ and $Q$ in $\overline{\cal D}$ it seems that in the quantum theory we can detect not only the presence of a charge in $D$, which is expected by Gauss's law already at the classical level, but we can even identify the flavor of the charged particle, i.e. the value of the index i in the interior of $D$. A similar argument in the gravitational case was discussed in \cite{Papadodimas:2013jku, Papadodimas:2017qit} for black hole states and in \cite{Chowdhury:2020hse} around empty AdS.

The reason we were able to get information beyond the total charge in $D$ is that in the vacuum the fields have non-trivial entanglement, on which the non-vanishing 2-point function \eqref{chargedetect} depends. When we act with the unitary containing the Wilson line, the Wilson line disturbs the pattern of entanglement in such a way that it breaks the symmetry between the fields $\phi_i$ and we can detect from $D'$ the flavor of the excitation in $D$

This suggests a way to avoid the issue and succeed in hiding the flavor of charge in $D$: we start with the analogue of a split state in the U(1) gauge theory, see the discussion in \cite{Donnelly:2017jcd}, and then create the charged excitation in $D$ by acting with the same unitary. In that case there is no entanglement bewtween $D$ and $D'$ in the matter sector and hence \eqref{chargedetect} will vanish making it impossible to tell from measurements in $D'$ what is the type of charged excitation in $D$.\footnote{A more mundane way to hide the charge is to add "screening charges" in the buffer region, but here we want to discuss how information can be localized even though a Wilson line extends all the way to infinity.} This requires creating the charged excitation on top of a split state, which has typical energy scaling like \eqref{energysplit}, rather than the ground state.

\subsection{Classical and Quantum Gravity}

First we notice that in non-perturbative quantum gravity we do not expect to be able to localize information in space: holography and AdS/CFT suggest that the fundamental degrees of freedom in quantum gravity are not local, but rather lie at the boundary. Moreover there is strong evidence that an ingredient towards the resolution of the black hole information paradox is that the naive factorization of the Hilbert space in space-like separated subregions may not be true in the underlying theory of quantum gravity.

On the other hand at the classical level in General Relativity we do have an exact notion of locality and information can be localized, as we will discuss below. An interesting question, which is the main focus of this paper, is to understand the fate of locality at the level of perturbative quantum gravity.

\subsubsection{On the initial value problem of general relativity}

In General Relativity the initial value problem is formulated by starting with a spacelike slice $\Sigma$ and specifying the data ${\cal C} = (h_{ab},K_{ab})$ where $h_{ab}$ is the intrinsic metric and $K_{ab}$ the extrinsic curvature of $\Sigma$. If we have matter then the values of the fields and their normal derivatives need to be specified.  Initial data related by spatial diffeomorphisms on the slice $\Sigma$ are gauge-equivalent and have to be physically identified. In general relativity there is one more subtlety: even if we have two initial data on the slice $\Sigma$ which are not related by a spatial diffeomorphism, they may still correspond to the same physical solution in space-time. This is related to the freedom of choosing the initial slice $\Sigma$ in    space-time and diffeomorphism invariance in full space-time.

Admissible initial data, which can be extended into a solution of the Einstein equations must obey the following constraints
\be
\label{grcon1}
\qquad R+(K_a^a)^2-K_{ab}K^{ab} = 16\pi G \rho
\ee
\be
\label{grcon2}
\nabla^aK_{ab}-\nabla_b K_c^c = -8 \pi G J_b \,,
\ee
where $R$ is the Ricci scalar of $h_{ab}$ on $\Sigma$, the covariant derivatives are with respect to $h_{ab}$ on $\Sigma$, $n^a$ is the unit normal to $\Sigma$ and $\rho = T_{ab}n^a n^b$ and $J_b= -h_b^c T_{ca}n^a$.

We now want to address the question of localization of information in classical general relativity, as formulated in subsection \ref{subsecclass}. A theorem, see for example \cite{Hawking:1973uf, Wald:1984rg}, settles question $B$ for pure general relativity: if we have two admissible initial data which agree, up to spatial diffeomorphism, on a part $D'$ of $\Sigma$, then the corresponding solutions will agree, up to a space-time diff, on the development of $D'$. This continues to be true in the presence of matter provided certain reasonable conditions are satisfied. This shows that in general relativity signals propagate at most at the speed of light: if we modify the initial data only in the region $D$, then the signals will propagate in the causal future of $D$.

Then we come to question $A$, that of localizing information on compact regions on $\Sigma$: to what extent is it possible to find two initial data  satisfying the constraints \eqref{grcon1}, \eqref{grcon2}, which agree on $D'$ but differ on $D$?\footnote{Here we need to keep in mind that even if the initial data differ on $D$ they may correspond to the same solution in space-time, as they may correspond to two different choices of the slice $\Sigma$ in the same space-time solution.} The equations \eqref{grcon1} and \eqref{grcon2} are 
non-linear and of elliptic nature, though underdetermined. Understanding the space of solutions of the constraint equations is an interesting problem which has been studied extensively in the literature. Here we summarize some relevant points:

\begin{enumerate}
    \item {\bf Gravitational Gauss law:} in asymptotically flat or AdS space-times, the energy and other conserved charges are defined at space-like infinity. The constraints of general relativity relate these asymptotic charges to contributions from excitations in the interior of space-time. For example, in the Newtonian limit the constraint equations reduce to the gravitational analogue of Gauss's law
    $$
    \Box\phi =4 \pi G\rho \,.
    $$
    As in electromagnetism this implies that the initial data in region $D'$ know about the total mass enclosed in $D$. 
    
\item { \bf Existence of localized deformations:} it is possible to find many solutions of the constraint equations which look the same in the domain $D'$ but differ on $D$. For example, if we restrict our attention to spherically symmetric solutions, Birkhoff's theorem implies that there is a large number of solutions of \eqref{grcon1} and \eqref{grcon2} which all look like the Schwarzschild metric of mass $M$ in $D'$ but differ in $D$. Examples include static, interior, star-like geometries supported by matter or more generally spherically symmetric, time-dependent collapsing geometries of mass $M$. More generally, it has been shown \cite{Corvino:2003sp} that under reasonable conditions a compact patch $D$ of a solution of the constraints \eqref{grcon1} and \eqref{grcon2} can be glued to a boosted, Kerr solution in $D'$ of appropriate mass, angular momentum, momentum and center of mass position. The existence of a large number of solutions,  which all look exactly the same in $D'$ demonstrates that it is possible to localize information in classical general relativity.

\item {\bf Comments on the vacuum:} For asymptotically AdS geometries, if a solution looks like empty AdS in $D'$\footnote{Here we assume that $D$ is compact so $D'$ includes the region near space-like infinity.} then it is guaranteed to be empty AdS in $D$ as well. In other words, starting with the vacuum it is not possible to modify the initial data in $D$ into a new solution, without at the same time modifying the solution in $D'$.

\end{enumerate}

\subsubsection{Diff-invariant observables in classical GR}

We now consider the question of defining local diff-invariant observables in gravity. This is a long-standing problem which is subtle even at the classical level. Let us consider general relativity, possibly coupled to other fields, defined with certain asymptotic boundary conditions at infinity (for example asymptotically flat or AdS) or on a closed manifold of fixed topology. We denote by $\overline{\cal X}$ the space of solutions of the equations of motion, in any possible coordinate system. On this space we have the action of the group Diff of diffeomorphisms\footnote{If the space-time is non-compact along space we only consider {\it small} diffeomorphism, i.e. those which become trivial fast enough at infinity.}. Solutions related by a diffeomorphism are physically identified and we introduce
\be
{\cal X} = \overline{\cal X}/{\rm Diff} \,.
\ee
We can think of a diff-invariant observable as a function which has definite values on points of ${\cal X}$. However, we do not demand an observable to be necessarily defined on the entire space of solutions ${\cal X}$. Instead we will allow observables to possibly have a limited domain of definition. Hence a diff-invariant observable is a map
\be
\label{diffobs}
A: U \subset {\cal X} \rightarrow {\mathbb R} \,,
\ee
where $U$ is an open subset of ${\cal X}$. Such observables can also be expressed as functions on $\overline{\cal X}$ which must obey $\overline{A}(s) = \overline{A }(f_*s)$, where $s$ denotes a solution in some coordinate system and $f_*$ the action of a diffeomorphism.

In order for a diff-invariant observable to be local we need to impose additional conditions. To formulate these conditions it is useful to introduce the Peierls bracket $\{A,B\}$ between two diff-invariant observables \cite{Peierls:1952cb}, which is a covariant generalization of the Poisson bracket. To compute the value of $\{A,B\}$ we consider a modification of the action as $S \rightarrow S+ \epsilon A$ and compute the difference of the first order change of observable $B$ on the perturbed solutions with advanced $(+)$ and retarded $(-)$ boundary conditions. The Peierls bracket is defined as\footnote{The first order solutions are not unique due to diffeomorphism invariance, however the ambiguity drops out when computing the change of the diff-invariant observable $B$.}
\be
\label{peierls}
\{A,B\} = \delta_A^- B - \delta_A^+B \,.
\ee
It can be shown that the Peierls bracket has similar properties as the Poisson bracket, for example linearity, antisymmetry and the Jacobi identity, and in fact coincides with the Poisson bracket if a Hamiltonian formalism is introduced. One of the advantages of the Peierls bracket is that we do not need to pass to the Hamiltonian formalism which is somewhat complicated due to the constraints. Notice that to define the Peierls bracket of two observables $A,B$ they must have a common domain of definition on ${\cal X}$ and the bracket will be generally a non-trivial function on this overlap.

We would like to define diff-invariant observables which can be associated to {\it points} in space-time with the property that if two such observables are associated to space-like separated points the corresponding Peierls bracket must vanish. The difficulty in doing this is that in order to define an observable we need to define it at least in an open neighborhood around a state as in \eqref{diffobs}, so we need some prescription for following "the same point", on which the candidate diff-invariant observable will be localized, as we move on the space of solutions ${\cal X}$. General covariance implies that there is no canonical way to keep track of the point as we change the state.

If the space-time has a well-defined boundary we can find prescriptions which select a point in space-time for each solution in ${\cal X}$ {\it relationally} with respect to the boundary. For example in AdS we can define a diff-invariant observable which {\it seems} to be localized at a point by considering a radial geodesic at right angle from a specific point on the boundary, moving a fixed regularized distance along it and measuring the value of a scalar quantity, for example a scalar field or a scalar combination of the curvature, at the resulting point. This gives a map from the space of solutions ${\cal X}$ to ${\mathbb R}$, so it is a diff-invariant observable. Notice however that the location of the resulting point depends on the entire geometry along the geodesic, all the way from the boundary. Changing the metric anywhere along this geodesic will move the resulting point. Hence the value of the observable will not strictly depend on local data near the point. Similarly, if we act with one of the asymptotic symmetries the boundary starting point will move and also the resulting bulk point will move. This implies that the Peierls brackets of this candidate observable with the boundary charges, or other observables along the geodesic will be non-zero, even though these regions are space-like separated. Hence this relational observable is not really local.

Another way to define diff-invariant observables is to consider a complete gauge fixing scheme. Then observables in the particular gauge labeled by a space-time coordinates are automatically diff-invariant. However they will generally have non-local Peierls brackets, since the assignment of a coordinate value to a point in space-time in the particular gauge, will generally depend on the solution everywhere.

Additional difficulties arise  in space-times without boundaries, for example in de Sitter space. A boundary is an (asymptotic) part of the spacetime where gravity is not dynamical anymore. This is why we can for example anchor geodesics to the boundary, and define relational diff-invariant observables. Without a boundary, there is no part of the space-time where gravity is turned off, and consequently no place to anchor geodesics.

\subsubsection{State-dressed observables}\label{sdob}

If we consider a solution that is sufficiently complicated it is possible to specify points, and hence define local diff-invariant observables, by using {\it features of the state}.  We emphasize that these observables will not have all the desired properties over the entire space of solutions ${\cal X}$, so these observables have certain aspects of state-dependence as discussed around \eqref{diffobs}.  One approach based on this idea was studied by DeWitt \cite{DeWitt:1962cg}, building on \cite{PhysRev.111.1182,Bergmann:1960wb}. For a $D$-dimensional space-time we start by identifying $D$ scalar quantities $Z^a,a=1,...,D$. These can be combinations of curvature invariants and other scalars formed by the fields of the theory. We could try to fix a coordinate system by using these $D$-scalars as coordinates.  We can use this intuition to introduce candidate local diff-invariant observables of the form
\be
\label{dewitt}
\phi(Z_0^a) = \int d^Dx \,\phi(x) \,\delta^{D}(Z^a(x)- Z^a_0) \,{\rm det}{\partial Z \over \partial x} \,.
\ee
Here $Z^a$ are the $D$ scalar quantities introduced above and $\phi$ is any other scalar combination of the fundamental fields of the theory. Similar constructions can be done for fields with tensor indices. 

Some comments are in order:

\begin{enumerate}

\item For a general space-time which is in-homogenous, and for certain choices of the values $Z_0^a$, the delta function in \eqref{dewitt} will click on a finite number of points, so the quantity above is well-defined and finite. In symmetric space-times it will either not click at all, hence the observable will be zero, or an infinite number of times so the observable will be ill-defined. This shows that \eqref{dewitt} is a quantity which is defined only on part of the phase space. This is in accordance with our expectation that state-dressed observables have to be  state-dependent \eqref{diffobs}.

\item Suppose that the observable \eqref{dewitt} is well defined on a state $s$ and a neighborhood $U$ of the space of solutions ${\cal X}$ around it. It is clear that, at least at the classical level, this observable is diff-invariant, i.e. a well defined map $\phi(Z_0^a) :U\subset {\cal X}\rightarrow {\mathbb R}$ and hence a good observable according to the definition \eqref{diffobs}.

\item One can show that under certain conditions, observables \eqref{dewitt} are also {\it local}. If we have a state $s$ on which two such observables $\phi(Z^a_A),\phi(Z^b_B)$ are well defined, with the property that the delta functions click at single points $A, B$ and that these points are space-like separated with respect to the metric of $s$, then the corresponding observables have vanishing Peierls brackets $\{\phi(Z^a_A), \phi(Z^b_B)\}=0$, see \cite{Dewitt:1999nrz} for a review. This follows from the causality properties of linearized Green's functions appearing in \eqref{peierls} around the solution $s$. Notice that if two points $A,B$ are spacelike separated on a solution $s$, then there is a small enough neighborhood of $s$ in which they remain space-like separated. Hence their Peierls bracket will vanish in this entire neighborhood.

\item This shows that, as long as we accept that observables  may be defined only locally on the phase space of solutions, it is possible to find local, diff-invariant observables in classical general relativity around states which are complicated enough. These are also the interesting states, i.e. those containing bulk observers who want to study physics in their environment. 

\item Similar ideas are useful in cosmology, where the value of a scalar field can be used as clock \cite{Page:1983uc,Kuchar:1991qf, Isham:1992ms}.

\end{enumerate}

The next question is whether it is possible to define similar observables at the quantum level. Aspects of this question were discussed in \cite{Giddings:2005id} and \cite{Marolf:2015jha}, where it was conjectured that there is a quantum version of these observables which retain their locality properties to all orders in the $\hbar$ expansion, even though they are not expected to be local at the non-perturbative level. Various difficulties are encountered at the quantum level including the question of the renormalization of the composite operators \eqref{dewitt}, establishing diffeomorphism invariance at the quantum level and the role of Poincare recurrences which will generally introduce infinite copies where the delta function will have support  \cite{Marolf:2015jha}. In this paper we provide support in favor of this conjecture by finding observables with certain similarities in spirit to \eqref{dewitt} directly in CFT language. This has the advantage that any object built directly in the CFT is by construction diff-invariant.

\subsubsection{A time-band in AdS}

We now specialize to a setup that will allow us to make contact with AdS/CFT. We consider geometries that are asymptotically AdS$_{d+1}$ and we consider a short time-band ${\cal T}_{-\epsilon,\epsilon}$ on the boundary in global coordinates, defined as the set of points $(-\epsilon,+\epsilon)\times {\mathbb S}^{d-1}\,\,,\,\,\epsilon>0$, where the first interval refers to the time coordinate $t$. Near the boundary we can select a Fefferman-Graham coordinate system where the fields, for example the metric and a scalar of mass $m^2$, have the behavior
\be
ds^2 = {dr^2 \over r^2}+r^2(-dt^2 + d\Omega_{d-1}^2) + r^{2-d}g_{\mu\nu}(r,x)\,\, dx^\mu dx^\nu\qquad g_{\mu\nu}(r,x) = g^{(0)}_{\mu\nu}(x) + g^{(2)}_{\mu\nu}(x) r^{-2} + ...  \notag  
\ee 
\be
\phi = r^{-\Delta}(\phi^{(0)}(x) + \phi^{(2)}(x) r^{-2}+...) \,,
\ee
where $x=(t,\Omega_{d-1})$ and $\Delta={d\over 2} + \sqrt{{d^2\over 4}+m^2}$. Here we consider normalizable states so the growing modes, which would be dual to sources in the CFT, are set to zero\footnote{We only assume that the sources are zero in the time band ${\cal T}$, they could be turned on in the far past in order to prepare a state.}. The Fefferman-Graham coefficients $g^{(0)}_{\mu\nu}(x), \phi^{(0)}(x)$  are diff-invariant observables and are labelled by boundary coordinates\footnote{The subleading coefficients are fixed by the equations of motion in terms of the leading ones.}.  This set of observables includes the asymptotic charges, for example the ADM Hamiltonian can be computed as
\be
\label{adsadm}
H \propto\int_{{\mathbb S}^{d-1}} d\Omega^{d-1} g_{00}^{(0)}(x) \,.
\ee
We focus on these Fefferman-Graham observables restricted in the time band ${\cal T}_{-\epsilon,\epsilon}$.
This set of observables is closed under Peierls brackets and form a Poisson algebra ${\cal A}$. Notice that in this algebra we do not include observables which would be finite distance under Poisson flow, otherwise flowing by finite distance with $H$ would take us out of the time-band, see also the discussion in \cite{Marolf:2008mf}. 

Starting with the classical theory, we ask whether we can find observables localized deep in the interior of AdS which are space-like with respect to the time-band and which have vanishing Peierls brackets with observables in the time-band algebra ${\cal A}$. These candidate observables are to be defined as in \eqref{diffobs}, in particular they need to be defined on a neighborhood $U \subset {\cal X}$ of a solution $s\in U$ and not necessarily on the entire space of solutions ${\cal X}$.
        
It is clear that observables defined relationally with respect to the boundary, or with a gauge fixing condition which makes use of the boundary, do not satisfy these conditions. Due to their gravitational Wilson lines they will have non-vanishing Peierls brackets with the Hamiltonian and other charges on the boundary \cite{Giddings:2018umg,Donnelly:2016rvo}. Such observables generally change the energy of the state, which due to the gravitational Gauss law can be measured in the time band ${\cal T}_{-\epsilon,\epsilon}$ by \eqref{adsadm}. Another point of view is that such observables identify a point in the bulk, and in particular {\it a moment in time}, relationally with respect to the boundary. Thus an infinitesimal motion in time of the starting point on the boundary is  translated via the relational prescription into an infinitesimal time motion of the corresponding bulk point. Then the Peierls bracket of the candidate bulk observable with $H$ generates time-derivatives of the point in the bulk and is non-vanishing.

The discussion of the previous subsection implies that if we start with an asymptotically AdS$_{d+1}$ solution $s$ of the bulk equations which is complicated enough, then we can define diff-invariant observables of the form \eqref{dewitt} in a neighborhood of $s$ so that they have vanishing Peierls bracket with all elements of the time-band algebra ${\cal A}$ including charges like the Hamiltonian \eqref{adsadm}. Such observables do not change the total energy of the state but instead they rearrange the energy, "absorbing" from the background solution the amount of energy they themselves create. These observables select a point in the bulk, and a moment in time, by using features of the state.

In what follows we will provide evidence that the same conclusions are true in perturbative quantum gravity. We will proceed by translating the question in CFT language and using the AdS/CFT correspondence.

\section{Holographic setup} \label{sectionAdSCFT}

In this paper, we will study the question of locality in quantum gravity in the context of the AdS/CFT correspondence. A question we would like to understand is how certain bulk subregions are encoded in the boundary CFT. There are cases where this is well understood. For example, the bulk dual of a boundary subregion is known as the entanglement wedge, which is the bulk region extending between the boundary subregions and the relevant Ryu-Takayanagi surface extending in the bulk \cite{Ryu:2006bv}. This correspondence between parts of the boundary and bulk is known as subregion-subregion duality  \cite{Czech:2012bh,Almheiri:2014lwa,Jafferis:2015del}, and it is worthwhile to mention that in general, the entanglement wedge of a boundary subregion is much larger than its causal wedge (the part of the bulk contained by lightrays shot from the causal developments of the boundary subregion).

Subregion-subregion duality and entanglement wedge reconstruction utilizes the organization and entanglement of CFT degrees of freedom organized spatially. We will be interested in rather different bulk subregions, which lie deep down in the bulk and never extend to the boundary CFT. What is the CFT dual of a causal diamond located deep near the center of AdS? The answer to this question remains elusive, and in particular it is understood that in general, these bulk regions do not correspond to the entanglement wedge of any boundary subregion. There have been previous attempts to understand the CFT mapping of such regions, see for example  \cite{Balasubramanian:2013rqa,Balasubramanian:2013lsa,Myers:2014jia,Headrick:2014eia} which attempt to assign a meaning to the entropy of a general closed codimension-2 spatial curve in AdS. Here we will follow a different approach by focusing on the algebra of single-trace operators \cite{Banerjee:2016mhh}.

We will start by reviewing some basic but relevant features of AdS/CFT, before turning to a discussion of the class of states that we will be considering throughout this paper and their salient properties.

\subsection{Gravitional states in AdS, large diffeomorphisms and asymptotic symmetries \label{sec:charges}}

We will be interested in gravitational solutions which are asymptotically AdS$_{d+1}$. We have in mind an embedding in a top-down setup with a holographic dual CFT, like $\mathcal{N}=4$ SYM at strong coupling, on ${\mathbb S}^{3}\times {\mathbb R}$ and the $N$-scaling we indicate in most of the paper refers to this theory. However for most of the discussion the details of the embedding in string theory, the extra fields, as well as the presence of a compact internal manifold are not important unless explicitly stated.

Solutions to the bulk equations of motion can be thought of as {\it states} in the dual CFT. If we think of a bulk geometry described by a Penrose diagram, the diagram really represents the entire time-history of the state. We can take the state to live at $t=0$ on a boundary Cauchy slice, and the portion of the geometry relevant to describing the state is an initial data surface given by a bulk Cauchy slice (or the Wheeler-de Witt patch associated to the boundary Cauchy slice). To view these geometries as states of the dual CFT, it is important that the bulk fields have a fall-off corresponding to normalizable modes with vanishing CFT sources.\footnote{If these states are prepared by a Euclidean path-integral \cite{Skenderis:2008dh,Botta-Cantcheff:2015sav,Marolf:2017kvq,Belin:2018fxe}, sources can be turned on in the Euclidean past which prepares the state, but it is important that they vanish as $t_E\to0$ for the geometries to be interpreted as states in the undeformed CFT.}

We want to consider semi-classical solutions with non-trivial bulk geometries, i.e. where backreaction is strong. The corresponding CFT states $\rsz$, which we take to be pure, have large energies which scale as 
\be \label{energyofstate}
\bra{\Psi_0}H\rsz \sim \mathcal{O}(N^2) \,,
\ee
and as we will see, they will generally also have an energy variance of the same order. We will also consider perturbative excitations of the quantum fields on top of the background geometry. These excitations add/subtract quantum particles which change the energy by an $O(N^0)$ amount, and whose backreaction on the geometry is thus generally small.

Geometries of this type will often be macroscopically time-dependent, such that the initial data on a bulk Cauchy slice changes as we perform time-evolution of the state. This has consequences for the variance of the energy, as we will now see. Any  state $\rsz$ can be expanded in the basis of CFT energy eigenstates as
\be \label{statedef}
\rsz =\sum_i c_i |E_i\rangle \,.
\ee
The time-dependence of the bulk geometry implies that such states will have energy variance
\be
\label{vardef}
(\Delta H)^2 \equiv \lsz H^2 \rsz - \lsz H\rsz^2 \sim \mathcal{O}(N^2) \,.
\ee
To see this, consider the inequality 
\be
{1\over 2}|\langle [H,A]\rangle| ={1\over 2} \braket{\partial_t A} \leq \Delta H \cdot\Delta A \,,
\ee
where in the first equality we assumed that the operator $A$ is not explicitly time-dependent. Then we have
\be \label{deltaHbound}
\Delta H \geq {1\over 2} \frac{\braket{\partial_t A} }{\Delta A} \sim O(N) \,,
\ee
where we have used large $N$ factorization for the operator $A$. This shows that provided there is macroscopic time-dependence (the classical vev of $A$ changes at leading order), the variance of the energy scales at least as $N^2$.\footnote{Note that if the variance is parametrically larger than $O(N^2)$, the state may no longer have a good semi-classical interpretation. An example would be a superposition of black holes of different masses.} Some bulk geometries we will consider are macroscopically time-dependent, but only inside the horizon. In this case, we cannot use the argument above, but we still expect the variance to be of order $N^2$. It is interesting to ask whether the variance is a quantity that can be extracted from the semi-classical geometry alone. In general, we expect that the quantum state of the fields in the bulk is important as well. We discuss this further in Appendix \ref{app:variance}.

There are various types of explicit constructions of states of this kind. There are states prepared by Euclidean path integral with sources for single-trace operators \cite{Skenderis:2008dh,Botta-Cantcheff:2015sav,Marolf:2017kvq,Belin:2018fxe}. These states should be interpreted as coherent states of the quantum gravitational dual, which are labelled by phase-space points corresponding to initial data\footnote{It appears that one may not construct arbitrary initial data this way, see \cite{Belin:2020zjb}. This will not affect our construction and for states prepared by a Euclidean path integral, we should simply keep in mind that we have access to a restricted class of initial data.}. There are also states prepared by a boundary state of the CFT, further evolved by some amount of Euclidean time \cite{Kourkoulou:2017zaj, Almheiri:2018ijj,Cooper:2018cmb,Miyaji:2021ktr}. The bulk interpretation of these states is that they correspond to black hole geometries with End-of-the-World branes sitting behind the horizon. This is an example where the bulk geometry is macroscopically time-dependent, but only behind the horizon. Similarly, for two-dimensional CFTs, we can construct pure states by performing the path integral over a surface of higher topology, for example half a genus-2 surface, see \cite{Marolf:2017vsk}. These geometries are also macroscopically time-dependent behind the horizon, but instead of having a brane behind the horizon, they have topology. Finally, it is worth noting that there are semi-classical geometries that also preserve supersymmetry, the most famous of which are the LLM geometries \cite{Lin:2004nb}. In these cases, one can obtain a better understanding of the dual CFT states. We will come back to these geometries in section \ref{sec:examples}.

As usual in gravity, we should identify solutions which are related by {\it small diffeomorphisms}, i.e. diffeomorphisms that vanish near the AdS boundary. There is also a class of large diffeomorphisms, which are compatible with the boundary conditions imposed in the definition of our theory of AdS gravity. This set of diffeomorphisms forms what is called the {\it asymptotic symmetry group}. In the case of AdS$_{d+1}, d\geq 3$ this is the conformal group $SO(2,d)$, while for $d=2$ it gets enhanced to the Virasoro group \cite{Brown:1986nw}. When acting on a given bulk solution these large diffeomorphisms will generally transform the geometry into a new state, which is physically distinguished from the previous one, unless of course the original state happens to be invariant under the symmetry. We will later also discuss solutions with two asymptotic boundaries, such as the eternal black hole in AdS, in which case the asymptotic symmetry group is larger. Let us now discuss the various elements of the asymptotic group/conformal group:

\begin{itemize}
    \item {\bf Time translations:} One particular class of states we will discuss are those with semiclassical time-dependence in the bulk, for example a state corresponding to the gravitational collapse of a star. In this case large diffeomorphisms corresponding to asymptotic time translations transform the state as $\rsz \rightarrow e^{-i H t}\rsz$. The initial data corresponding to $\rsz$ is not the same as that of $e^{-i H t}\rsz$. Our end goal will be to provide local operators whose gravitational dressing is done towards a feature of the state. If the state is time-dependent then we can select a moment in time by using the features of the state, as opposed to the boundary time coordinate. On the other hand if the state is static, then the only way to identify a moment in time is by dressing to the boundary. This is why it will be important for us to consider time-dependent states.

    \item {\bf SO(d) rotations:} If the state breaks $SO(d)$, then asymptotic rotations transform it to a new state. In this case we can use the features of the state to identify the angular location of a point. On the other hand, if the state is $SO(d)$ invariant it will generally not be possible and at best we can obtain an operator smeared over the bulk angular coordinates, or alternatively we can fix the angular location by dressing to the boundary.
    
    \item {\bf AdS boosts:} The Lorentzian conformal group acting on ${\mathbb S}^{d-1} \times {\mathbb R}$ has another $2d$ generators which correspond to ``boosts'' in various directions. These can be realized as $d$ non-independent copies of an $SL(2,\mathbb{R})$ algebra, see for example \cite{Freivogel:2011xc}. Any state with finite energy cannot be annihilated by Hermitian combinations of these generators, which we show in Appendix \ref{appendixboosts}. The only state which is annihilated by these generators is the global vacuum and any other state will necessarily transform under the action of these boosts\footnote{States with infinite energy like the AdS-Rindler vacuum could also potentially be annihilated by some boost generators.}. Therefore, in any non-trivial state, we can fix the radial position of an operator without referring to the boundary.

\end{itemize}

In a top-down setup, the gravity dual may have an internal manifold, like the $S^5$ in the context of $\mathcal{N}=4$ SYM. In such cases, we would need to break the R-symmetry to localize a bulk operator in the internal space, which would be a straightforward generalization of our construction.

\subsection{Locality in AdS}   

We are now ready to discuss locality in quantum gravity with asymptotically AdS boundary conditions. We would like to understand whether one can define local observables and whether we can localize information deep in the center of the AdS.

The presence of the AdS boundary allows us to define one natural class of diff-invariant observables: The fields in AdS can be expanded in a Fefferman-Graham expansion. The coefficients of this expansion are themselves diff-invariant observables, which are dressed to the boundary since the Fefferman-Graham gauge is chosen with respect to the boundary. Let us call these observables FG-observables. For example, the AdM Hamiltonian is one particular observable in this class. In perturbative quantum gravity, we can also consider the expectation values of these observables as well as their higher-point correlation functions. As we will discuss below, if we want to stay within the regime which can be described by semi-classical gravity we may need to restrict the complexity of the correlators (for example the number of operator insertions in the correlation function). We emphasize again that all these observables are dressed with respect to the boundary. In particular, they will generally not commute with the Hamiltonian or the other charges described in the previous section.

The question we would like to address is the following. If we start with a state with a semi-classical geometric description, is there a way to modify the state in the interior of AdS, without modifying any of the correlators of FG-observables localized in a short time-band of the boundary? If the answer is yes, this means we can localize information since an observer living near the boundary will have no way to know whether or not we modified the state. Rather than trying to come up with bulk objects that achieve this goal, we will address this question directly in the dual CFT. This has the following advantage: any object built out of CFT degrees of freedom is necessarily diff-invariant and non-perturbatively well defined. Provided the object acts in the right away, we can be assured that the construction is fully consistent.

\subsection{The CFT description and the time band algebra}

Consider a large $N$ holographic CFT which is dual to semi-classical general relativity coupled to matter fields. In the large $N$ limit, we can define the algebra ${\cal A}$ generated by single-trace operators in a time-band ${\cal D}_{t_1,t_2}$, where we allow products of single-trace operators where the number of factors is arbitrary but scales like $O(N^0)$.\footnote{Notice that at finite $N$ the true algebra in a time-band would be the same as the full CFT algebra. However, in the large $N$ limit, a natural hierarchy emerges between "small products" of single-trace operators and the rest of the algebra, which allows us to consider the notion of a time-band algebra. At large but finite $N$, this is not a true algebra, since it does not close, but it defines a particular set of operators, which are a subset of all of the operators in the CFT.}
This was originally discussed in \cite{Banerjee:2016mhh}, inspired by the earlier work \cite{Papadodimas:2012aq, Papadodimas:2013wnh, Papadodimas:2013jku}. In \cite{Banerjee:2016mhh} it was proposed that the algebra ${\cal A}$ can be thought of as being dual to the causal wedge of the region ${\cal D}_{t_1,t_2}$ in the bulk (see Fig. \ref{fig1}). This picture also suggests that the algebra ${\cal A}$ has a commutant which can be idenfitied with a spacelike-separated causal diamond in the interior. Algebras of this type have received attention recently \cite{Leutheusser:2021qhd, Leutheusser:2021frk, Witten:2021unn,Chandrasekaran:2022eqq,Leutheusser:2022bgi}.

The work \cite{Banerjee:2016mhh} studied this setup for states which are small perturbations around the AdS vacuum. The geometry of AdS is homogeneous and featureless since it is a maximally symmetric space. As already discussed in the previous section, this makes the definition of local diff-invariant observables challenging. We would like to revisit the time-band algebra, this time in cases where the bulk state has features, which in particular are time-dependent. This means the state must be highly excited as can be seen for example from its energy \eqref{energyofstate}.

At infinite $N$ the problem can be understood in terms of QFT on a curved and in general time-dependent background. In particular, gravitational backreaction of the quantum fields can be ignored and one does not need to talk about gravitational dressing, which is a form of backreaction. In this case, the existence of the commutant is obvious because we are in a QFT situation. Note that if the Hamiltonian (which is always an element of the time band algebra) is normalized appropriately\footnote{A useful normalization is $h={1\over N}(H-\lsz H\rsz)$, which ensures that $\lsz h^2 \rsz \sim O(N^0)$.}, its commutator with the other single-trace operators is suppressed by $1/N$ and thus vanishes when $N$ is infinite.

At the level of $1/N$ corrections, the existence of the commutant is less obvious. Backreaction must now be taken into account and the gravitational Gauss law can spoil the commutator between $H$ and the other operators of the time-band algebra. For example, the standard way to write bulk fields in terms of CFT operators is the HKLL construction \cite{Banks:1998dd, Bena:1999jv,Hamilton:2005ju,Hamilton:2006az, Hamilton:2006fh, Hamilton:2007wj,Heemskerk:2012mn}
\be
\label{hkll}
\hkll(t,r,\Omega) = \int_{\rm bdry} dt' \,d\Omega'_{d-1} K(t,r,\Omega; t',\Omega') {\cal O}(t',\Omega') \,,
\ee
where $K$ is related to a Green's function of the Klein-Gordon operator on the appropriate bulk geometry. This operator is defined purely within the CFT so it is manifestly diff-invariant. To leading order at large $N$, it acts as a bulk field and commutes with other bulk fields at spacelike separation. Notice however that in order to define the kernel $K$ we have to choose a coordinate system in the bulk. As we already mentioned, this gauge choice is defined by making use of the asymptotic boundary, and an HKLL operator is thus dressed to the boundary. Because of this, the commutator between an HKLL operator and the Hamiltonian will not vanish at subleading orders in the $1/N$ expansion.

The physical origin of this effect is the gravitational Gauss law: acting with \eqref{hkll} will generally create or destroy a particle in the bulk, thus changing the energy of the state, which can be immediately measured at spacelike infinity by $H$. One can try to correct the HKLL operators at higher orders in $1/N$ by mixing it with other single- and multi-trace operators, see \cite{Kabat:2011rz,Kabat:2012av,Heemskerk:2012mn}, but  the commutator with the Hamiltonian is universal and generally cannot be removed in this way. It is also possible to think about the dressing in terms of (smeared) gravitational Wilson lines connecting the bulk operator to the boundary, which make it diff-invariant at the price of making it non-local \cite{Anand:2017dav,Castro:2018srf,Chen:2019hdv,Giddings:2019hjc}. The commutator with $H$ is nonzero because $H$ picks up the contribution of the Wilson line.

This raises the question of whether the algebra ${\cal A}$ still has a commutant at subleading orders in $1/N$. The main goal of this paper is to provide evidence for the existence of such a commutant. We will do so by identifying a class of operators that are gravitationally dressed with respect to {\it features of the state}, rather than dressed to the boundary. In particular, these operators will have vanishing commutators with the Hamiltonian, to all orders in $1/N$. In this paper, we will focus mostly on ensuring that bulk operators have a vanishing commutator with the Hamiltonian (and the other charges), but it would be important to extend our construction to all single-trace operators in ${\cal D}_{t_1,t_2}$. We given an alternative argument for the existence of a commutatant to all orders in $1/N$ in section \ref{otconop}.

The existence of a commutant for ${\cal A}$ in $1/N$ perturbation theory would imply that  information can be localized in regions of the bulk and is not visible from the boundary at the level of perturbative quantum gravity\footnote{See  \cite{Marolf:2008mf,Donnelly:2017jcd,Bousso:2017xyo,Donnelly:2018nbv,Jacobson:2019gnm,Giddings:2020usy,Chowdhury:2020hse, Chowdhury:2021nxw, Giddings:2021khn} for other discussions of localization of information in perturbative quantum gravity, with varying conclusions.}. We are now ready to formulate the concrete goal that we will achieve in this paper.

\subsection{Formulating the main goal \label{sec:ourgoal}}

Our goal is to improve the locality properties of \eqref{hkll} by moving the gravitational dressing from the boundary to the state. From a technical point of view, we will find CFT operators $\drop$ which obey two properties:

\begin{enumerate}
\item  $[Q_i,\drop]=0$ to all orders in $1/N$, for all asymptotic charges $Q_i\in SO(2,d)$.
\item The correlators of $\drop$ agree with those of $\Phi_{\textrm{HKLL}}$ to leading order in the large $N$ expansion, on the code subspace of $\rsz$.
\end{enumerate}
In taking the large $N$ limit  it is important to 
track how various effects scale with $N$. As we will see, our new operators $\drop$ have vanishing commutator with $Q_i$ to all orders in the $1/N$ expansion, but have a non-vanishing commutator at the level of $e^{-N^2}$ corrections. 

In what follows we will first focus on ensuring a vanishing commutator of $\hat{\Phi}$ with the Hamiltonian $H$  to all orders in $1/N$ and then discuss the generalization to the other charges in $SO(2,d)$.
    
As we will see, our construction will not work for $\rsz=\ket{0}$. Technically, this is because the vacuum does not comply with the properties \eqref{energyofstate} and \eqref{vardef}. Physically, it is because the AdS vacuum has no feature that we can use to attach the dressing of our local operator. Note that this is in line with the results of \cite{Chowdhury:2021nxw}, where a protocol to reconstruct the bulk state from correlators in the time-band was discussed.

\subsection{Time-shifted states and return probability}

We will now present the main technical tool that will enable us to define state-dressed operators: the return probability. Let us start with a state $\rsz$ satisfying the properties \eqref{energyofstate} and \eqref{vardef}. We define the following one-parameter family of states
\be
|\Psi_T\rangle = e^{-i T H} \rsz \qquad T\in {\mathbb R} \,.
\ee
In the bulk, the states $|\Psi_T\rangle$ are related to $\rsz$ by a large diffeomorphism, i.e. one that does not vanish near the boundary and induces a boundary time-translation. It is important to emphasize that they are {\it different} quantum states, even though they are related by a symmetry. If we think about the  phase space of gravity in AdS, the family of states correspond to different phase space points, just like a particle moves on phase space as a function of time in classical mechanics. From the bulk perspective, if $\rsz$ was a coherent state, we can also think of $|\Psi_T\rangle$ as coherent states. 

We would now like to consider the overlap of such states. In particular, we would like to study the overlap
\be \label{overlapdef}
\lsz \Psi_T\rangle \,.
\ee
Thinking of these states as coherent states is useful to gain intuition about such overlaps. For the simple harmonic oscillator, the overlap of two coherent states is
\be
\braket{\alpha|\beta}=e^{-\frac{1}{\hbar} f(\alpha,\beta) } \,,
\ee
for a very simple quadratic function $f$. For states on the gravitational phase space, recalling that $\hbar\sim G_N \sim 1/N^2$, we expect
\be
\label{decaybulk}
\lsz\Psi_T\rangle = e^{-N^2 f_0(T)} \,,
\ee
for a function $f_0$ whose real part is positive. In the gravitational setting, it is not straightforward to directly compute $f_0(T)$ from the phase space information, see \cite{Papadodimas:2015jra} for a discussion on nearby states. There is a general way to compute $f_0(T)$ based on a Euclidean preparation of the states \cite{Belin:2018fxe}, but it requires some effort (in particular solving the non-linear Einstein equations). The computation of $f_0(T)$ directly from the information on an initial data slice, which specifies the point on phase-space, is an interesting problem.\footnote{Similarly, we do not know of a gravitational argument that guarantees that the real part of $f_0(T)$ is positive, which must be the case if the geometries have a state interpretation in the dual CFT. We comment on this further in the discussion.}

It is also instructive to think about the overlap from a microscopic point of view. In the CFT, the overlap is given by
\be
\label{decaybdry}
\lsz\Psi_T\rangle = \sum_i |c_i|^2 e^{-i T E_i } \,.
\ee
Note that there are $e^{S(E)}$ terms here, each of size $e^{-S(E)}$. The suppression \eqref{decaybulk} must therefore come from the summation over a large number of phases.

If the bulk state has no periodicities in time, we expect the real part of $f_0(T)$ to increase as we increase $T$. However, this increase will not continue forever. We will shortly give an estimate of the time-average of \eqref{decaybdry}, and argue that the decay will saturate at some point. Physically, the non-trivial overlaps \eqref{decaybdry}  imply that it is not correct to think that all the states $|\Psi_T\rangle$ are independent, see also \cite{Papadodimas:2015xma, Papadodimas:2015jra,Chakravarty:2020wdm} for related discussions. In particular, even if the bulk state is not macroscopically periodic, there will still be a microscopic periodicity of the state due to Poincare recurrences, that will happen at very large $T\sim \mathcal{O}(e^{e^{N^2}})$. Throughout this paper, we will be interested in much earlier time scales so it will be sufficient for us to treat the states $|\Psi_T\rangle$ as quasi-orthogonal since all overlaps will be exponentially small.

We will also need to define the notion of the code subspace. Starting with the state $\rsz$ we define the  code subspace as
\be
\label{code0}
{\cal H}_{ 0} = {\rm span}\{\rsz, {\cal O}(t,\Omega) |\Psi_0\rangle,...,{\cal O}_1(t_1,\Omega_1)...{\cal O}_n(t_n,\Omega_n)\rsz\} \,,
\ee
\be
\label{codet}
{\cal H}_{T}= {\rm span}\{|\Psi\rangle_T,   {\cal O}(t,\Omega) |\Psi_T\rangle,...,{\cal O}_1(t_1,\Omega_1)...{\cal O}_n(t_n,\Omega_n)|\Psi_T\rangle\} \,,
\ee
with the corresponding projector $P_T$. The projectors $P_0$ and $P_T$ are simply related by time-evolution, i.e. we have
\begin{equation}
 P_T = e^{-i TH } P_0 e^{i TH}   \,,
\end{equation}
and in particular, we emphasize again that  $P_T \neq P_0$. In what follows, it will be convenient to work with real quantities rather than the overlap \eqref{overlapdef}, and we are now ready to define the return probability.

\subsection{The return probability \label{subsec:ra}}

We now ready to examine the $T$-dependence of the overlap \eqref{decaybdry} in more detail. As explained above, it is more convenient to work with a real quantity so let us define \textit{the return probability}
\be
\label{radef}
R(T) := |\lsz e^{-i T H} \rsz|^2 \,.
\ee
It is similar to the spectral form factor  (the two coincide when $\rsz=\ket{\textrm{TFD}}$ and $H=H_L+H_R$). Recently, the spectral form factor has been extensively discussed in connection to the black hole information paradox and quantum chaos, see for example \cite{Cotler:2016fpe}. The time-scales of interest in that context are again late times such as $t\sim e^{N^2}$ (note this is much shorter than the Poincare recurrence time which is doubly exponential). Here again, we will be interested in much earlier time-scales.

In general, it is difficult to compute \eqref{radef}.  As we mentioned above, the overlaps can be computed from time-shifted coherent states in gravity but the best known technology to do so uses the Euclidean path integral and involves solving the non-linear Einstein's equations. Nevertheless, we can compute the very early time dependence using large $N$ factorization. We present this calculation in Appendix \ref{appendixdecay}. At early times, we have

\be
\label{earlyr}
R(T) = e^{-(\Delta H)^2 T^2} \,,
\ee
which is generally valid for times up to $T\sim O(N^{-1})$. For the purposes of this paper, we want to understand how the return probability behaves at time-scales $T\sim\mathcal{O}(1)$. Here, the decay does not follow from large $N$ factorization and it is in general not an easy task to compute it. 

In Appendix \ref{appendixdecay}, we review that for the TFD state, the return probability (which is the spectral form factor) decays as
\be
\label{decay}
R_{\textrm{TFD}}(T)=e^{-N^2 f_{\rm TFD}(T)},
\ee
where $f_{\rm TFD}(T)$ is $O(N^0)$ and for early times $T \sim O(N^0)\ll \beta$ behaves like $f_{\rm TFD}(T)\approx \alpha T^2$, where $\alpha$ is an $O(N^0)$ constant which depends on the temperature. This is an extremely fast decay, much faster than thermalization where the prefactor in the exponent is of order $N^0$, and shows that thermofield double states at different times orthogonalize exponentially fast. 

We expect similar behaviour for many other semi-classically time-dependent states, that is for timescales of $T\sim\mathcal{O}(1)$, we expect
\be
\label{raexpect}
R(T)\sim e^{-N^2 \tilde{f}_0(T)} \,,
\ee
for a positive and $O(N^0)$ function $\tilde{f}_0(T)$ which depends on the state $\rsz$. We expect that for small $T$ the function $\tilde{f}_0(T)$ starts quadratically, as in \eqref{earlyr}. Note that this fast decay is not a consequence of quantum chaos, as it can occur at weak coupling or even in free theories, provided they have a large number of degrees of freedom (see \cite{Chen:2022hbi} for a study of this question in weakly coupled $\mathcal{N}=4$ SYM). The difference between a free theory and a holographic one will manifest itself in the time-scale during which the exponentially small overlap remains valid. For free $\mathcal{N}=4$ SYM, the spectrum is integer spaced and so the return probability will be periodic with period $2\pi$, while in a chaotic theory it will take doubly exponentially long for the signal to return to unity.

The average late-time value of the return probability depends on whether the theory is chaotic or not. For a system with no degeneracies,\footnote{Systems like ${\cal N}=4$ SYM will have degeneracies due to superconformal symmetry. For example, for every primary, there are towers of descendants with degenerate energy levels. Nevertheless, the number of degenerate states is exponentially smaller than the number of all states, at least in the high-energy  sector of the theory, so the degeneracy only contributes a subleading effect.}
\be
\label{plat}
\overline{R} = \lim_{t_*\rightarrow \infty}{1\over 2t_*} \int_{-t_*}^{t_*} \,dT\,R(T) = \sum_i |c_i|^4 \,.
\ee
For the type of states we are considering, i.e. those with a large energy variance, this is exponentially small, and scales as $e^{-\alpha' N^2}$, where $\alpha'$ is an $O(1)$ constant which depends on the particular $\rsz$ we have picked. This value is often referred to as the plateau, especially in the context of the spectral form factor.

Between the initial decay \eqref{decay} and the plateau \eqref{plat}, there can be other regimes, which are particularly interesting in connection to quantum chaos \cite{Shenker:2013pqa,Saad:2018bqo}. For example, in the spectral form factor, the plateau is preceded by a ramp where the signal grows linearly. These effects will not be important for the present work, as we will only consider $\mathcal{O}(1)$ timescales. The crucial point we will exploit throughout the paper is that the signal is already exponentailly small in $N^2$ at those timescales.

The overlap \eqref{overlapdef} obeys the property
\be
\label{timetrans}
\langle \Psi_{t_0} |\Psi_{t_0+T}\rangle = \langle \Psi_0|\Psi_T\rangle \,.
\ee
This may appear trivial, but it means that even if the bulk geometry appears to be static at the semi-classical level, the return probability may still decay following \eqref{decay} if the state had a period of manifest bulk time-dependence in the far past. Said differently, the variance in energy which determines the decay is unchanged under time-evolution, so even if the 1-point functions have stabilized, the variance remains large. This observation is particularly relevant in the case of a black hole formed by gravitational collapse.

The exponential decay \eqref{decay} can be extended to more general correlators of the form $\lsz {\cal O}(t_1)\ldots {\cal O}(t_n) |\Psi_T\rangle$, where ${\cal O}$ are single-trace operators. We expect
\begin{equation}
\label{cordecay}
    \begin{split}
 \lsz {\cal O}(t_1)\ldots {\cal O}(t_n) |\Psi_T\rangle &= 
    F(T)  \lsz \Psi_T\rangle    \end{split} \,,
\end{equation}
where $F(T)$ is finite in the large $N$ limit and satisfies
\be
\label{cordecayb}
F(0) = \lsz {\cal O}(t_1)\ldots {\cal O}(t_n) \rsz\qquad,\qquad {d^kF(T)\over dT^k}|_{T=0} = O(N^0) \,.
\ee
To see the exponential decay we write \eqref{cordecay} as
\be\label{timdepcor}
 \lsz {\cal O}(t_1)\ldots {\cal O}(t_n) |\Psi_T\rangle = \frac{ \lsz {\cal O}(t_1)\ldots {\cal O}(t_n) |\Psi_T\rangle}{ \lsz \Psi_T\rangle} \lsz \Psi_T\rangle \,.
\ee
The second term in this product is responsible for the decay of the correlator. The first term is hard to evaluate from first principles, but in holography its meaning is clearer. In the bulk theory, it is computed by computing a correlation function on a background dictated by the Euclidean path integral with different sources on the northern and soutern hemisphere (corresponding to $\rsz$ and $\ket{\Psi_T}$, respectively). This correlator is $\mathcal{O}(1)$ and a smooth function of the background, which will generally change slowly with $T$, so we expect its time derivatives not to scale with $N$ as indicated in \eqref{timdepcor}. We check this statement in a few examples in section \ref{sec:examples}.

To sum up, any state in the code subspace \eqref{code0} has an exponentially small overlap with any state in the code subspace \eqref{codet}. This can be summarized by the relation
\be
\label{orthcode}
R_{code}(T) = \frac{1}{d_{code}} Tr[P_T P_0]
= \mathcal{O}(e^{-N^2\tilde{f}(T) })
\ee
where $d_{code}$ is the dimensionality of the code subspace, and for the time-scales we have discussed. The decay \eqref{orthcode} can be used in combination with other useful inequalities. For example, for a Hermitian operator ${\cal O}$ with eigenvalues $\lambda_i$, and if $[P_0,{\cal O}]=0$, we have
 $ |\lsz {\cal O} |\Psi_T\rangle|^2\leq \sqrt{\Tr[{\cal O}^4]} \sqrt{\Tr[P_T P_0]}$ and $|\lsz {\cal O} |\Psi_T\rangle|^2\leq {\rm max} (\lambda_i^2)\, \Tr[P_T P_0]$.

\subsection{Other asymptotic charges}\label{returnother}

More generally we can consider the change of the state by large diffeomorphisms corresponding to the other asymptotic symmetries of the theory, in the case of $AdS_{d+1}$ the conformal group $SO(2,d)$ with the generators we discussed in section \ref{sec:charges}. This leads us to define a natural generalization of the return probability
\be
\label{returngeneral}
R(g) = |\lsz U(g) \rsz|^2\qquad,\qquad g \in SO(2,d) \,,
\ee
where $U(g)$ is the unitary realizing the conformal transformation of the CFT on ${\mathbb S^{d-1}}\times {\rm time}$.

What can we expect for these overlaps? To start, let us suppose the state $\rsz$ breaks rotational $SO(d)$ symmetry at the classical level. By this, we mean that bulk dual geometry breaks the symmetry, which would be the case for some spherically asymmetric lump of matter. Take $J$ to be the angular momentum generator, then we expect that the variance of $J$ will be of $O(N^2)$ for such a state. Hence we expect that for small values of a rotation angle $\phi$ dual to $J$ we will have
\be
R(\phi) = e^{-(\Delta J)^2 \phi^2} = e^{- \kappa N^2 \phi^2} \,,
\ee
for $\kappa\sim \mathcal{O}(1)$. For more general angles, we expect
\be \label{decayphi}
R(\phi) = e^{-N^2 f_{\textrm{rot}}(\phi)} \,.
\ee
However, because angular momentum is quantized, we have
\be
R(\phi+2\pi) = R(\phi) \,,
\ee
hence the function $f_{\textrm{rot}}(\phi)$ has period $2\pi$. In this direction of the conformal group the return probability has a very short Poincare recurrence equal to $2\pi$. 

All in all we find that as we increase $\phi$ away from 0 the return probability $R(\phi)$ very quickly dips down to exponentially small values and stays there until the Poincare recurrence at $\phi=2\pi$. As we see from \eqref{decayphi}, for any fixed $\phi$ which is in the range $(0,2\pi)$, we have $R(\phi)$ being exponentially small in the large $N$ limit.

Of course if the state respects spherical symmetry then the return probability will not decay in the corresponding $SO(d)$ directions. It is worthwhile to discuss several distinct scenarios. In the simplest case, the state preserves the symmetry and is thus annihilated by the generators of rotations. The second simplest situation is the case where the symmetry is manifestly broken at the classical level (for example an asymmetric lump of matter). In this case, the breaking of the symmetry is manifest, and would be visible in the 1-point function of single-trace operators. There are also more subtle situations where the state breaks the symmetry classically in the bulk, but this may be invisible in the 1-point functions. An example of this are states by prepared by the path integral on higher genus surfaces in $d=2$, and have topology behind the horizon \cite{Marolf:2017vsk}.\footnote{The thermofield double also has this property. It breaks part of the rotational symmetry of the two CFTs, but the breaking is invisible in 1-point functions. It would be interesting to understand if this type of breaking always requires a horizon.}

Finally as discussed in section \ref{sec:charges}, we expect that semi-classical states also break the other conformal symmetries. We can get some intuition by considering a state dual to a conformal primary of dimension $\Delta$. In this case the return probability along one of the conformal boost directions is determined by a group theoretic computation
\be
R(s) = |\langle \Delta| e^{-i s K}| \Delta \rangle|^{2} = \left({1 \over \cosh^2 s}\right)^{2\Delta} \,.
\ee
For primary states with $\Delta \sim O(N^2)$, we get exponential decay of the form $e^{-N^2 f(s)}$ for any non-zero $s$. Notice that for the conformal boosts we do not expect any Poincare recurrence for large $s$, which in the case of primaries is obvious from the formula above, since such a transformation monotonically increases the energy of the state.

In the case of AdS$_3$ the asymptotic symmetry group is enhanced to Virasoro and similar statements hold for the flow of the state under more general large diffeomorphisms generated by $L_n,\overline{L}_n$.

To summarize, if we start with a state $\rsz$ which breaks all conformal symmetries at the level of the semi-classical geometry we expect that $R(g)$ defined in \eqref{returngeneral} will decay exponentially fast in all directions away from the identity element on the conformal group manifold.

\section{State-dressed operators}\label{sec:operators}

We are now in a position to introduce operators $\hat{\Phi}$ which satisfy the two properties described in section \ref{sec:ourgoal}, namely their commutator with the Hamiltonian and other asymptotic charges is zero to all orders in the $1/N$ expansion and they act like HKLL operators to leading order at large $N$ on the code subspaces $\{\mathcal{H}_{T},\; T\in(-t_{\star},t_{\star})\}$. Here $t_*$ is an order one (i.e. $N^0)$) time of our choice. We define the HKLL operator $\Phi$, \eqref{hkll}, in the $N\to\infty$ limit. In this limit the bulk is described by a quantum field theory on a curved spacetime and code subspaces for different $T$ will be strictly orthogonal to one another. In addition, $\Phi$ is a local bulk operator which commutes with all the boundary single-trace operators in the time band algebra, including the appropriately normalized Hamiltonian \cite{Kabat:2011rz,Hamilton:2006az}. But it will no longer be commuting once $1/N$ corrections are included. In particular, we will have
\be\label{phicom}
[\Phi, \frac{H-\langle H\rangle}{N}]=O(1/N)\neq0 \,.
\ee
Again, the physical reason behind this is that \eqref{hkll} is a diff-invariant operator that is dressed to the boundary. Note that for the naive HKLL operator \eqref{hkll}, the commutator with other single-trace operators will also be non-zero at order $O(1/N)$. For almost all single-trace operators, this can be removed order by order in $1/N$ by adding the appropriate corrections to $\Phi$ \cite{Kabat:2011rz}. However, these modifications will not be able to remove the non-vanishing commutator with the Hamiltonian \eqref{phicom}. Thus, to remove the gravitational dressing to the boundary CFT, a more sophisticated procedure is required.

We start by focusing on setting the commutator with the Hamiltonian to zero and discuss the extension to other asymptotic charges later. To this end, we introduce the following operator\footnote{Recall that $P_0$ is the projector on the code subspace of $\rsz$, and thus $[\Phi,P_0]=0$ in that code subspace. Therefore, we could have defined operators with the same action on the code subspace as \eqref{defop}, using a single projector on the left (or right) of $\Phi$. Even though the resulting operators would act in the same way on the relevant code subspace, the operators would not be exactly identical: they would have additional non-zero matrix elements associated to subspaces orthogonal to ${\cal H}_0$.}
\be
\label{defop}
\drop= c\int_{-t_*}^{t_*}dT\,\, e^{-i T H } P_0 \Phi P_0 e^{i T H } \,,
\ee
where $t_*$ is an $O(N^0)$ timescale of our choice, and $c$ is an overall normalization constant
\be
\label{normalization}
c^{-1} = \int_{-t_*}^{t_*} dT \lsz P_T \rsz \,.
\ee

As we will see, the projector $P_{0}$ will be key and will make $\drop$ act appropriately on the code subspaces. The range $(-t_{\star},t_{\star})$ determines the set of code subspaces on which $\drop$ acts in the desired fashion, and ultimately cannot be taken to be bigger than the time range where the exponential decay of the return probability (\ref{decay}) is valid. To make the operator (\ref{defop}) have the desired properties on as many states as possible, we can take this range to be the time range where the return probability decays exponentially, though this is not strictly necessary and a $t_*$ of $O(N^0)$ is sufficient. We also provide an alternative presentation of the operators in subsection \ref{sim}. In the following subsections, we will study the action of these operators in the relevant code subspaces, and will be particularly interested in their commutator with the Hamiltonian. 

\subsection[Vanishing commutator with H to all orders in 1/N]{Vanishing commutator with $H$ to all orders in $1/N$}\label{comH}

We now show that the operator \eqref{defop} has vanishing commutator with $H$ to all orders in $1/N$. We start by rewriting the commutator as
\be
[H,\drop] =-i{d\over ds}\big(e^{i s H } \drop e^{-is  H }\big) \Big|_{s=0},
\ee
and performing a change of variables, we find
\be
\begin{split}
\label{commutator}
[H,\drop]&=-i{d\over ds}\big(c\int_{-t_*-s}^{t_*-s}dT\,\, e^{-i T H } P_0 \Phi P_0 e^{i T H }\big)\Big|_{s=0}\\
&= i c (P_{t_*} \Phi_{t_*} P_{t_*} - P_{-t_*} \Phi_{-t_*} P_{-t_*} ) \,,
\end{split}
\ee
where we defined $\Phi_{t_*}=e^{-i H t_*}\Phi e^{i H t_*}$. Using the decay of the return probability through (\ref{orthcode}), we see that the commutator inserted inside a correlator of a small number of single-trace operators and evaluated on the state $\ket{\Psi_T}$ will give an exponentially small answer, since each of the two terms  in \eqref{commutator} give exponentially small numbers. This is valid for any $T$ as long as $|T|<t_{\star}$  and $|T|-t_{\star}\sim O(N^0)$.
Thus,
\be
\label{hcommut}
[H,\drop] = O(e^{-\gamma N^2}) \,,
\ee
 where $\gamma$ is positive and $O(N^0)$, proving property 1, defined in subsection \ref{sec:ourgoal}, for these operators. Note (\ref{hcommut}) is true for our set of code subspaces with $T$ constrained as above, but not for all states. For example, the commutator is not exponentially suppressed in the state $\ket{\Psi_{t_*}}$.

\subsection{Similar action as HKLL operators}

A vanishing commutator with the Hamiltonian is necessary but not sufficient. There are many CFT operators that commute with the Hamiltonian up to exponentially small corrections in $N^2$, but they will not have the same effect as acting with a local bulk operator, see for example footnote \ref{trivialun}. Therefore, we also need to show that the operator $\hat{\Phi}$ behaves in the same way as the HKLL operator \eqref{hkll} to leading order at large $N$ inside correlation functions of single-trace operators. For that we consider
\begin{equation}
    \begin{split}
\lsz&{\cal O} ...\drop ... {\cal O} \rsz =
        \\
= &c \int_{-t_*}^{t_*} dT\, \lsz {\cal O}...e^{-iT H } P_0 \Phi P_0 e^{i TH } ... {\cal O} \rsz
        \\
        =&c \int_{-t_*}^{t_*} dT\, \lsz {\cal O}...P_0 P_T (e^{-i TH } \Phi e^{iT H }) P_T P_0... {\cal O} \rsz .
    \end{split}
\end{equation}
In the last line, we have inserted two projectors $ P_0$, which we are free to do since the correlators is evaluated in the state $\rsz$. The integrand above corresponds to $\Tr P_T P_0$, up to some operator insertions that do not affect its general structure. From \eqref{orthcode} we see that the integrand will be exponentially suppressed as $|T|$ increases (and is not $O(1/N)$) because of the exponentially small overlap of the code subspaces. We can thus evaluate the integral by a saddle-point method controlled by the large $N$ limit. The dominant contribution comes from $T=0$\footnote{One might worry about the possibility of rapidly oscillating phases, such as the one in $\lsz \Psi_T\rangle$ displacing the location of the saddle point. Notice however that from \eqref{cordecay},\eqref{cordecayb} it follows that such rapidly oscillating  phases cancel between the bra and ket contribution.}. Using \eqref{cordecay} and \eqref{normalization} we have
\be
\lsz {\cal O}...\drop ... {\cal O} \rsz =
\lsz {\cal O}...\Phi ... {\cal O} \rsz  + O(1/N),
\ee
as desired. The $1/N$ corrections can be thought of coming from corrections to the leading saddle-point, and would be sensitive to the more detailed form of $F(T)$ in \eqref{cordecay}.

Notice that if we apply the operator $\drop$ to one of the time-shifted states, then as long as $|T| < t_*$, we find
\be\label{shiftedaction}
  \langle \Psi_T |{\cal O}...\drop ... {\cal O} |\Psi_T\rangle= 
  \\
   \langle \Psi_T| {\cal O}...(e^{-iT H } \Phi e^{i TH }) ... {\cal O} |\Psi_T\rangle  + O(1/N)
\ee
Thus in the code subspace $\mathcal{H}_{T}$, $\hat{\Phi}$ acts as $e^{-iT H } \Phi e^{i TH }$ to leading order at large $N$. We discuss the physical interpretation of this in the next subsection. To make this even more manifest, we can also write \eqref{defop} as
\begin{equation}
  \drop = c \int_{-t_*}^{t_*} dT~ P_T (e^{-i T H} \Phi e^{i T H })P_T . 
\end{equation}
Since we have shown that, to leading order at large $N$, $\hat{\Phi}$ and $\Phi$ have the same matrix elements on the entire code subspace it follows that higher point functions of $\hat{\Phi}$ will also agree at large $N$ with those of $\Phi$.  Consider for instance,
\be
\hat{\Phi}_{i}= c\int_{-t_*}^{t_*}dT\,\, e^{-i T H } P_0 \Phi_{i} P_0 e^{i T H }
\ee
where $\Phi_{i}\equiv\Phi(x_{i})$ is an HKLL operator located at a certain spacetime point $x_{i}$, then in the large $N$ limit
\be
\begin{split}
\lsz {\cal O}...\drop_{1}\drop_{2} ... \drop_{n} ... {\cal O} \rsz =& c^{n} \int_{-t_*}^{t_*} dT_{1}...dT_{n}\, \lsz {\cal O}... P_{T_{1}} (e^{-i T_{1}H } \Phi_{1} e^{iT_{1} H }) P_{T_{1}}P_{T_{2}} \\ &     (e^{-i T_{2}H } \Phi_{2} e^{iT_{2} H }) P_{T_{2}}...P_{T_{n}} (e^{-i T_{n}H } \Phi_{n} e^{iT_{n} H }) P_{T_{n}} ... {\cal O} \rsz \\ \approx &\;
\lsz {\cal O}...\Phi_{1}\Phi_{2} ... \Phi_{n} ... {\cal O} \rsz \,.
\end{split}
\ee
In addition, this implies that the commutator of $\hat{\Phi}_{i}$'s is the same as that of HKLL operators in the large $N$ limit. Two operators, $\hat{\Phi}(x_{i})$ and $\hat{\Phi}(x_{j})$, will have zero commutator at spacelike separated points whereas they have $O(1)$ commutator if they are timelike-separated. This is true even though these operators do not translate under commutation with the boundary Hamiltonian, up to exponentially small corrections in $N$. Nevertheless, they still have bulk space-time labels and  preserve the causal properties of HKLL operators in the large $N$ limit.

\subsection{Interpretation and comments}

We have just seen that to leading order in the large $N$ limit, the operator \eqref{defop} acts like the HKLL operator \eqref{hkll} in the appropriate code subspace. However, it commutes with $H$ to all orders in $1/N$.  The existence of these operators provides strong evidence that the algebra of single-trace operators in a short time band can have a non-trivial commutant when acting on time-dependent states of high energy. 

The vanishing of the commutator with $H$ should be interpreted as \eqref{defop} being gravitationally dressed not with respect to the boundary, but instead with respect to features of the bulk state, in particular its time-dependence. This can be seen by the fact that $\hat{\Phi}$ acts differently on different states. On the time-shifted states $\ket{\Psi_{T}}$ and their code subspaces, it acts as $e^{-iT H } \Phi e^{i TH }$. For example, imagine that in the state $\rsz$ we have a supernova explosion taking place at $t=0$ and we chose the operator \eqref{hkll} so that it acts right next to the explosion. In the state $|\Psi_T\rangle$ the explosion obviously takes place at $t=-T$. From equation \eqref{shiftedaction}, we can see that the operator $\drop$ will act again right next to the supernova explosion, even though the supernova is now at $t=-T$. Therefore, one and the same operator $\drop$ knows how to always act at the correct moment (right next to the explosion) for the entire family of states $|\Psi_T\rangle$, as long as $|T|< t_*$. The finiteness of $t_*$ indicates that there is still some residual boundary dressing, which however is not visible in pertubation theory\footnote{Similar remarks were made in \cite{Marolf:2015jha} for the DeWitt observables in AdS. One can certainly imagine increasing the size of $t^*$, but there is a potential limitation due to Poincare recurrences.}.

The property of being dressed with respect to features of the state is also present in the local observables one defines in general relativity, discussed in section \ref{sdob}. These state dressed observables are defined at points where a set of $D$ scalars, like the Ricci scalar or $\mathcal{R}_{\mu\nu\rho\sigma}\mathcal{R}^{\mu\nu\rho\sigma}$ where $\mathcal{R}_{\mu\nu\rho\sigma}$ is the Riemann tensor, 'click' with a certain set of numbers. The observables are labeled by these values and they are evaluated precisely where the scalars take those values in each state. Locality of these observables requires them to be defined only in some neighbourhood of a classical solution. In the same spirit, the operators discussed in this section are  also local for a certain family of code subspaces, see section \ref{comH}.

As mentioned earlier, if the spacetime is so symmetric that the scalars take the same values throughout the spacetime, then these classical observables are not well defined. Since every point in the spacetime is physically equivalent, it is reasonable that local observables are ill defined for these solutions. For this reason, the observables are state dependent. Similarly, it is not possible to apply the same logic discussed in the previous subsections to empty AdS, or other static states,  as there are no time-dependent features in the bulk that can be used as a 'clock' to define a moment in time where the operator acts. Technically, the return probability for such states does not exhibit the rapid decay \eqref{decaybulk}. We thus see a nice parallel between the classical and quantum situations.

The definition \eqref{defop} gives a bulk operator which is dressed with respect to features of the state, but in an implicit manner.     Going back to our example of a supernova explosion, one might guess that the dressing is with respect to the supernova and that one could in principle define a gravitational Wilson line between the operator and the supernova. But what if the state described instead two supernovas exploding at the same or different times? To which explosion would our operator be dressed to? The construction \eqref{defop} does not correspond to a dressing relative to a {\it specific} localized feature in the bulk, but rather to the overall time-dependence of the state. If we wanted to associate the dressing to a particular part of the bulk geometry then we would have to enlarge the set of code subspaces on which our operator correctly acts. For example, if our operator did not move under the time-translation of one of two supernovae, we would say that it is dressed to the other one. We hope to return to this question in the future, but see subsection \ref{subsec:entangledcfts} for some related remarks.

\subsection{A similarity transformation}\label{sim}

We briefly mention a variant of operators with properties similar to those of \eqref{defop}. We first define the shifted Hamiltonian\footnote{This shift is useful in order to avoid rapidly oscillating phases in the discussion below.} 
\be
\hat{H} = H - \lsz H \rsz {\mathbb I} \,.
\ee
Then we introduce
\be
V = {c\over \sqrt{2}}\int_{-t_*}^{t_*} dT e^{-i \hat{H} T} P_0 \,,
\ee
with $c$ given in \eqref{normalization}. We have
\be
\label{unitaryv}
V V^\dagger = {c^2\over 2}\int_{-t_*}^{t_*} dT \int_{-t_*}^{t_*} dT' e^{-i \hat{H} T} P_0 e^{i \hat{H} T'} \,,
\ee
where we used $P_0^2=P_0$. Following arguments similar to those of the previous subsection, we find that to leading order at large $N$, and when computing the matrix elements of \eqref{unitaryv} within the code subspace, the two integrals in \eqref{unitaryv} can be computed by a saddle point method, where the dominant saddle is $T=T'=0$. We then find that in this class of states and at large $N$
\be
V V^\dagger \simeq {\mathbb I}, \,\,\,\,\, \text{ and }\,\,\,\,\,\,\,  V^\dagger V \simeq {\mathbb I} \,.
\ee
in the sense that, within the code subspace $V$ behaves like a unitary, up to $1/N$ corrections.

Then we start with a boundary-dressed operator $\Phi$ and  define
\be
\label{defop2}
\hat{\Phi} = V \Phi V^\dagger \,.
\ee
Following similar arguments as before we can show that the operator \eqref{defop2} satisfies properties 1 and 2 of subsection \ref{sec:ourgoal}. To check the commutator of $\hat{\Phi}$ with $H$. We write
\be
\begin{split}
& [H,\hat{\Phi}] = -i {d\over ds} \left(e^{i \hat{H} s} V \Phi V^\dagger e^{-i \hat{H} s}\right)|_{s=0} \cr
& = -i {d\over ds}  {c^2\over 2}(\int_{-t_*-s}^{t_*-s} dT  e^{-i \hat{H} T}) P_0 \Phi P_0( \int_{-t_*-s}^{t_*-s} dT'e^{i \hat{H} T'})|_{s=0} \,,
\end{split}
\ee
which again localizes on boundary terms and is thus exponentially suppressed.

Second, to show that the leading large $N$ correlators of $\hat{\Phi}$ are the same as those of $\Phi$ we follow exactly the same reasoning as in the previous subsection, but now we will have two time-integrals. Each one of these time integrals will lead to a sharply suppressed Gaussian around $T=T'=0$ and can be evaluated by saddle-point at large $N$, reproducing the desired result.

\subsection{Other asymptotic charges}

More generally we need to make \eqref{hkll} commute with all boundary symmetry generators corresponding to asymptotic symmetries. For asymptotically AdS$_{d+1}$ space-times this is the conformal group $SO(2,d)$ and we consider a generalization of the form
\be
\label{allcharges}
\drop = c\int_{B} d\mu(g) U(g)P_0 \Phi P_0 U(g)^{-1},
\ee
where now
\be
c^{-1}=\int_{B} d\mu(g) \lsz U(g)P_0 U(g)^{-1} \rsz \,.
\ee
Above, $d\mu(g)$ is the Haar measure on $SO(2,d)$ and $B$ is a reasonably sized connected submanifold of $SO(2,d)$ containing the identity. If the variance of all charges is $\mathcal{O}(N^2)$, the size of this ball should be $\mathcal{O}(N^0)$, but not too big in the periodic directions. The commutator with conformal generators will then be given by operators in the code subspace of states $U(g_*)|\Psi_0\rangle$, where $g_*$ lies on the boundary $\partial B$. For the construction to work in this generalization we must make sure that the overlaps
\beq
\label{genoverlap}
R(g) = |\langle \Psi_0|U(g)|\Psi_0 \rangle|^2,
\eeq
decay exponentially in the geodesic distance of $g$ from the identity. As discussed in subsection \ref{returnother} we expect this to be true for states which break all symmetries at the semiclassical level\footnote{For compact symmetries, such as rotations, $R(g)$ will have recurrences every $2\pi$. Hence along the compact directions we take $g_* \sim O(1)< 2\pi$.}. The quantity $R(g)$ is an interesting generalization of the return probability \eqref{radef} that would be interesting to study further.

\section{A more general argument for the commutant}\label{otconop}

The operators \eqref{allcharges} constructed in the previous section commute with the asymptotic charges to all orders in $1/N$, however they commute with the other single-trace operators in the time-band generally only to leading order in $1/N$. To identify a commutant for the time-band algebra ${\cal A}$, the operators \eqref{allcharges} have to be improved. In this short section we outline a somewhat different argument suggesting that it is indeed possible to find a commutant to all orders in $1/N$. We caution the reader that the argument that follows is based on certain assumptions which seem physically plausible, but for which a rigorous proof is still lacking. A more careful treatment for the existence of a commutant (as well as a mathematically precise definition of the time-band algebra to all orders in $1/N$ in the first place) would be desirable.

Let us start with a standard HKLL operator $\Phi$. We also introduce the notation $q_i={Q_i-\langle Q_i\rangle \over N}$ for  where $Q_i$ denotes any of the asymptotic $SO(2,d)$ charges and ${\cal O}_j$ a general single-trace operator in the time-band. Our goal is to find an operator $\hat{\Phi}$ which has the following properties:
\begin{enumerate}
    \item $[\hat{\Phi},q_i] = 0$ and $[\hat{\Phi},{\cal O}_j]=0$ for all $q_i \in SO(2,d)$ and ${\cal O}_j \in {\cal A}$, to all orders in $1/N$.
    \item To {\it leading order} at large $N$ the correlators of $\hat{\Phi}$ with $q_j,{\cal O}_i$ must be the same as those of $\Phi$. In particular this means that for single-trace operators ${\cal O}_i$ outside the time-band we generally expect $[{\cal O}_i,\hat{\Phi}] = O(N^0)$.
\end{enumerate}
The first condition is obvious. The second condition is necessary in order to ensure that the operator $\hat{\Phi}$ acts in the expected way, at least to leading order at large $N$, and creates particles  that can be detected with an $O(1)$ effect by operators outside the time-band when light rays from the diamond hit the boundary.

Here we remark that in order for the two conditions to be mutually consistent, it is important that we impose the second condition only to leading order at large $N$. The point is that $[q_i, \Phi] = O(1/N)$ hence when looking at leading order correlators it is indeed consistent to demand simultaneously that i) $\hat{\Phi}$ commutes with $q_i$ and that ii) $\hat{\Phi}$ acts like $\Phi$. However, when moving on to subleading corrections we have a non-vanishing commutator $[q_i,\Phi]$ hence we cannot impose both conditions at the same time. We choose to impose that our operators $\hat{\Phi}$ continue to commute with $q_i$ to all orders in $1/N$, but we allow their correlators to depart from those of $\Phi_i$ at subleading orders in $1/N$.

We now define the desired operators $\hat{\Phi}$ by specifying how they act on the code subspace ${\cal H}_0$. Earlier we defined the code subspace as the space generated by acting on $\rsz$ with single-trace operators, which are not necessarily restricted in the time-band. However, by an analogue of the Reeh-Schlieder theorem\footnote{This was discussed in \cite{Banerjee:2016mhh} for the case of empty AdS and at large $N$. We believe that a similar result should hold for more general heavy states and even when taking $1/N$ corrections into account, but it would be interesting to develop a more careful proof.} we expect that for reasonable bulk states $\rsz$ the code subspace ${\cal H}_0$ can also be generated by acting on $\rsz$ with only elements of the time-band algebra ${\cal A}$ 
\be \label{spanH}
{\cal H}_0 = {\rm span}\{{\cal A} \rsz\} \,.
\ee
We now define the action of the operator $\hat{\Phi}$ on the code subspace by the following conditions
\be
\label{altdefop}
\hat{\Phi} A \rsz = A \Phi  \rsz \,, \qquad \forall A\in {\cal A} \,.
\ee
This set of linear equations, one for every element of the small algebra ${\cal A}$, defines the action of $\hat{\Phi}$ on the code subspace, in a way which satisfies the desired properties as we will see below.

Notice that these equations can also be represented as follows: we first select a basis of linearly independent elements $A_i$ of the algebra ${\cal A}$. then we define the matrix of 2-point functions
\be
\label{2pointcode}
g_{ij} = \lsz A_i^\dagger A_j \rsz \,.
\ee
From \eqref{spanH}, it follows that the set of states $|i\rangle = A_i \rsz$ form a (possibly over-complete) basis of the code subspace. Since $\hat{\Phi}$ is an operator on the code subspace it can be written as
\be
\label{expandoperator}
\hat{\Phi} = K^{ij}|i\rangle \langle j| = K^{ij} A_i\rsz\lsz A_j^\dagger \,.
\ee
for an appropriate choice of $K^{ij}$. To find the matrix $K$, we start with the desired relation \eqref{altdefop} written as
\be
\hat{\Phi} A_l \rsz = A_l \Phi \rsz \,,
\ee
then we replace $\hat{\Phi}$ with \eqref{expandoperator} and multiply from the left with $\lsz A_k^\dagger$ to get
\be
g_{jl} \,g_{ki}\,K^{ij}  = \lsz A_k^\dagger A_l \Phi \rsz \,.
\ee
If the set of states $|i\rangle= A_i \rsz$ are linearly independent then the matrix $g_{ij}$ is positive definite and invertible. In that case we can solve for $K$ as
\be
\label{solvek}
K_{ij} = g^{i k} g^{jl}  \lsz A_k^\dagger A_l \Phi \rsz \,,
\ee
where $g^{ij}g_{jk}=\delta^i_k$. When \eqref{solvek} is replaced in expression \eqref{expandoperator}, we find an explicit solution of the desired equation \eqref{altdefop}.

We emphasize that the necessary ingredient to arrive at \eqref{solvek} was the linear independence of the states $A_i\rsz$, which is equivalent to the statement that there is no non-vanishing operator in ${\cal A}$ which annihilates the state $\rsz$. We discuss this condition in the following subsection.

\subsection{On the consistency of the defining equations}

Before checking that the operators $\hat{\Phi}$ defined by \eqref{altdefop}, or equivalently via \eqref{expandoperator},\eqref{solvek}, have the desired properties, we need to check that equations \eqref{altdefop} are self-consistent linear equations. The only possible source of inconsistency is the following: if there was an element $A\neq 0$ of the time-band algebra ${\cal A}$ such that
 $
 A \rsz =0
 $, this could potentially be a problem since we would then have $A \rsz=0$, while in general $A \Phi \rsz\neq 0$. Then the equation \eqref{altdefop} would imply $0 = A \rsz = A \Phi \rsz \neq 0$ which is a contradiction. Relatedly, $g_{ij}$ defined in \eqref{2pointcode} would not be invertible and we would not be able to get to \eqref{solvek}.

We will now show that this situation does not arise, that is
\be
\label{sepcondition}
A\rsz \neq 0 \qquad \forall A\in {\cal A}\,\,,\,\, A\neq 0 \,.
\ee
We will prove this by first proving that at large $N$ \eqref{sepcondition} is true and then we will argue that $1/N$ corrections cannot change the conclusion. 

We have been working under the assumption that the time-band is short enough, which means that in the bulk there will be a region which is space-like relative to the time band. In the large $N$ limit, where gravitational backreaction is turned off, operators inside that region (for example usual HKLL operators) commute with all elements of the algebra ${\cal A}$, including the appropriately normalized asymptotic charges $q_i$. Hence, in the large $N$ limit the algebra ${\cal A}$ has a non-trivial commutant ${\cal A}'$. We want to argue that this commutant continues to exist when $1/N$ corrections are taken into account, provided that the state $\rsz$ has non-vanishing variance of $O(N^2)$ under the asymptotic charges.

Assuming that at large $N$ the theory in the bulk behaves like usual QFT on a curved background, we expect that an analogue of the Reeh-Schlieder theorem will hold for the commutant ${\cal A}'$, which means that we can generate the code subspace ${\cal H}_0$ by acting on $\rsz$ with elements of ${\cal A'}$.

Suppose now that there was an element $A$ of the time-band algebra ${\cal A}$ which annihilated the state $\rsz$. Then for any element $a' \in {\cal A}'$ we have
\be
A a' \rsz= a' A \rsz = 0 \,.
\ee
Since states of the form $a'\rsz$ generate ${\cal H}_0$ we conclude that the operator $A$ has vanishing matrix elements in ${\cal H}_0$ at large $N$. From this we can not immediately conclude that $A=0$ as an operator when $1/N$ corrections are included. For example, for $\rsz=|0\rangle$ the normalized $SO(2,d)$ generators $q_i = {Q_i \over N}$ have vanishing matrix elements at large $N$, since they annihilate $|0\rangle$ and commute with all other operators. However they are non-vanishing operators at order $1/N$. If $A$ is a non-vanishing operator which has vanishing matrix elements at large $N$ on ${\cal H}_0$ then it means that it acts as a central element at large $N$. Here we make an additional assumption, that the only central elements are the $SO(2,d)$ generators $q_i$ and their functions.  Since, by assumption, the state $\rsz$ has non-trivial variance under these generators, we conclude that it cannot be annihilated by a non-trival $A$.

Let us assume now that we have a state of the form $A\rsz$ which has finite (i.e. $O(N^0)$) positive norm at large $N$. Including $1/N$ corrections will generally modify the norm of this state, but it will do so by corrections suppressed by powers of $1/N$. Since the previous argument established that the leading large $N$ norm of the state $A\rsz$ is a finite positive number, perturbative $1/N$ corrections cannot make it vanish. Hence we expect property \eqref{sepcondition} to be true to all orders in $1/N$ perturbation theory.

We emphasize that the fact that we cannot annihilate the state by the time-band algebra ${\cal A}$ relies on the fact that we have restricted our attention to small products of single-trace operators. As discussed in a related context \cite{Papadodimas:2013jku,Banerjee:2016mhh}, if we consider the {\it full} algebra of operators in the time-band we can find sufficiently complicated combinations which can annihilate the state\footnote{For example, consider a state $|\Psi\rangle$ with $\lsz \Psi\rangle=0$. Then the (complicated) operator $|\Psi\rangle\langle \Psi|$ annihilates $\rsz$.}.

Finally, as should be clear from the above, if the state $
\rsz$ has very small or vanishing variance in the asymptotic charges then \eqref{sepcondition} fails and it is not possible to define operators obeying \eqref{altdefop}.

\subsection[Proof that operators have the desired properties]{Proof that $\hat{\Phi}$ has the desired properties}

Having established that equations \eqref{altdefop} are consistent, we argue that the operator $\hat{\Phi}$ has the desired properties.

First it is obvious by \eqref{altdefop} that the operator $\hat{\Phi}$ has vanishing commutators with elements of ${\cal A}$. To see that consider $A_1\in A$ and a general state in the code subspace which can be written as $A_2\rsz$, with $A_2\in {\cal A}$. Then we have
\be
[\hat{\Phi},A_1] A_2 \rsz = \hat{\Phi} (A_1 A_2)\rsz - A_1 (\hat{\Phi} A_2 \rsz) = A_1 A_2 \Phi \rsz - A_1 A_2 \Phi \rsz = 0 \,,
\ee
where in the second equality we used \eqref{altdefop}.  Since this is true for all $A_2$, we find
\be
[\hat{\Phi},A_1] = 0 \qquad \forall A_1 \in {\cal A} \,,
\ee
where it should be understood that this equation holds on the 
 relevant code subspace.

Second, we will show that {\it to leading order at large $N$}, the operator $\hat{\Phi}$ acts like the HKLL operator $\Phi$. To see this, consider an arbitrary matrix element on the code subspace. Two general states of the code subspace can be written as $A_1\rsz,A_2\rsz$. Then we have
\be
\lsz A_1^\dagger \hat{\Phi} A_2 \rsz= \lsz A_1^\dagger A_2 \Phi\rsz = \lsz A_1^\dagger \Phi A_2\rsz  + \lsz A_1^\dagger [\Phi,A_2]\rsz \,.
\ee
In the first equality we used \eqref{altdefop}. Now, the operator $A_2$ is some combination of single-trace operators in the time band, as well as the normalized $SO(2,d)$ generators $q_i$. All of these operators have commutators with $\Phi$ which are suppressed by powers of $1/N$. Hence the last term in the equation above is suppressed. All in all, we find
\be
\lsz A_1^\dagger \hat{\Phi} A_2 \rsz = \lsz A_1^\dagger \Phi A_2\rsz  + O(1/N) \,,
\ee
which establishes the desired result. This ensures that large $N$ correlators of $\hat{\Phi}$ are the same as $\Phi$.

We emphasize that the operators defined in this section are not exactly the same as the operators \eqref{defop} discussed earlier. For example, unlike \eqref{defop} the operators \eqref{altdefop} were defined to act only on the code subspace ${\cal H}_0$ of $\rsz$ and not on the code subspace ${\cal H}_T$ for $T = O(N^0)$. Also, the commutator of \eqref{defop} with the Hamiltonian is of order $e^{-N^2}$ while it is exactly zero, within the code subspace, for the operators \eqref{altdefop}.

\section{Examples \label{sec:examples}}

In this section we consider various examples. Our primary focus will be on examining the validity of equations \eqref{raexpect}, \eqref{cordecay},\eqref{cordecayb}, on which the construction of our operators relies.

\subsection{Coherent states}

In general, we are interested in time-dependent semi-classical geometries. Many of these states can be thought of as bulk coherent states. We will discuss the overlap of these states closely following \cite{Belin:2018fxe}. In the CFT, these states are prepared by a Euclidean path integral
\begin{equation}
\label{coherent}
   \ket{\Psi}=Te^{-\int_{t_{E}<0}dt_{E}d^{d-1}x \; \phi_{b}(t_{E},x){\cal O}(t_{E},x)} \ket{0} \,,
\end{equation}
where ${\cal O}$ is a single-trace operator dual to a supergravity field,  and the source is scaled appropriately so that it leads to states with non-trivial gravitational backreaction, i.e. the expectation value of the energy and variance of this state will scale like \eqref{energyofstate} and \eqref{vardef}.

In the large $N$ limit the overlap of two such states can be computed by a Euclidean gravitational path integration which in the semi-classical limit can be approximated by a saddle point computation. For example, the norm of the state is
\begin{equation}\label{12.3}
    \braket{\Psi|\Psi}\approx e^{-I_{grav}(\lambda_{b})} \,,
\end{equation}
where $\lambda_{b}$ is the following boundary condition for the bulk field
\begin{equation}
\lambda_{b}=
\begin{cases}
    & \phi_{b}(t_{E},x),\; t_{E}<0 \\
     & \phi^{\star}_{b}(-t_{E},x), \; t_{E}>0 \,,
\end{cases}
    \end{equation}
and $I_{grav}(\lambda_{b})$ is the on-shell gravitational action in the presence of the sources specified above. 

Generalizing to two states $\ket{\Psi_{1}}$ and $\ket{\Psi_{2}}$, the normalized inner product between them is
\begin{equation}
\label{sugrainner0}
    \mathcal{R}= {\left|\langle \Psi_1|\Psi_2\rangle \right|^2 \over \langle \Psi_1|\Psi_1\rangle\langle \Psi_2|\Psi_2\rangle} \,,
 \end{equation}
which at large $N$ can be computed by a supergravity saddle-point computation
\begin{equation}
\label{sugrainner}
    \mathcal{R}\approx \exp\left[-2\Re(I_{grav}(\Tilde{\lambda}))+I_{grav}(\lambda_{1})+I_{grav}(\lambda_{2})\right] \,,
\end{equation}
where the supergravity solutions have the boundary sources $\tilde{\lambda}$, $\lambda_{1}$ and $\lambda_{2}$ which take the following form
\begin{align}
\label{boundarycondgrav}
    \tilde{\lambda}=
\begin{cases}
    & \phi_{2}(t_{E},x),\; t_{E}<0 \\
     & \phi^{\star}_{1}(-t_{E},x), \; t_{E}>0,
\end{cases} &&
\lambda_{i}=
\begin{cases}
    & \phi_{i}(t_{E},x),\; t_{E}<0 \\
     & \phi^{\star}_{i}(-t_{E},x), \; t_{E}>0,
\end{cases}
\end{align}
where $i=1,2$\footnote{The sources $\phi_{2}(t_{E},x)$ and $\phi^{\star}_{1}(-t_{E},x)$ should decay sufficiently fast at the $t=0$ surface such that the states are normalizable. This also implies that the bra and ket preprations of different states can be smoothly glued to each other.}. 

Notice that in each of the terms of \eqref{sugrainner}, the gravitational on-shell action is proportional to ${1\over G_N}\sim N^2$. Since quantum mechanically we need $\mathcal{R}\leq 1$,
we find that the following inequality has to be satisfied
\be
\label{inequality}
2\Re(I_{grav}(\Tilde{\lambda}))\geq I_{grav}(\lambda_{1})+ I_{grav}(\lambda_{2}) \,,
\ee
for the on-shell value of solutions of the Einstein plus matter equations, for any choice of sources of the form \eqref{boundarycondgrav}. If the two sources are different, we expect a strict inequality. It would be interesting to explore this inequality directly from the gravitational point of view. We discuss this further in the discussion.

We now move on to the computation of the return probability for states of the form \eqref{coherent} after a small (not $N$-dependent) time evolution. That is, we take the time-evolved state, $\ket{\Psi(T)}=e^{-iHT}\ket{\Psi}$, and consider the following quantity
\begin{equation}
    R(T)= {\left|\langle \Psi(0)|\Psi(T)\rangle \right|^2 \over \langle \Psi(0)|\Psi(0)\rangle\langle \Psi(T)|\Psi(T)\rangle} \,.
\end{equation}
To apply the general formalism described above, we need to analyze how the Euclidean sources $\phi_0$ preparing the state $|\Psi(0)\rangle$ need to be modified to $\phi_T$, in order to prepare $|\Psi(T)\rangle$. From a technical point of view computing $\phi_T$ in terms of $\phi_0$ is not straightforward, as it requires a solution of the Einstein equations. Nevertheless, we can in principle compute the return probability using  \eqref{sugrainner0} and \eqref{sugrainner} with a modified source
\begin{align}
\label{boundarycondgravb}
    \tilde{\lambda}=
\begin{cases}
    & \phi_{T}(t_{E},x),\; t_{E}<0 \\
     & \phi^{\star}_{0}(-t_{E},x), \; t_{E}>0,
\end{cases} &&
\lambda_{T}=
\begin{cases}
    & \phi_{T}(t_{E},x),\; t_{E}<0 \\
     & \phi^{\star}_{T}(-t_{E},x), \; t_{E}>0 \,.
\end{cases}
\end{align}
Thus we get     
\begin{equation}
 R(T) = \exp\left[-2\Re(I_{grav}(\tilde{\lambda})) +I_{grav}(\lambda_0)+I_{grav}(\lambda_t)\right],    
\end{equation}  
and this is exponentially suppressed in the semi-classical limit because of the $1/G_{N}\sim N^2$ coefficient in the gravitational action and the condition \eqref{inequality}.

\subsection{Thermofield double state}\label{sec:tfd}

We now consider the thermofield double state
\begin{equation}
\label{tfdstate}
    \ket{{\rm TFD}}= \frac{1}{\sqrt{Z(\beta)}}\sum_{n} e^{-\frac{\beta E_{n}}{2}} \ket{E_{n}}_{L} \otimes \ket{E_{n}}_{R} \,,
\end{equation}
where the $\ket{E_{n}}$'s are the energy eigenstates and $Z(\beta)$ is the partition function at inverse temperature $\beta$. In the strong coupling limit, for temperatures below the Hawking-Page temperature, the state is dual to two entangled thermal AdS geometries, while for temperatures higher than the Hawking-Page temperature, it is expected to be dual to the eternal black hole in AdS \cite{Maldacena:2001kr}. This geometry has two asymptotically AdS boundaries, on the "left" and the "right", hence the asymptotic symmetry group is $SO(2,d)_L \times SO(2,d)_R$. The state \eqref{tfdstate} is invariant under certain combinations of the asymptotic charges, for example we have
\be
(H_R-H_L)\ket{{\rm TFD}}=0\qquad {\rm but} \qquad (H_R+H_R) \ket{{\rm TFD}}\neq 0
\ee
and similarly for the other charges. In this case we can generalize the return probability to include all possible large diffeomorphisms on the two sides
\be
R(g_1,g_2) = |\langle{\rm TFD}|\, U_L(g_L) \, U_R(g_R)\, |{\rm TFD}\rangle|^2\qquad,\qquad g_{L/R}\in SO(2,d)_{L/R} \,.
\ee
We expect $R(g_L,g_R)$ to rapidly decay along certain directions but remain constant along others due to the symmetries of the state \eqref{tfdstate}.

In what follows we focus on a particular class of deformations, corresponding to evolving with $H_L+H_R$. This gives what is usually called the spectral form factor (SFF) defined as
\be
\label{bhsff}
R(t) = |\langle{\rm TFD}| e^{-i {T\over 2}(H_L +H_R)} |{\rm TFD}\rangle|^2 = \left |\frac{Z(\beta+iT)}{Z(\beta)}\right|^{2} \,,
\ee 
which was introduced in the context of the eternal AdS black hole in \cite{Papadodimas:2015xma} and studied in detail in \cite{Cotler:2016fpe}.

We are interested in studying \eqref{bhsff} above the Hawking-Page temperature for small times, i.e, $T\sim O(1)$. One way to proceed is by computing $Z(\beta)$ and then analytically continuing $\beta\rightarrow \beta +i T$. If we are above the Hawking-Page temperature $Z(\beta)$ can be estimated by the Euclidean AdS-Schwarzschild black hole saddle point
 \begin{equation}
     Z(\beta)\approx e^{-I_{BH}(\beta)} \,,
 \end{equation}
where $I_{BH}(\beta)$ is the on-shell action on the Euclidean black hole background.  For example, we find 
\begin{align}
  I_{BH}(\beta) = -\frac{\pi^{2}}{2G_{N}\beta} \,\,(\text{for $AdS_{3}$}) &&  I_{BH}(\beta)= \frac{\beta }{G_{N}}g(r_H) \,\, (\text{for $AdS_{5}$}) \,,
\end{align}
where we have set the AdS radius $\ell_{\textrm{AdS}}=1$ and $r_H$ is the horizon radius, while 
\be
g(r_H)=\frac{V_{3}}{8\pi}(-r_H^{4}+r_H^{2})
\ee
where $V_{3}$ is the dimensionless volume associated with the metric on a unit sphere. For the $AdS_{5}$ case, $r_H\approx \pi/\beta$ for small real $\beta$. A detailed discussion of the action can be found in \cite{Emparan:1999pm}. The central charge of the CFT$_{2}$ is $c=3/2G_{N}$ and the rank for the gauge group of the dual four dimensional $SU(N)$ $\mathcal{N}=4$ super Yang Mills theory is given by $N^{2}=\pi /2G_{N}$.

For small $T$ the complexified partition function $Z(\beta+iT)$ will be given in terms of the analytic continuation of the above actions. Thus for $T\ll \beta$, one gets the following for AdS$_{3}$,
\begin{equation}
   R(T)\approx e^{-\frac{2\pi^{2}}{\beta^{3}}c\;T^{2}},
    \end{equation}
which is exponentially small in the large central charge limit\footnote{There will be additional terms suppressed in $T^{2}/\beta^{2}$ which will not affect the exponential decay in the large $c$ limit as long as $t$ is smaller than $\beta$.}. Similarly for AdS$_{5}$, we find that $Z(\beta)\sim e^{\frac{\pi N^{2}}{\beta^{3}}}$ in the high temperature limit. Again for $T\ll \beta$, we have
\begin{equation}
    R(T)\approx e^{-\frac{12\pi}{\beta^{5}}N^{2}T^{2}},
    \end{equation}
As $T$ becomes larger and approaches $T\sim\beta$, the dominant saddle point will no longer be the black hole, as the analytically continued action can start to compete with thermal AdS. In addition the analytically continued black hole saddle point corresponds to a geometry with a complex metric, and as $T\sim O(\beta)$ this metric becomes 'unallowable' according to the criteria of \cite{Kontsevich:2021dmb}, see also \cite{Witten:2021nzp}. Thermal AdS becomes the dominant saddle point before the metric becomes not allowable \cite{Chen:2022hbi}.

An exponential decay of $R(T)$ in $N$ is to be expected even when $T\sim \beta$, since in this case the thermal AdS saddle dominates and, $|Z(\beta+iT)|^{2}$ $\sim e^{\tilde{g}(T)/\beta^{3}}$ where $\tilde{g}$ is $O(N^0)$ periodic function of time. Thus, the numerator of \eqref{bhsff}  $|Z(\beta+iT)|^{2}$ is $N^0$  while the denominator is $O(e^{N^2})$ leading to an exponentially suppressed $R(T)$.

\subsection[Weakly coupled, large N gauge theories]{Weakly coupled, large $N$ gauge theories}

It is interesting to consider the behavior of the SFF at small, or even vanishing 't Hooft coupling $\lambda$. In this case the bulk dual is stringy and moreover at $\lambda=0$, the spectrum of the dual CFT is (half)-integer-spaced and thus not chaotic at all. Nevertheless the decay \eqref{raexpect} is still valid for a certain time-scale, even in the free theory. This was discussed in detail in \cite{Chen:2022hbi}. For concreteness, we consider the partition function of free ${\cal N}=4$ SYM on ${\mathbb S}^3\times {\mathbb R}$, where the sphere has unit radius. It has the form \cite{Sundborg:1999ue,Aharony:2003sx}
\begin{equation}
\label{freepart}
    Z(\beta)=\int\mathcal{D}U\,\,e^{\sum_{R}\sum_{m}^{\infty}\frac{1}{m}z^{R}_{m}(\beta)\chi_{R}(U^{m})} \,,
\end{equation}
where $\mathcal{D}U$ is the invariant Haar measure on the gauge group normalized to one, $\chi_{R}$ is character in the representation $R$ and 
\begin{equation}
    z^{R}_{m}(\beta)= \sum_{R_{i,B}=R}e^{-m\beta E_{i}}+(-1)^{m+1}\sum_{R_{i,F}=R}e^{-m\beta E_{i}} \,,
\end{equation}
where the first sum is over bosonic states and the sum in the second term is over fermionic states.

The behavior of the SFF $\left|{Z(\beta+iT) \over Z(\beta)}\right|^2$, as well as of the microcanonical analogue  $Y_{E,\Delta E}(T)$, based on the analytic continuation of \eqref{freepart} was discussed in \cite{Chen:2022hbi}.

Even at $\lambda=0$ the SFF obeys \eqref{raexpect}, though in this case the Poincare recurrence time is very short, i.e. $4\pi$.\footnote{In our conventions conformal dimensions in the free theory are half-integers.} While in this limit the bulk theory does not admit a semiclassical gravitational description, we could still apply the procedure \eqref{defop} to identify operators with vanishing commutators with the Hamiltonian to all orders in $1/N$, though now they do not have a nice bulk interpretation.\footnote{To start with, the HKLL procedure cannot be implemented at subleading orders in $1/N$ due to the many stringy fields present in the bulk. Therefore, the issue of non-commutativity with the Hamiltonian does not stand out like it does in the case of Einstein gravity.} In doing so, we would need to be careful to take $t_*$ to be a short $O(1)$ time-scale which is less than $4\pi$.

Here we notice that similar results have been derived for the analytically continued superconformal index \cite{Choi:2022asl}, which can be thought of as the SFF for 2-sided eternal supersymmetric AdS black holes.

\subsection{Perturbative states around empty AdS }

We now briefly discuss the return probability for perturbative states around empty AdS. We want to consider states which have a large number of particles, but still small enough so that we can ignore gravitational backreation. We can get some useful estimates by considering a thermal gas of particles in AdS$_{d+1}$. These are dual to a gas generated by single-trace operators in the CFT. Suppose we have low-lying single-trace operators with conformal dimension $\Delta_i$. For simplicity we consider only scalars and we take the radius of AdS$_{d+1}$ to be 1. Then the partition function of single-particle states $z(\beta)$ and the multi-trace Fock-space partition function are respectively
\be
z(\beta) = \sum_i {e^{-\beta \Delta_i} \over (1-e^{-\beta})^{d}}\qquad,\qquad Z(\beta) = \exp\left[\sum_{n=1}^\infty {1\over n} z(n \beta)\right] \,.
\ee
It is now straightforward to do the analytic continuation 
\be
Z(\beta + i T) = \exp{\sum_{n=1}^\infty\sum_i {e^{-(n\beta+in T)\Delta_i}\over(1-e^{-n\beta+i nT})^d}} \,.
\ee
For scalar BPS operators dual to SUGRA modes, $\Delta_i$ is integer. Then it is obvious that the SFF  $R(T)=\left|{Z(\beta+iT) \over Z(\beta)}\right|^2$ has periodicity $T =T+2\pi $, as expected. What we want to estimate is the decay rate of the SFF at early times, and how close to 0 the SFF drops between the recurrences.

First we notice that the partition function factorizes to a product over $\Delta_i$. Hence we can study the behavior of a given $\Delta_i$ and we drop the sum over $i$.
If we first take the small $\beta$ limit, before analytically continuing, we find
\be
\label{hightgff}
  Z(\beta) \sim \exp\left[\zeta(d+1) {1\over \beta^d} \right] \,.
\ee
Using this approximation  we find that for early times
\be
R(T) \sim e^{-{d(d+1)\zeta(d+1)\over \beta^{d+2} } T^2} \,.
\ee
As expected the decay is controlled by the variance of $H$. Of course if we use the high temperature approximation \eqref{hightgff} to perform the analytic continuation, then we do not see the recurrences. At high temperature the SFF starts decaying quite rapidly, stays close to zero for a while and then goes back to 1 every $T=2\pi \times {\rm integer}$. To find an estimate of how closely it approaches zero  it is convenient to evaluate it at $T=\pi$. Suppose that the conformal dimension is an even integer. Then we find
\be
\label{platfree}
R(\pi) = { \exp\left[2\sum_{n=1}^{\infty} {1\over n} {e^{-n \beta \Delta} \over (1-(-1)^n e^{-n \beta})^d} \right]\over \exp\left[2\sum_{n=1}^{\infty} {1\over n} {e^{-n \beta \Delta} \over (1- e^{-n \beta})^d}\right]}
\sim e^{-(2-2^{-d})\zeta(d+1){1\over \beta^d}-{1\over 2^d}\log{\beta \Delta \over 2}}
\ee
So we see significant suppression at small $\beta$, though of course, the suppression does not scale like $e^{-N^2}$.

We expect a similar qualitative behavior for $R(T)$ for generic pure states of similar energy profile as the states studied above (namely high energy states whose energy scales as $O(N^0)$): they will have recurrences every $2\pi$, but the return probability will quickly decay to small values for $0<t<2\pi$. If we use \eqref{defop} for such states, with $t_*\sim O(1) < \pi$, then the commutator with $H$ will be suppressed by a factor of the order of \eqref{platfree} rather than $e^{-N^2}$. Note that this is not small enough, since the commutator we are trying to cancel is $\mathcal{O}(1/N)$, which in the large $N$ limit is much smaller than the suppression controlled by \eqref{platfree}.

\subsection{LLM geometries}

An interesting class of semiclassical states with  AdS$_{5}\cross S^{5}$ asymptotics in type IIB supergravity are the LLM geometries \cite{Lin:2004nb}. These are dual to ${1\over 2}$-BPS states in ${\cal N}=4$ SYM.  While these geometries do not break all of the asymptotic symmetries, they do provide a useful toy model where we can study in detail the behavior of the return probability as a function of time. 

The  ${1\over 2}$-BPS states in $\mathcal{N}=4$ SYM on $S^{3}\cross R$ are states that preserve 16 of the 32 supersymmetries of the theory in addition to the bosonic symmetries $SO(4)\cross SO(4) \cross R$ where $R$ corresponds to the Hamiltonian  $H-\hat{J}$ where $H$ is the Hamiltonian and $\hat{J}$ an $R$-symmetry generator. These states correspond to operators that lie in the $(0,J,0)$ representation of the $SU(4) \sim SO(6)$ R-symmetry and they saturate a unitarity bound for their conformal dimension. It is illuminating to consider the $\mathcal{N}=1$ vector and three chiral multiplet decomposition of the $\mathcal{N}=4$ theory. In this case the  scalars of the chiral multiplets are organized into $Z^{j}=\phi^{j}+i\phi^{j+3}$, where $j=1,2,3$, which are in the adjoint representation. We will focus on $j=1$ from now on without loss of generality. Then the states we are interested in correspond, via the state-operator map, to  single-trace operators of the form $\Tr(Z^{n_{i}})$, as well as multi-trace operators of the form $\Pi_{i} (\Tr(Z^{n_{i}}))^{r_{i}}$ 
\cite{Corley:2001zk,Lin:2004nb,Berenstein:2004kk}.

Since these operators saturate the unitarity bound $\Delta = J$, they correspond to the lowest Kaluza-Klein mode of $Z$ on $S^{3}$. This mode has a harmonic oscillator potential due to its conformal coupling to the curvature of $S^{3}$. Thus we are interested in gauge invariant states of the matrix $Z$ in a harmonic potential \cite{Berenstein:2004kk}. The ground state of this model, corresponding to empty AdS, is given by a Gaussian wave function
\begin{equation}\label{13.1}
    \Psi_{vac} = C e^{-\frac{1}{2} N^{2} \tr(Z^{2})} \,,
\end{equation}
 where $C= (\pi/N)^{-N^{2}/4}$ and we introduce the notation $\tr Z={1\over N}\; \Tr Z ={1\over N} \sum^{N}_{i}\nu_{ii}$. Fluctuations with operators with $\Delta=J \ll N$ will be small excitations around the ground state. As discussed in  \cite{yaffe1982large} excitations with $\Delta=J \sim N^{2}$ will be other coherent states which are given by
 \begin{equation}\label{13.2}
    \Psi = C [J(Z)]^{1/2}e^{- N^{2} \tr(\frac{1}{2}\phi(Z)^{2}-i \psi(Z))} \,,
\end{equation}
 parameterized by two functions $\phi(Z)$ and $\psi(Z)$ which are monotonically increasing and arbitrary functions of $Z$, respectively. $J[Z]$ is the Jacobian given by $\det[\partial\phi(Z)_{ij}/\partial Z_{kl}]$.

It is well known that one can describe such a system by $N$ fermions in a harmonic potential \cite{brezin1978planar}. In the large $N$ limit, states in such a system can be thought of as droplets in a two dimensional phase space, where for example a circular droplet corresponds to the ground state of the system \cite{brezin1978planar,Jevicki:1979mb,Shapiro:1980vx}. The precise connection between the functions $\phi(Z)$ and $\psi(Z)$ and the droplet picture on the phase space will be discussed in the next subsection.

In the bulk, the LLM solutions correspond to 10 dimensional geometries of asymptotically $AdS_{5} \cross S^{5}$ spacetimes, see appendix \ref{llm}, that are completely determined by a function $z$ on a two dimensional surface. In particular, specifying whether $z$ takes value $1/2$ or $-1/2$ at each point on this plane completely specifies the full bulk solution. This is in parallel with the two dimensional fermionic phase space mentioned earlier where the fermion takes occupation number $1$ (black) or $0$ (white) at each point in the phase space, giving droplet of a given shape. For instance, in the fermionic picture the ground state is a circular droplet of a certain radius, say $r_{0}$.
It corresponds in the bulk is to the empty $AdS_{5}\cross S^{5}$.

Fluctuations with operators of $\Delta=J \ll N$ correspond to having ripples in the edge of the circular droplet and corresponds to having gravitons propagating in the $AdS_{5}\cross S^{5}$ background. While operators of energy $\Delta= J \sim N$ correspond to giant gravitons in the bulk. Operators with $\Delta=J \sim N^{2}$ correspond to other bulk geometries and different shapes of droplets in the fermionic phase space \cite{Lin:2004nb,Berenstein:2017abm,Berenstein:2017rrx}. The geometries will not be time translation invariant (rotational invariant in the fermionic picture) in general\footnote{There are also static configurations, concentric circles for example \cite{Lin:2004nb}}, but they are invariant under $t \rightarrow t + 2\pi$.

The goal here is to consider a certain geometry that breaks time translation invariance and compute its return probability for short time scales. In the fermionic picture this corresponds to a droplet that breaks the rotational invariance, an ellipse for instance. In the matrix quantum mechanics picture it is easy to compute the return probability, evolving (\ref{13.2}) with the quadratic Hamiltonian and computing the square of the inner product. But first, we need to review the dictionary between the two pictures.  

\subsubsection{Computation of the return probability}

The way the matrix quantum mechanics picture and the fermionic picture are related will be obvious once we diagonalize the matrix $Z$ and express it in terms of the eigenvalues ($\mu_{i}$), where the Jacobian becomes 
\begin{equation}
   J(Z)=\prod_{i}^{N}\phi^{'}(\mu_{i})\prod_{i\neq j}\frac{\phi(\mu_{i})-\phi(\mu_{j})}{\mu_{i}-\mu_{j}}  \,,
\end{equation}
which is 1 for the vacuum. In the large $N$ limit, the Gaussian measure $dZ \exp(-N^{2} \tr(Z^{2}))$ will reduce to the well known Wigner semi-circle distribution for the density of eigenvalues \cite{1959Potf},
\begin{equation}
    d\varrho(\mu)=\frac{1}{\pi}(2-\mu^{2})^{1/2}\Theta(2-\mu^{2})d\mu.
\end{equation}
Let us now introduce new variables to parameterize the coherent states in the large $N$ limit, $w(\mu):=d\varrho(\phi(\mu))/d\mu$ which is the density of eigenvalues and $v(\mu):=\psi(\mu)$. These parameters are canonical conjugates of one another\footnote{Note that the two variables are not totally independent and $w(\mu)$ has to satisfy a constraint, in particular $\int d\mu \; w(\mu)=1$.}, that is their Poisson bracket is the Dirac delta function. In the large $N$ limit, the appropriately renormalized Hamiltonian ($h_{cl}$) can also be written in terms of $w$ and $v^{'}=dv/d\mu$ and thus an action can be written for these variables \cite{yaffe1982large,Dhar:1992rs, Dhar:1992hr,Dhar:1993jc}. In particular,
\begin{equation}\label{6.30}
    h_{cl}=\frac{1}{2}\int\;d\mu \;w(\mu)(v^{'}(\mu)^{2}+\frac{\pi^{3}}{3}  w(\mu)^{2}+\mu^{2}).
\end{equation}
Coming back to the two dimensional phase space picture, we consider a blob centered at the origin. We assume the horizontal direction (x-axis) represents the $q$ variable of the phase space, which we take to be the eigenvalues ($\mu$). Consider a vertical line crossing the blob. Assuming that the blob has a simple geometry without folds, this vertical line intersects the boundary of the blob twice. We parametrize these points by $p_{\pm}(\mu)$ respectively. Then, the density of eigenvalues for any $\mu$ is proportional to $(p_{+} - p_{-})(\mu)$. Computing the kinetic energy of fermions for a given $d\mu$ by  integrating $p^{2}/2$ from $p=p_{-}$ to $p=p_{+}$ and matching this to the kinetic part of (\ref{6.30}), we get
\begin{equation}\label{13.19}
  p_{\pm} = \pm \pi w+ v^{'}.
\end{equation}
This has also been mentioned in the context of $c=1$ string theory in \cite{Polchinski:1991uq,Ginsparg:1993is,Das:1992dm,Das:2004rx}. Note that for the vacuum (i.e. the empty AdS$_{5} \cross S^{5}$ geometry), $p_{\pm}=\pm(2-\mu^{2})^{1/2}\Theta(2-\mu^{2})$ and $v'=0$.

Since we are looking for a time dependent geometry, we need a blob in fermion phase space that breaks the rotational symmetry. The simplest non trivial modification of (\ref{13.19}) is to take $v$ to be quadratic\footnote{Translated circle blobs will not correspond to physical geometries when the gauge group is $SU(N)$, since the centre of the blob is fixed by imposing the condition $\Tr(Z)$=0.}. In this case we have
\begin{equation}
p_{+}(\mu)= (2-\mu^{2})^{1/2}\Theta(2-\mu^{2})+2\mu.    
\end{equation}
 This can be seen to be half of a tilted ellipse, which combined with an appropriate $p_{-}$ gives the full elliptic blob. This will evolve non trivially under rotation and the corresponding geometry will be a time dependent one. This geometry, together with the five form, can be found using the mapping discussed earlier, by first solving for $z(x_{1},x_{2},y)$ then inserting it into (\ref{13.5}), (\ref{13.6}), (\ref{13.7}) and (\ref{13.8}).

Now we proceed with the computation of the return probability for this state. We go back to (\ref{13.2}) and consider a state $\Psi(0)$ with  $\phi=Z$ and $\psi=v=Z^{2}$ and after evolving it, compute the 
overlap
\begin{equation}
 \label{13.21}
\langle \Psi(0)|\Psi(T)\rangle=
     \int dZ \; \Psi(Z,0)^{*}\Psi(Z,T) 
 \end{equation}
The state we are interested has the form 
\begin{equation}
    \Psi(Z,0) = \left(\frac{\pi}{N}\right)^{-N^{2}/4} e^{-\frac{1}{2} N^2(1-2i) \tr(Z^{2})}=\prod_{i,j}\varphi(\nu_{ij})\text{  ,   where } \varphi(\nu)=\left(\frac{\pi}{N}\right)^{-1/4} e^{-\frac{N}{2}(1-2i)\nu^{2}} \,.
\end{equation}
Since we are dealing with matrix quantum mechanics with a quadratic potential, each matrix element evolves independently and governed by the usual harmonic oscillator propagator
\begin{equation}\label{13.24}
    \varphi(\nu,T)=\int d\nu^{'}K(\nu^{'},\nu,T)\varphi(\nu^{'}) \,,
\end{equation}
where 
\begin{equation}
   K(\nu^{'},\nu,T) = \sqrt\frac{N}{2\pi i \,\text{sin}T}
   \exp\left[\frac{iN}{2 \text{sin}T}((\nu^{2}+(\nu^{'})^{2})\text{cos}T-2\nu \nu^{'})\right] \,.
\end{equation}
for $t<\pi$. We can then compute the overlap
\begin{align}
      \langle \Psi(0)|\Psi(T)\rangle= [z(T)]^{N^{2}}, &&  z(T)=\int d\nu\; \varphi^{\star}(\nu,0)\varphi(\nu,T) 
\end{align}
Following \eqref{13.24} we find
\begin{align}
   \varphi(\nu,T )= \left(\frac{N}{\pi \mathcal{X}}\right)^{1/4}e^{-N \mathcal{Y}\nu^{2}} \text{ and},  &&  \Psi(Z,T) = \left(\frac{N}{\pi \mathcal{X}}\right)^{-N^{2}/4} e^{-N^2 \mathcal{Y} \tr(Z^{2})} \,,
\end{align}
where $\mathcal{X}$ and $\mathcal{Y}$ are periodic functions of time given by,
\begin{equation}\label{XandY}
  \begin{split}
    \mathcal{X}(T)&=( \text{cos} T +(2+i) \text{sin}T)^{2} \\
    \mathcal{Y}(T)&=\frac{1}{2}\left( \frac{(1-2i)\;\text{cos}T+i\;\text{sin} T}{(i+2)\;\text{sin} T+ \text{cos} T} \right) \,.
  \end{split} 
\end{equation}
Thus, 
\begin{equation}
   z(T)=\int d\nu\; \varphi^{\star}(\nu,0)\varphi(\nu,T) =  \mathcal{A}^{1/2} \,,
\end{equation}
where
\begin{equation}\label{A}
    \mathcal{A}=\frac{1}{3i\; \text{sin}T +\text{cos} T} \,.
\end{equation}
It can be checked that $z(T)$ is $1$ when $T=0$ and 
\begin{equation}\label{6.47}
    |z(T)|^{2}=|\mathcal{A}| = \left( \frac{1}{9\;\text{sin}^{2} T + \text{cos}^{2} T}\right)^{1/2} \,.
\end{equation}
Since $|z(T)|^{2}\leq 1$, $R(T)$ is an exponentially decaying function in the large $N$ limit, for small times, but a periodic function in $ T = \pi$.
That is
\begin{equation}
    R(T)=|\langle \Psi(0)|\Psi(T)\rangle|^2 =e^{-N^{2}F(T)} \,,
\end{equation}
where the function $F(T)=- 2\text{log}|z(T)|$ is zero at $T=0$, and increases to the local maximum $F(T=\pi/2) = \text{log }9$  and goes back to zero at $ T = \pi$. Thus in the time scales we are interested in, in particular $T< \pi/2 $, the square of the inner product (\ref{13.21}) which is the return probability of a given LLM semi classical geometry in the large $N$ limit, is exponentially suppressed in $N^{2}$ as expected. Note that the return probability is periodic in $\pi$, which is due to the symmetry of the particular state considered. In general, the period will be $2\pi$.

We can also compute the overlap of states in different code subspaces built upon $\Psi(Z,0)$ and $\Psi(Z,T)$. The simplest is the inner product of the states $\Psi(Z,0)$ and $\Tr(Z^{2n})\Psi(Z,T)$ which can be written as
\be
\begin{split}
\langle \Psi(0)|\,\tr(Z^{2n})\,|\Psi(T)\rangle &\equiv\int dZ\; \Psi^{\star}(Z,0) \tr(Z^{2n})\Psi(Z,T)\\
& = \left(\frac{\pi}{N\mathcal{X}^{1/2}}\right)^{-N^{2}/2}\int  dZ\; \tr(Z^{2n})\,e^{-{\mathcal{S}\over 2}N^2\;\tr(Z^{2})} 
\end{split}
\ee
where $\mathcal{S}= (1+2i)+2\mathcal{Y}$. Following (\ref{timdepcor}), we can rewrite the above integral as
\be
\langle \Psi(0)|\,\tr(Z^{2n})\,|\Psi(T)\rangle = \langle \Psi(0)|\Psi(T)\rangle \, \frac{\int   dZ\; \tr(Z^{2n})\,e^{-{\mathcal{S}\over 2}N^2\;\tr(Z^{2})} }{\int dZ\; \,e^{-{\mathcal{S}\over 2}N^2\;\tr(Z^{2})}}
\ee
The second factor corresponds to an expectation value in a Gaussian matrix model. Keeping only planar diagrams at large $N$ we find
\be
\langle \Psi(0)|\,\tr(Z^{2n})\,|\Psi(T)\rangle\simeq \langle \Psi(0)|\Psi(T)\rangle \, \frac{C_{n}}{\mathcal{S}^{n}}
\ee
where $C_{n}=\frac{1}{n+1}\binom{2n}{n}$ are the Catalan numbers. 

Similarly, for multi-trace operators the overlap  can be computed and using large $N$ factorization we get
\be
\langle \Psi(0)|\,\prod_{i}^k \tr(Z^{2n_i})\,|\Psi(T)\rangle\simeq \langle \Psi(0)|\Psi(T)\rangle \, \frac{\prod_{i}^{k}C_{n_{i}} }{\mathcal{S}^{n}},
 \ee
where $n=n_{1}+...+n_{k}$. 

Thus, as long as $n$ does not scale with $N$ the correlator will still be exponentially suppressed, otherwise the periodic coefficient can spoil the exponentially decaying behaviour. This is to be expected since in such cases the dimension of the multi-trace operators will be order $N$ and they will not be just small fluctuations of the background and can, in principle, evolve the state back in time to $T=0$. In any case, our code subspace is constructed by the action of multitrace operators whose dimension is finite in the large $N$ limit, i.e, $n$ is an $O(1)$ number.

\subsection{Kourkoulou-Maldacena states in SYK model}\label{kmsec}

The SYK model is a quantum mechanical model of $N$ Majorana fermions
interacting with random interactions
which
is given by the Hamiltonian
\begin{equation}
H = \sum_{i k l m} j_{iklm}~\psi_i \psi_k \psi_l \psi_m    \,,
\end{equation}
where $\psi_i$ are the Majorana fermions $ \{ \psi_i ,\psi_j\} = \delta_{ij}$, and the coupling $ j_{iklm}$ has drawn from the distribution 
\begin{equation}
    P(j_{iklm}) \sim \exp( - N^3 j_{iklm}^2 / 12 J^2) \,,
\end{equation}
leading to disorder average of 
\begin{equation}
    \overline{j_{iklm}} =0 , \qquad \overline{
    j_{iklm}^2} = \frac{3! J^2}{N^3}.
\end{equation}

In a particular realization of the couplings, we consider pure states which are obtained  by using the Jordan-Wigner transformation and combining pairs of Majorana fermions into qubit like operators and choosing states with definite eigenvalues for the $ \sigma _3$ components of all qubits. 
These states are denoted by $ \ket{B_s}$, where $ s = (s_1, s_2, ..., s_{N/2})$ with $s_k = \pm 1$, and they satisfy the relations below
\begin{equation}
    S_k \ket{B_s} = s_k \ket{B_s},
\end{equation}
where $ S_k = \sigma _3^k /2 \equiv 2 i~ \psi^{2k-1} \psi^{2k}$ is the spin operator.
By choosing all possible combinations of the $\{s_k\}$'s we get a basis of the Hilbert space whose dimension is $ 2^{N/2}$ ($N$ is an even integer number). We further evolve these states over some distance $ l$ in Euclidean time in order to get low energy states $ \ket{B_{s,l}} = e^{- l H } \ket{B_s}$ which we will refer to as Kourkoulou-Maldacena (KM) states. To stay in the low-energy regime where the SYK model exhibits conformal invariance we take $ 1 \ll l J \ll N $ \cite{Kourkoulou:2017zaj}. 

As discussed in \cite{Kourkoulou:2017zaj} the KM states can be thought of as a toy model of pure black hole microstates which are out of equilibrium and which contain excitations behind the horizon. Hence they are states which exhibit time-dependence and our general formalism should be applicable. We start by discussing the behavior of the return probability for these states.

\subsubsection{Analytical computation of the return probability at large $N$}

We start with the normalization of the KM states. In the large $N$ limit, due to the approximate $O(N)$ symmetry of the theory it can be shown \cite{Kourkoulou:2017zaj} that
\begin{equation}
    \langle B_{s,l} \ket{B_{s,l}} = \bra{B_s} e^{- 2 l H} \ket{B_s}
    = 2^{- N/2} Z(\beta) \,,
\end{equation}
where $ \beta =2 l$ \cite{Kourkoulou:2017zaj}.
The return probability then in the large $N$ limit is given by
\begin{equation}
    R(T) = \Big|\frac{\langle B_{s,l}| e ^ {-iHT} \ket{B_{s,l}} }{\langle B_{s,l} \ket{B_{s,l}} }\Big|^2 = \Big|\frac{Z(\beta + i T)}{Z(\beta)}\Big|^2.
\end{equation}
In a low temperature expansion, the partition function can be estimated \cite{maldacena1604comments} using the Schwarzian approximation to be
\begin{equation}\label{12}
    Z(\beta) \propto  \frac{e^{2 \sqrt{2} \pi ^2 \alpha _S \frac{N}{\beta J}}}{(\beta J)^{ 3/2}}.
\end{equation}
Using \eqref{12}  we find for the return probability 
\begin{equation}
 \label{RSYK}
         R(T)= \frac{1}{(1+ \frac{T^2}{\beta ^2})^{3/2}}
         e^{- (4 \sqrt{2} \pi ^2 \alpha _S \frac{N}{J\beta ^3}) T^2} \,,
 \end{equation}
 which is compatible with \eqref{raexpect}, after we take into account the different $N$-dependence in the SYK model vs ${\cal N}=4$ SYM.

We can now try to test the more general decay of the inner product between states in time-shifted code subspaces \eqref{cordecay}. Let us denote the unit-normalized KM states as
\be
|\widehat{B}_{s,l}\rangle  = {|B_{s,l}\rangle \over \sqrt{\langle B_{s,l}|B_{s,l}\rangle}} \,,
\ee
and denote their time-dependence as $|\hat{B}_{s,l}(T)\rangle = e^{-i H T}|\widehat{B}_{s,l}\rangle$. We consider an operator $A(t)$ which is a simple combination of the fermions, so that the state $A(t) |\hat{B}_{s,l}\rangle$ is in the code subspace. Then we write
\be
\langle \widehat{B}_{s,l}(0)| A(t) |\widehat{B}_{s,l}(T)\rangle = \langle \widehat{B}_{s,l}(0)|\widehat{B}_{s,l}(T)\rangle \times 
{\langle B_{s,l}(0)| A(t) |B_{s,l}(T)\rangle \over \langle B_{s,l}(0)|B_{s,l}(T)\rangle}  \,.
\ee
Let us focus on the last ratio. We can rewrite it as
\be
{\langle B_{s,l}(0)| A(t) |B_{s,l}(T)\rangle \over \langle B_{s,l}(0)|B_{s,l}(T)\rangle}  = {\langle B_s|    e^{-(l+i {T \over 2})H}  A(t-{T \over 2}) e^{-(l+i {T \over 2}) H}|B_s\rangle \over \langle B_s| e^{-(l+i {T \over 2})H}  e^{-(l+i {T \over 2}) H} |B_s\rangle}  \,,
\ee
which depends holomorphically on $l+i {T\over 2}$, so we can evaluate if by analytic continuation. All in all we find
\be
\label{kmcor}
\langle \widehat{B}_{s,l}(0)| A(t) |\widehat{B}_{s,l}(T)\rangle 
= \langle \widehat{B}_{s,l}(0)|\widehat{B}_{s,l}(T)\rangle \times  \left[\langle \hat{B}_{s,l}(0)| A(t-{T\over 2})|\hat{B}_{s,l}(0)\rangle\right]_{{l\rightarrow l + i {T \over 2}}} \,.
\ee
At large $N$ and for flip-invariant operators \cite{Kourkoulou:2017zaj} we can also write this as
\be
\langle \widehat{B}_{s,l}(0)| A(t) |\widehat{B}_{s,l}(T)\rangle 
= \langle \widehat{B}_{s,l}(0)|\widehat{B}_{s,l}(T)\rangle \times  \langle A(t-{T\over 2})\rangle_{\beta}|_{\beta\rightarrow \beta + i T} \,,
\ee
where in the last term we first compute the thermal 1-point function  $\langle A(t-{T\over 2})\rangle_{\beta}$ as a function of $\beta$ and then analytically continue $\beta$.

As an example, we consider the case where $A =\psi^k (t) \psi^k(t')$ (no summation over $k$ implied). Following \cite{Kourkoulou:2017zaj} we have for real time and large $N$
\begin{equation}\label{15}
       \langle \widehat{B}_{s,l}(0)| \psi^k(t) \psi ^k(t') \,\,|\widehat{B}_{s,l}(0)\rangle=  G_\beta (t-t ') \,,
        \end{equation}
where, for $t>t'$, we have
\be
\label{thermalkm}
G_{\beta} (t-t') = \frac{\pi ^{1/4}}{\sqrt{2 \beta J}} \frac{e^{-i \pi /4}}{\sqrt{\sinh[\pi (t - i \epsilon) / \beta]}} ,
\ee
Therefore, using \eqref{kmcor} we get
\begin{equation}
\label{sykex1}
     \langle \widehat{B}_{s,l}(0)| \psi^k(t) \psi ^k(t') |\widehat{B}_{s,l}(T)\rangle   =
     \langle \widehat{B}_{s,l}(0)|\widehat{B}_{s,l}(T)\rangle
      \,\,G_{\beta + iT}(t-t') \,,
      \end{equation}
where the last term can be computed as the analytic continuation of \eqref{thermalkm}.

Similarly for $A=\psi^{2k-1}(t) \psi ^{2k}(t') S_k $ we have \cite{Kourkoulou:2017zaj}
        \begin{equation}
     \langle \widehat{B}_{s,l}(0)| \psi^{2k-1}(t) \psi ^{2k}(t') S_k |\widehat{B}_{s,l}(0)\rangle = 
         -2i s_k G_\beta (t) G_\beta(t') + O(1/N),
\end{equation}
hence
\begin{align}
\label{sykex2}
&\langle \widehat{B}_{s,l}(0)| \psi^{2k-1}(t) \psi ^{2k}(t') S_k |\widehat{B}_{s,l}(T)\rangle =  \langle \widehat{B}_{s,l}(0)|\widehat{B}_{s,l}(T)\rangle
     \times\\& \times  \left[
         -2i s_k G_{\beta+i T}(t-{T\over 2}) G_{\beta+iT}(t'-{T\over 2}) + O(1/N)\right] \,.
\end{align}
The examples \eqref{sykex1} and \eqref{sykex2} are consistent with our general expectations, see \eqref{cordecay} and \eqref{cordecayb}.

\subsubsection{Some numerical checks}

\begin{figure}
  \centering
  \begin{subfigure}[b]{0.32\linewidth}
    \includegraphics[width=\linewidth]{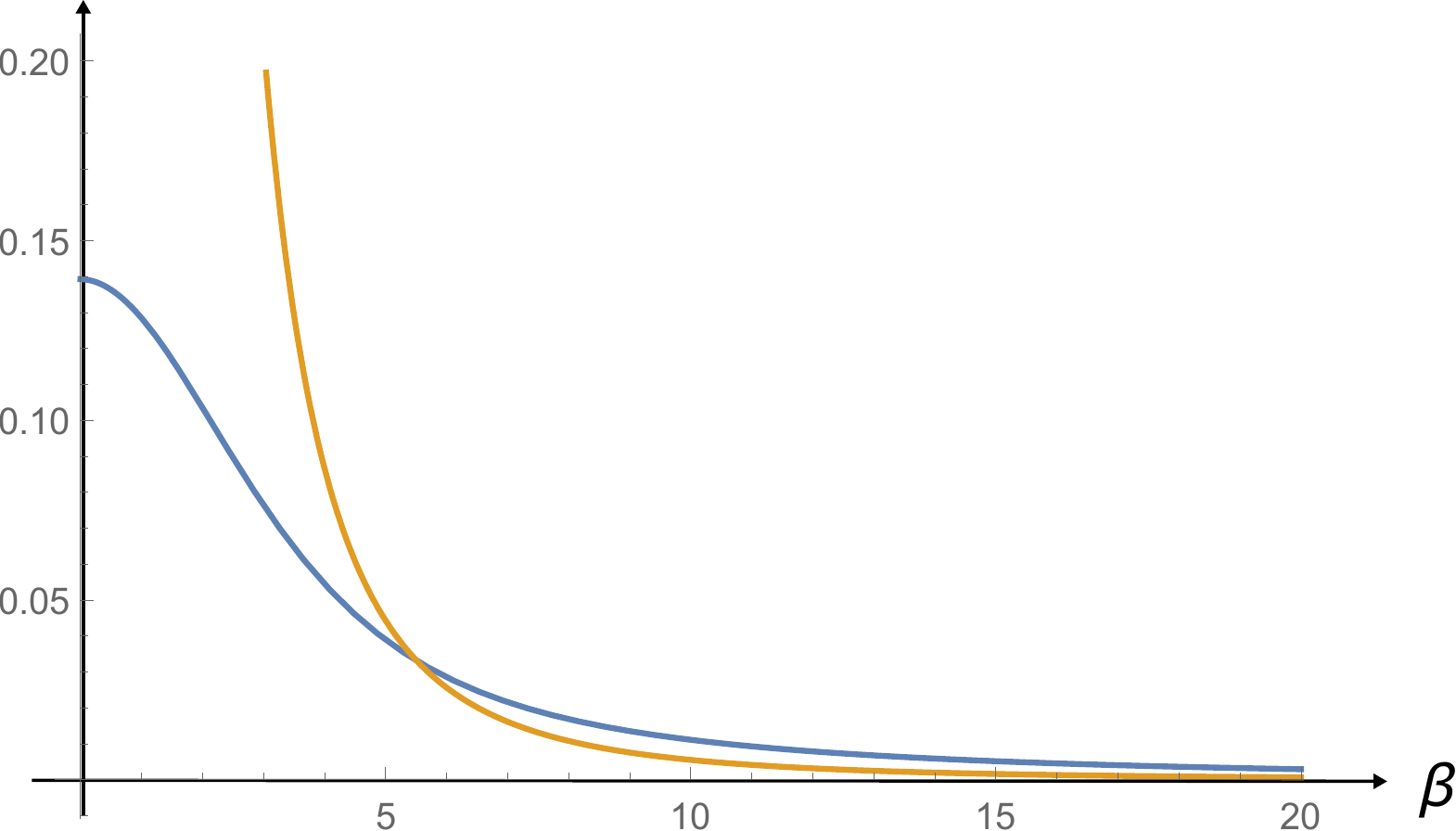}
    \caption{N = 14}
  \end{subfigure}
  \begin{subfigure}[b]{0.32\linewidth}
    \includegraphics[width=\linewidth]{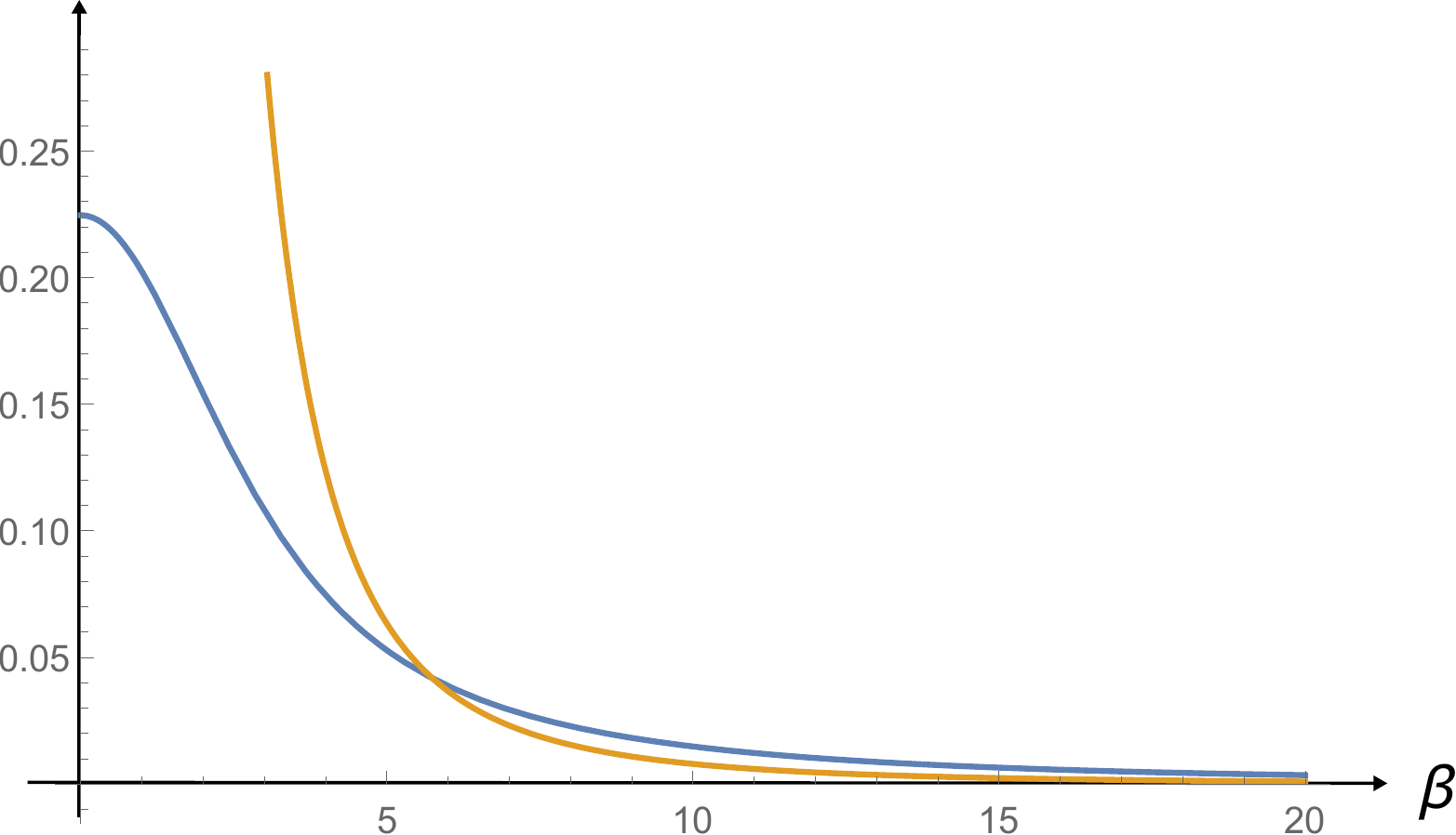}
    \caption{N = 20}
  \end{subfigure}
  \begin{subfigure}[b]{0.32\linewidth}
    \includegraphics[width=\linewidth]{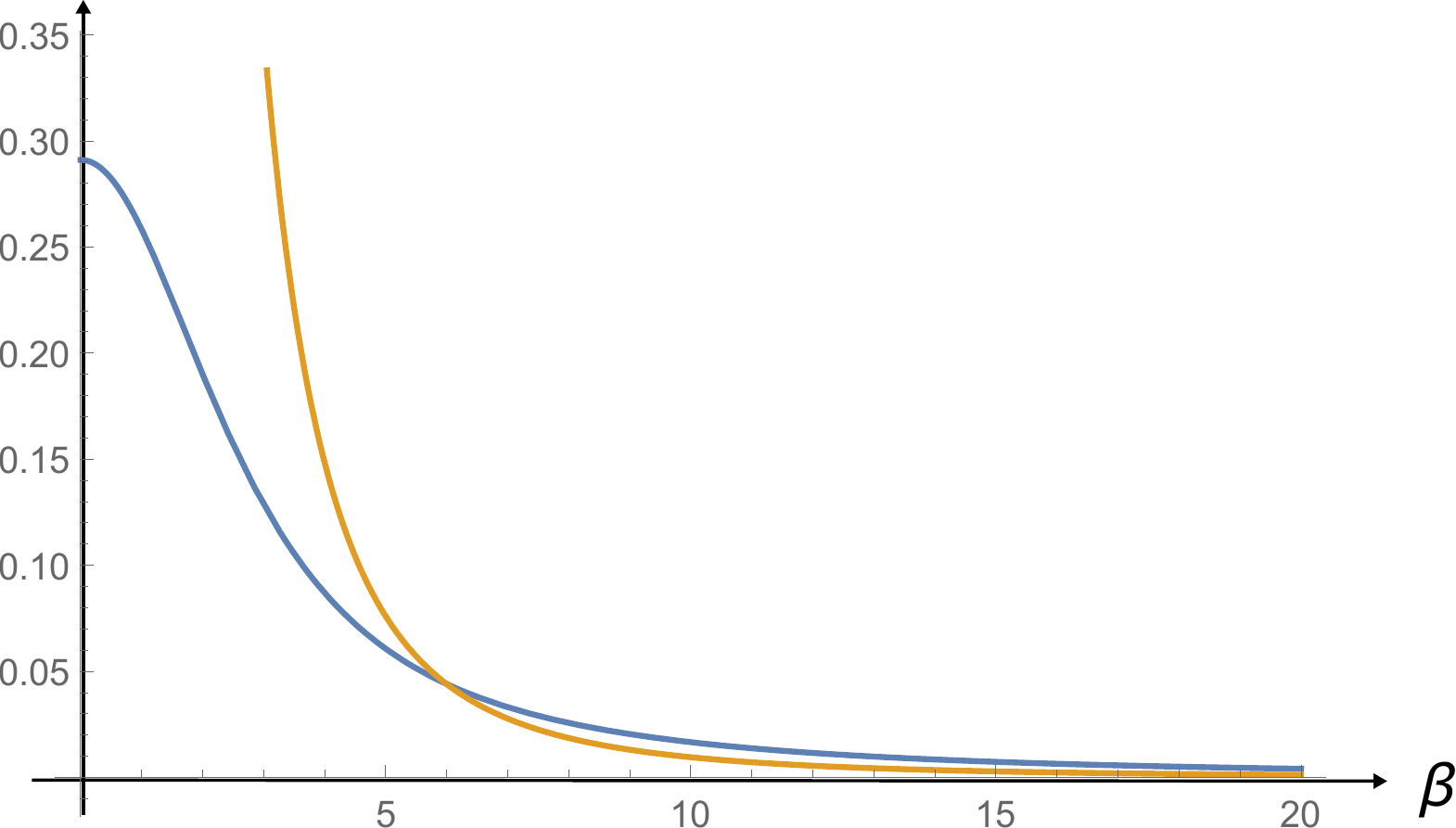}
    \caption{N = 24}
  \end{subfigure}
  \caption{The blue lines are the numerical results for the variance of Hamiltonian as a function of $\beta$ while the yellow ones are the Schwarzian approximation $\Delta H^2= 0.396 N/ \beta ^3.  $}
  \label{fig2}
\end{figure}

\begin{figure}[h!]
  \centering
    \includegraphics[width=0.5\linewidth]{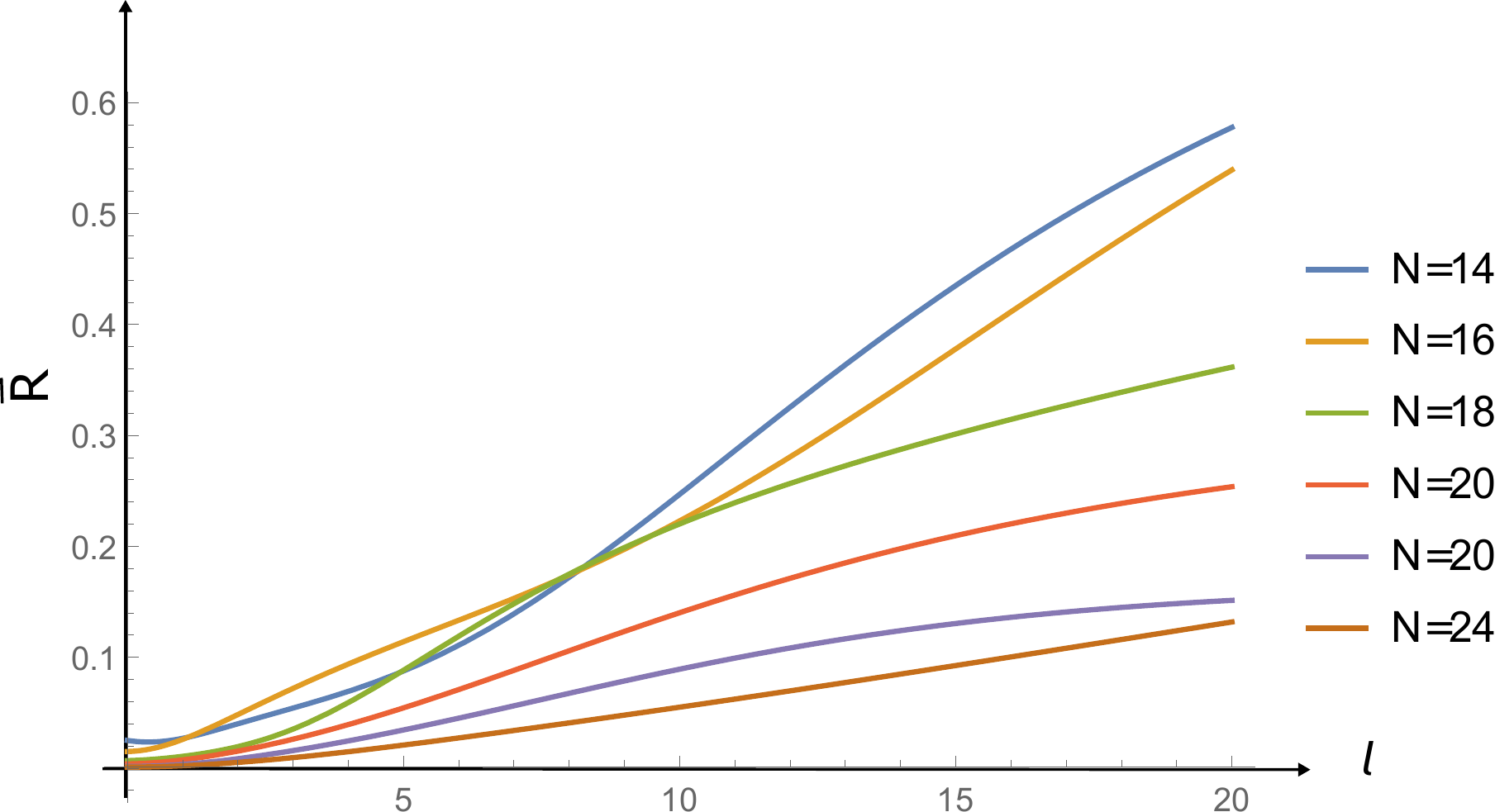}
\caption{The plateau height $\Bar{R}$ as a function of $ l = \beta /2$\,. \,\,  }
  \label{fig4}
\end{figure}

In this subsection we perform some simple numerical checks of  \eqref{cordecay} and \eqref{orthcode}, as well as the behavior of the operators \eqref{defop} for KM states in the SYK model. The first step is to select an appropriate value for the inverse temperature $\beta=2l$. The early time decay of the return probability is
\be
\label{sykdecay}
R(T) = e^{-\Delta H^2 T^2} \,.
\ee
Earlier we used the Schwarzian approximation to compute the partition function \eqref{12} from which we can also get the variance
\be
 \Delta H^2 = 4 \sqrt{2} \pi^2 \alpha_S \frac{N}{ \beta^3} = 0.396  \frac{N}{\beta^3} \,.
 \ee
 We compare this result with a numerical computation of the variance $\Delta H^2$ for a KM state constructed from $ \ket{B_s}=\ket{+--...-}$. This is shown in Figure \ref{fig2}. In Figure \ref{fig4}, we show the value of the plateau for the KM state, as defined in \eqref{plat} for various values of $N$ and $\beta$.  For the range of values of $N$ we are interested in, we can take the inverse temperature to be $\beta=5$, which is the value we will use in what follows.

\begin{figure}[h!]
  \centering
    \includegraphics[width=0.5\linewidth]{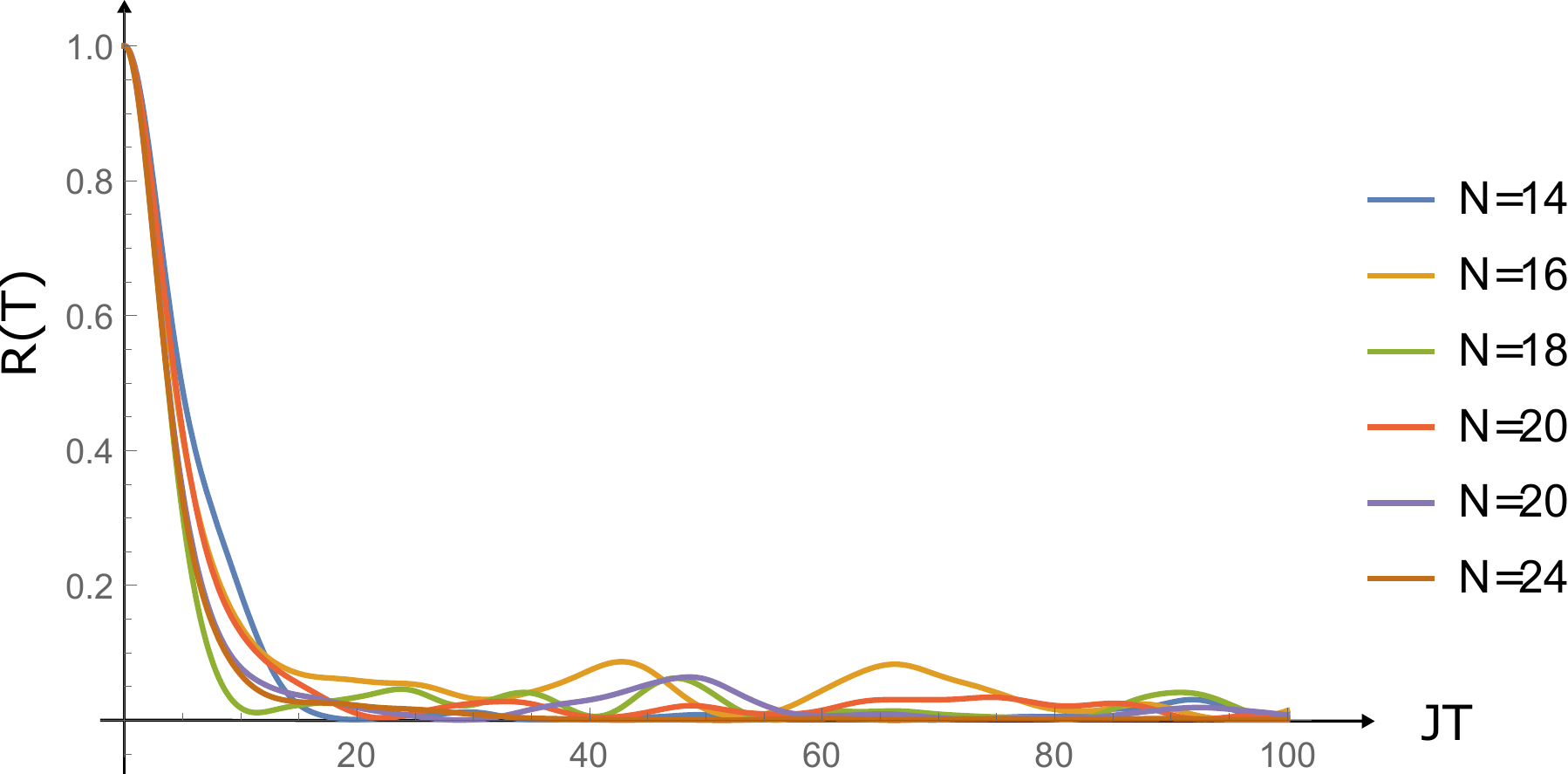}
  \caption{Return probability as a function of $T$ for different values of N}
  \label{fig6}
\end{figure}

In Figure \ref{fig6} we can see the return probability as a function of $t$ for different values of $N$ for the corresponding KM state. As discussed in subsection \ref{subsec:ra}, we expect that the overlap between any state in the code subspace at $t=0$ will and the one at $t=T$ will also decay exponentially fast. We can encode the overlap between all such pairs of states by
\be
R_{code}(T)={1\over d_{code}}\Tr[P_T P_0] \,.
\ee
For the numerical computation we need to make some choice about the code subspace. One condition is that the dimension $d_{code}$ of the code subspace should satisfy $ d_{code} \ll 2 ^{N/2}$. As an example, and for the purpose of the numerical computation, we can define the code subspace as 
\begin{equation}\label{codes}
    \mathcal{H}_{code} = \text{span} \{ {\cal O}_1^{i_1} ... {\cal O}_k^{i_k} |B_s\rangle; i_j = 0,1\} \,,
\end{equation} for some choice of the operators ${\cal O}_i$. Here $D_{code} = 2^{k} $ the value of $k$ should be such that $D \ll 2^{N/2}$. Note that the states in \eqref{codes} are generally not orthonormal but it is easy to write a projector on the code subspace in terms of elements of this basis, see  \cite{Bahiru:2022ukn} for a related discussion.

\begin{figure}
  \centering
  \begin{subfigure}[b]{0.32\linewidth}
    \includegraphics[width=\linewidth]{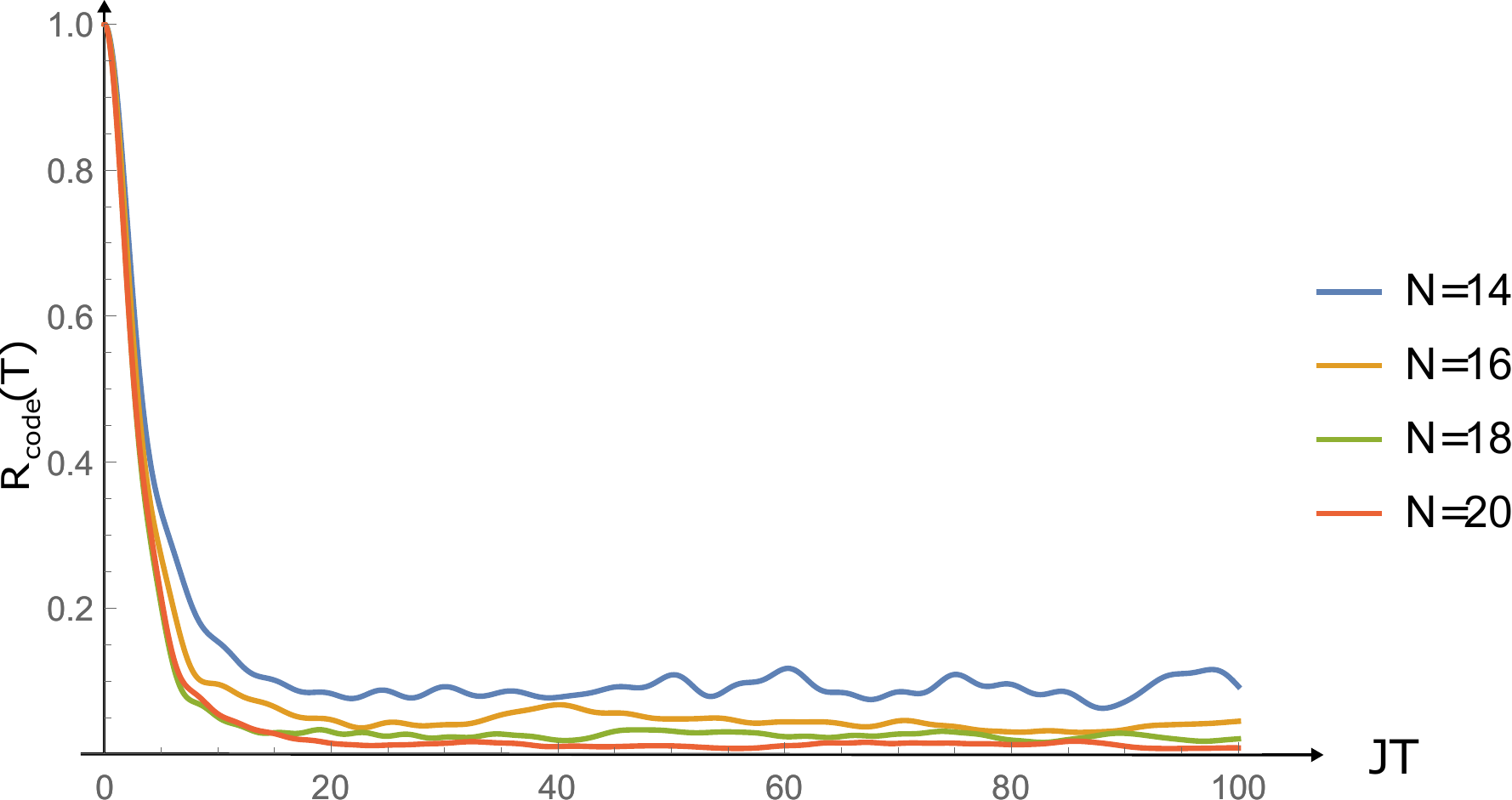}
    \caption{}
  \end{subfigure}
  \begin{subfigure}[b]{0.32\linewidth}
    \includegraphics[width=\linewidth]{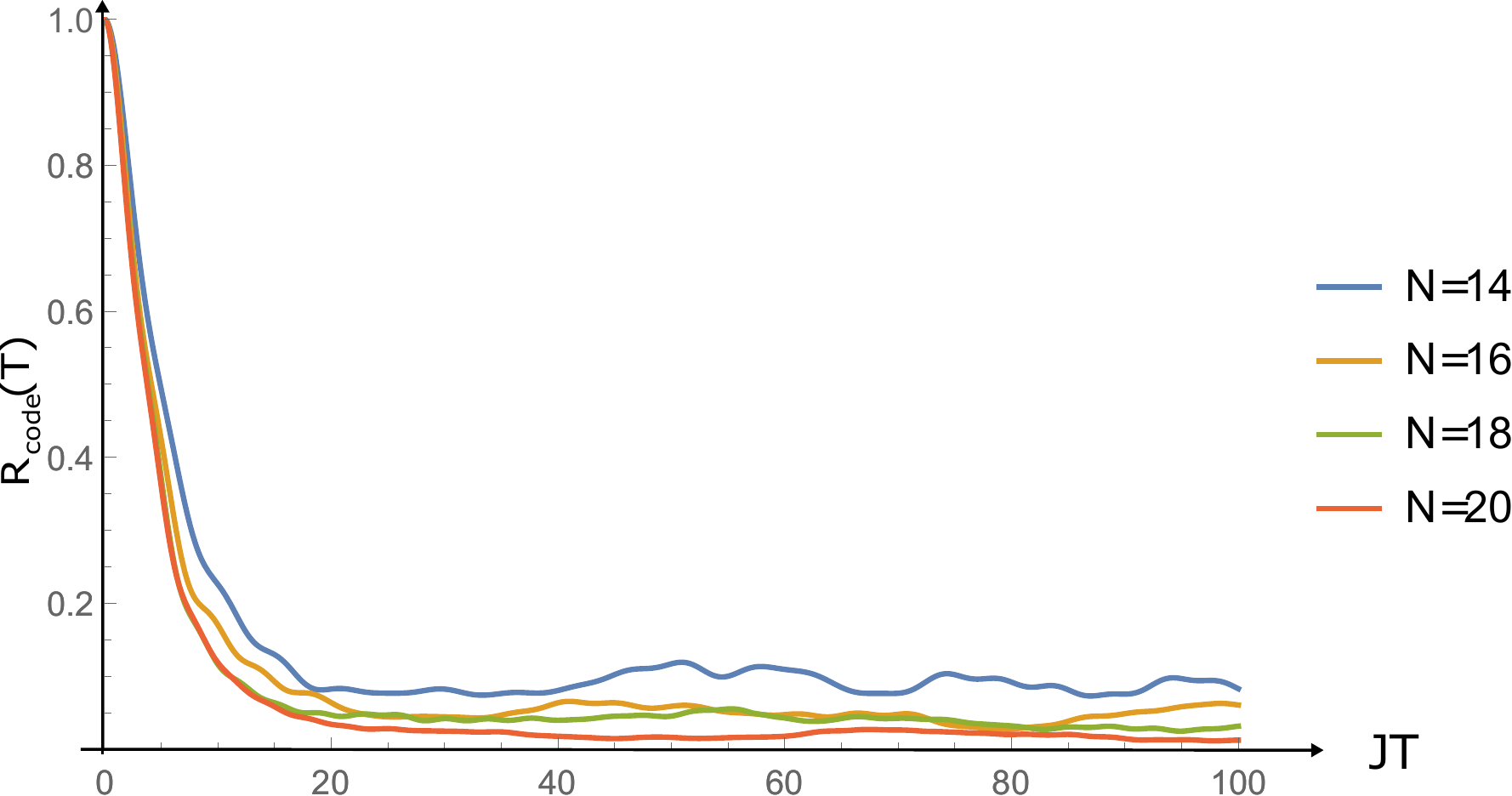}
    \caption{}
  \end{subfigure}
    \begin{subfigure}[b]{0.32\linewidth}
    \includegraphics[width=\linewidth]{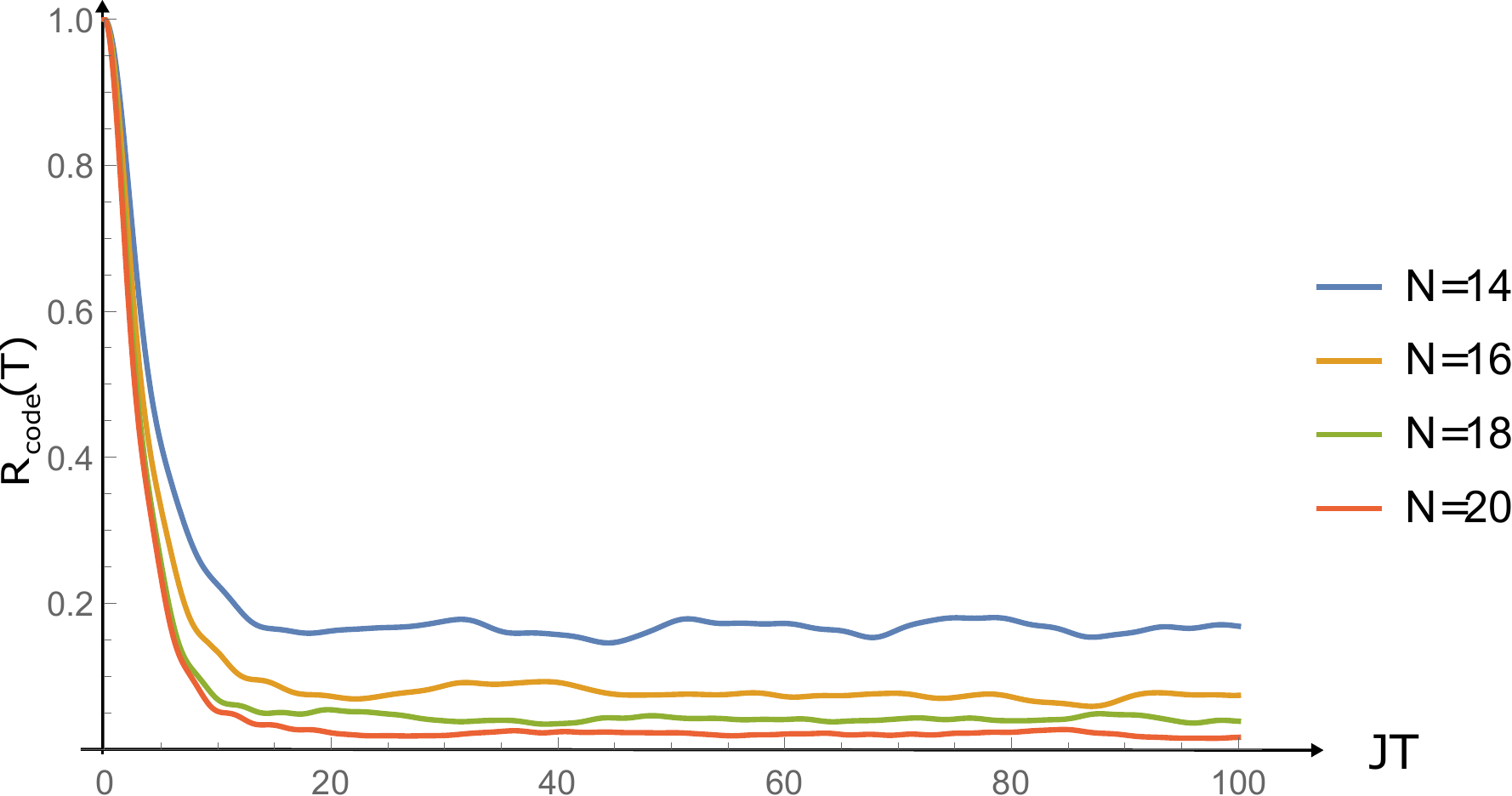}
    \caption{}
  \end{subfigure}

  \caption{$ R_{code}(T)$ as a function of $T$ for three different examples of codesubspaces in the form of \eqref{codes}.}
  \label{fig7}
\end{figure}

In Fig. \ref{fig7}, we see plots of the behavior of  $ R_{code}(T)$ as a function of time for some specific choices of such a code subspace:

\begin{itemize}
    \item \textbf{a :}
     the dimension of the code subspace is $ D =8$ and the operators are chosen to be 
     $$ {\cal O}_1 = \psi_1(t=0), \qquad\qquad {\cal O}_2 = \psi_1(t=0.1), \qquad\qquad {\cal O}_3 = \psi_1(t=0.5).$$
    \item \textbf{b :}
         the dimension of the code subspace is $ D =8$ and the operators  are chosen to b
     $$ {\cal O}_1 = \psi_1(t=0), \qquad\qquad {\cal O}_2 = \psi_1(t=0.1), \qquad\qquad {\cal O}_3 = h.$$
    \item \textbf{c :}
         the dimension of the code subspace is $ D = 16$ and the operators are chosen to be 
     $$ {\cal O}_1 = \psi_1(t=0), \quad\quad {\cal O}_2 = \psi_1(t=0.1), \quad\quad {\cal O}_3 = \psi_1(t=0.5),\quad\quad {\cal O}_4 = \psi_1(t=1).$$
\end{itemize}
where in case (b) the operator $h$ is the normalized Hamiltonian
\begin{equation}
    h= \frac{1}{\sqrt{N}} (H - \langle H \rangle).
\end{equation}

\begin{figure}
  \centering
    \begin{subfigure}[b]{0.45\linewidth}
    \includegraphics[width=\linewidth]{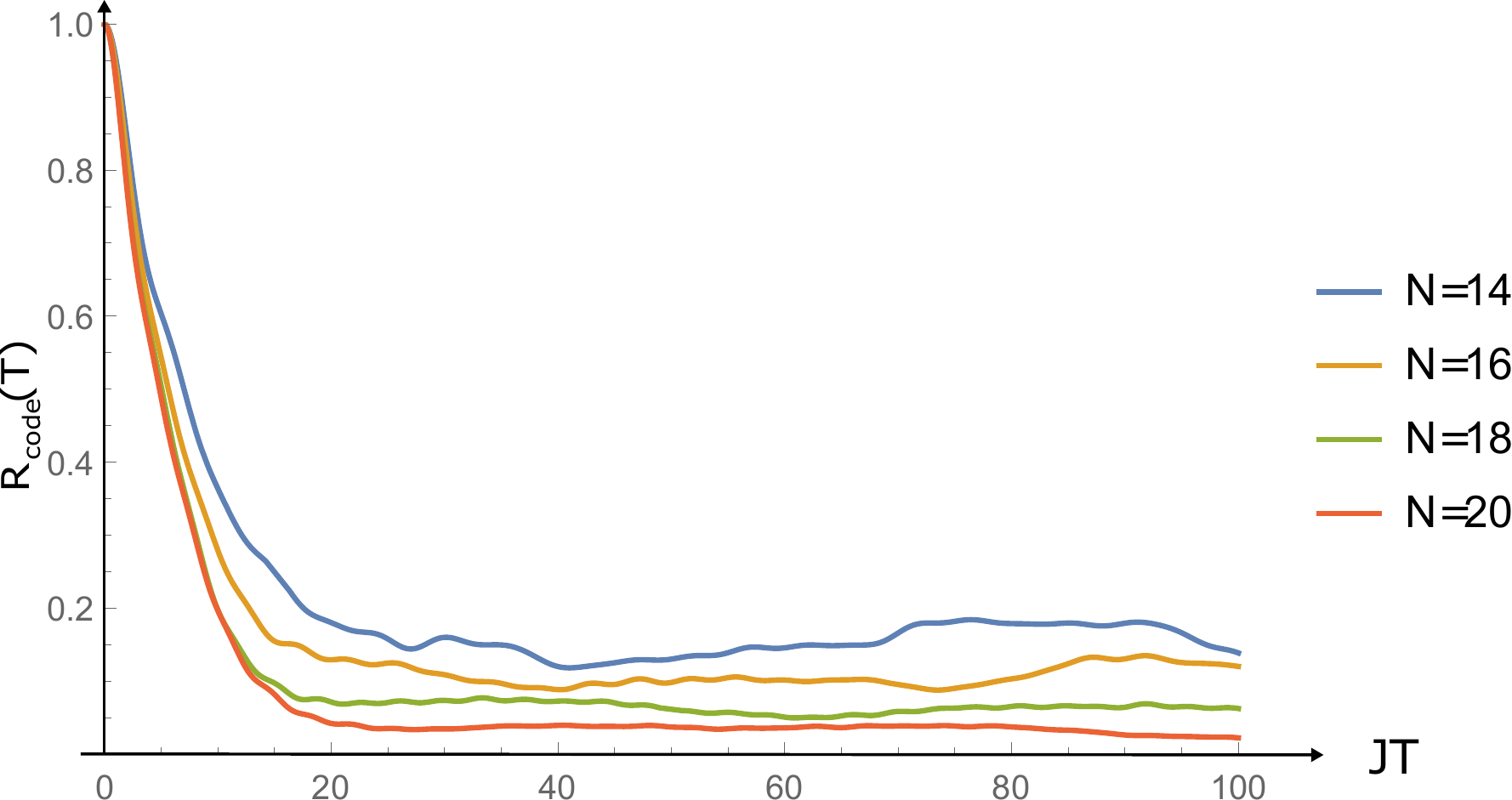}
    \caption{}
  \end{subfigure} 
    \begin{subfigure}[b]{0.45\linewidth}
    \includegraphics[width=\linewidth]{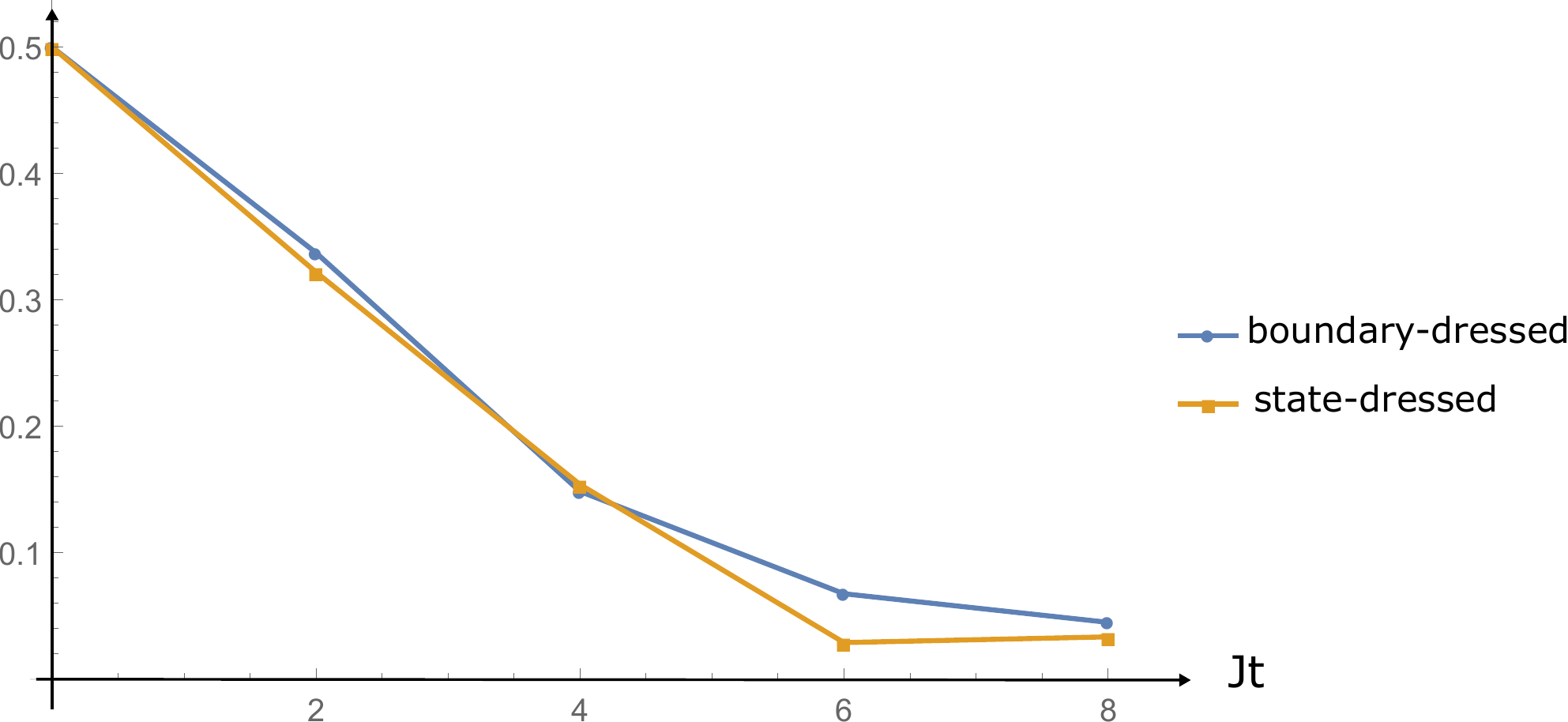}
    \caption{}
    \label{fig8.b}
  \end{subfigure}
  \caption{ Results for the code subspace \eqref{code, dress}. (a) $ R_{code}(T)$ as a function of $T$. (b) The blue line is  $\langle \psi_3(0)\psi_3(t)\rangle$ as a function of $t$, while in the case of the yellow line, $\psi_3(0)$ is replaced by the dressed operator obtained from our proposal. Here $N$=20.}
  \label{fig8}
\end{figure}

We finally check that the operator \eqref{defop} has similar correlators as the boundary-dressed operator. We take the code subspace as
\begin{equation}\label{code, dress}
    \mathcal{H}_{code} = \text{span} \{|B_s\rangle, {\cal O}_1|B_s\rangle,...  {\cal O}_k|B_s\rangle, h|B_s\rangle, h {\cal O}_1|B_s\rangle,...  h {\cal O}_k|B_s\rangle \},
\end{equation}
where the dimension of the code subspace is $d_{code} = 2(k+1) \ll 2^{N/2}$. In Fig. \ref{fig8}, we plot the result for the case of $ k=5$ and where the operators chosen to be
 $$ {\cal O}_1 = \psi_1(t=0), \quad {\cal O}_2 = \psi_1(t=2), \quad {\cal O}_3 = \psi_1(t=4) \quad {\cal O}_4 = \psi_1(t=6), \quad {\cal O}_5 = \psi_1(t=8)$$
for $ N=20$ ($ d_{code} =12 \ll 2^{10}$) are plotted. One can see from Fig.\ref{fig8.b} that the state-dressed operator for $\psi_3$ has approximately the same correlation function as the original one.

\subsection{Holographic boundary states}

The KM states discussed in the previous section can be thought of as certain a-typical black hole microstates in the context of SYK/AdS$_2$. Interesting analogs in higher dimensional examples of AdS/CFT can be found by considering boundary states in CFTs \cite{Almheiri:2018ijj,Takayanagi:2011zk,Karch:2000gx}. A boundary state characterizes boundary conditions which can be imposed on a boundary of space-time on which the CFT lives. For each allowed boundary condition, we can evolve the state along the Euclidean time to suppress the high-energy contributions and obtain a state of finite energy which is called a regularized boundary state of the CFT.
 
For holographic theories, the CFT path integral maps onto the gravity path integral. Therefore, we will be able to make use of the AdS/CFT correspondence to deduce the corresponding geometries if we can choose a state for which we can understand a gravity prescription for dealing with the boundary condition at the initial Euclidean time. As discussed in \cite{Cooper:2018cmb}, we can describe  boundary states by  starting with the TFD state of two CFTs labeled by L and R
\begin{equation}
    \ket{{\rm TFD}(\beta /2)} = \frac{1}{Z} \sum_i e^{-\beta E_i /4} \ket{E_i}_L \otimes \ket{E_i}_R,
\end{equation}
and then project the TFD state onto some particular pure state $ \ket{B}$ of the left CFT. As a result we obtain a pure state of the right CFT given by 
\begin{equation}
     \ket{\Psi_{B,\beta}} = \frac{1}{Z} \sum_i e^{-\beta E_i /4} \langle B\ket{E_i} \ket{E_i}.
\end{equation}

If the temperature is high enough, the TFD state is dual to the maximally extended AdS-Schwarzschild black hole in the bulk. The geometry which is dual to these regularized boundary states is expected to contain a significant portion of the left asymptotic region.
Therefore, in a holographic CFT, this class of regularized boundary states can be regarded as microstates of a single-sided black hole. 
These black hole microstates can be thought of as black holes with  end of the world (EOW) branes on the left side.\footnote{Proving from first principles that  boundary states dual to EOW branes exist is far from trivial. It has been investigated from a bootstrap perspective in \cite{Reeves:2021sab}, where it was suggested that such boundary states must be extremely fine-tuned. In \cite{Belin:2021nck}, the full classification of boundary states in large $N$ symmetric orbifolds was carried out, and typical boundary states are not of this form.} Generally the EOW brane configuration is time-dependent at the macroscopic level. Hence these are states with energy and energy variance compatible with \eqref{energyofstate} and \eqref{vardef}, so we expect to be able to apply our construction and define operators \eqref{defop}. As we will discuss in the next section, one way to think of them is that the gravitational dressing has been moved over to the EOW brane.

\subsubsection{Computation of the return probability and correlators}

First we define unit-normalized boundary states
\begin{equation}
 \ket{\widehat{B}_a(0)}= \frac{e^{- \frac{\beta H}{4}}\ket{B_a}}{\sqrt{\bra{B_a} e^{-\frac{\beta H}{2}}\ket{B_a}}}    \,.
\end{equation}
Then we want to show that return probability of a boundary state
\begin{equation}
    R(T)= |\langle \widehat{B}_a(0)\ket{\widehat{B}_a(T)}|^2 \,,
\end{equation}
decays exponentially fast at early time. For boundary states in holographic 2d CFTs we have \eqref{5}
\begin{equation}\label{bcftnorm}
     G(\beta)=  \langle B_a | e^{- \frac{\beta H}{2}}| B_a \rangle \simeq e^{\frac{\pi ^2 c}{6 \beta}}.
\end{equation}
where we have taken the CFT to be defined on a spatial circle of length $2\pi$. For small $T$ we have 
\begin{equation}
\label{ra2dbcft}
R(T)=    \frac{|G(\beta + 2 i T)|^2}{|G(\beta)|^2}  
\simeq e^{- \frac{4\pi^2 c}{3 \beta ^3} T^2}.
\end{equation}
The energy variance of the boundary state can be easily computed from \eqref{bcftnorm} and we find
\begin{equation}
\label{varbcft}
\begin{split}
     \Delta H ^2 = \langle H^2 \rangle - \langle H \rangle ^2 = 
   \frac{4\pi^2 c}{3 \beta ^3},
\end{split}
\end{equation}
so the initial decay \eqref{ra2dbcft} is, not surprisingly, consistent with \eqref{earlyr}, \eqref{raexpect} and \eqref{varbcft}.

 In higher dimensional cases we can read from \eqref{11}
\begin{equation}\label{7}
    G(\beta)= e^{\frac{\alpha _d}{\beta ^{d-1}}},  
\end{equation}
thus 
\begin{equation}
\label{rabcft}
    R(T)=\frac{|G(\beta + 2 i T)|^2}{|G( \beta)|^2} \simeq \exp\left[- \frac{\alpha_d}{\beta ^{d+1}} 4 d (d-1) T^2\right].
\end{equation}
We can again check that 
\begin{equation}
\begin{split}
     \Delta H ^2 = \langle H^2 \rangle - \langle H \rangle ^2 
          = \frac{\alpha_d}{\beta ^{d+1}} 4 d (d-1),
\end{split}
\end{equation}
which is compatible with \eqref{rabcft}.

We now proceed with checking that the other states in the code subspace around a boundary state are orthogonal to the time evolved code subspace. Consider for example the state ${\cal O}(t,x)|\widehat{B}_a\rangle$. Following similar reasoning as in subsection \ref{kmsec} we can show that
\begin{equation}
\label{bcft1}
|\langle \widehat{B}_a(0)| {\cal O}(t,x) |\widehat{B}_a(T)\rangle|^2 =  \langle \widehat{B}_a(0)| {\cal O}(t,x) |\widehat{B}_a(T)\rangle \,
\langle {\cal O}(t-{T\over 2},x) \rangle_{\beta \rightarrow\beta + 2 i T}.
\end{equation}
where $\langle \widehat{B}_a(0)| {\cal O}(t,x) |\widehat{B}_a(T)\rangle = \frac{G_a(I, \beta + 2 i T)}{G_a(I, \beta)}$.
More generally
\begin{multline}
\label{bcft1a}
\langle \widehat{B}_a(0)|{\cal O}(t_1,x_1){\cal O}(t_2,x_2)...{\cal O}(t_n,x_n) |\widehat{B}_a(T)\rangle
 =\\
  \langle \widehat{B}_a(0)| {\cal O}(t,x) |\widehat{B}_a(T)\rangle
    \langle {\cal O}(t_1-{T\over 2},x_1) {\cal O}(t_2-{T\over 2},x_2) ... {\cal O}(t_n-{T\over 2},x_n) \rangle_{ \beta \rightarrow\beta + 2 i T}. 
\end{multline}
Thus, as long as the analytical continuation of the correlation function in $ \beta$ does not introduce any surprising $N$-dependent factors we will get the expected behavior \eqref{cordecayb}. We now check this condition for low-point functions in 2d boundary states.

Here we assume that for a holographic CFT, and if we are working in the large $N$ limit, the 1-point function of light conformal primaries can be computed by a method of images. Then for a 1-point function of a scalar primary ${\cal O}$ with dimension $\Delta$ on a boundary state we have 
\begin{equation}
    \langle \widehat{B}_a(0)| {\cal O}(t,x)|\widehat{B}_a(0)\rangle =  \frac{A_{\cal O}}{(\frac{\beta}{\pi} \cosh [\frac{2 \pi}{\beta} t])^\Delta} .
    \end{equation}
    for some constant $A_{\cal O}$  which depends on the boundary state $a$ and the operator ${\cal O}$.
After the analytic continuation necessary for \eqref{bcft1} we find
\begin{equation}
\label{bcft2}
     \langle O(t- {T \over 2},x)\rangle _{\beta \rightarrow \beta + 2iT}  =
      \frac{A_{\cal O}}{(\frac{(\beta + 2iT)}{\pi} \cosh [\frac{2 \pi}{(\beta + 2iT)} (t-{T\over 2})])^\Delta} .
\end{equation}
 Hence we notice that the results \eqref{bcft1},\eqref{bcft2} are consistent with our general expectations \eqref{cordecay},\eqref{cordecayb}.

We can also check 2-point functions, which we can compute in the large $N$ limit. First we compute the 2-point function on the boundary state, using the method of images
\begin{equation}
\label{bcft3}
    \begin{split}& 
      \langle \widehat{B}_a(0)|  {\cal O}(t_1,x_1) {\cal O}(t_2,x_2)|\widehat{B}_a(0)\rangle  = 
       \\
    &\sum_{n=-\infty}^{+\infty}\frac{1}{\big|\frac{\beta}{\pi} \sinh(\frac{\pi}{\beta} [(x_1 - x_2+2\pi n) - (t_1 - t_2)])\big|^{2\Delta}} \pm
     \frac{1}{\big|\frac{\beta}{\pi} \cosh(\frac{\pi}{\beta} [(x_1 - x_2+ 2\pi n) - (t_1 + t_2)])\big|^{2\Delta}},  
    \end{split}
\end{equation}
After the analytic continuation necessary for \eqref{bcft1a} we find from \eqref{bcft3} that 
 we do not notice any unexpected behavior of this part of the correlator as $T$ increases, so the result \eqref{bcft1a} is dominated by the decay of the return probability, and is consistent with our expectations \eqref{cordecay},\eqref{cordecayb}.

\section{Black Hole microstates}\label{sec:bhs}

One question which is particularly interesting is whether we can apply our construction to black hole microstates. We have already mentioned in section \ref{sec:charges} that there are various classes of black hole microstates, some of which have macroscopic time dependence and some of which do not. We will now discuss these various cases in more detail and interpret our operators for these types of states.

\subsection{States with macroscopic time-dependence}

We will start with the simplest situation: states with macroscopic time-dependence. This can be visible outside the horizon, for example black holes in the presence of infalling matter. Alternatively it can be that the geometry appears to be static outside the horizon but there is no corresponding Killing isometry in the interior. As the first case is more straightforward, we focus on the second case. Two examples of such states are boundary states of the CFT, corresponding to end-of-the-world branes inside the horizon, which have already been discussed in the previous section. A second example is states prepared by the Euclidean path integral on some surface of higher topology. The dual geometries have topology behind the horizon, and are often referred to as geons \cite{Louko:1998hc,Guica:2014dfa,Marolf:2017vsk}. It is worth re-emphasizing that both of these states are usually prepared by the Euclidean path integral and are in fact very a-typical states, even if the CFT 1-point functions are very close to those in a thermal state (or said differently, even if the classical  geometry is exactly that of a black hole outside the horizon). 

Both of these examples involve pure states $\rsz$ that have a large energy variance, of order $N^2$, such that the return probability will decay as \eqref{decay}. We can thus apply our construction to build local operators that are not dressed to the boundary CFT. The interpretation is that the operators are dressed with respect to the time-dependence of the interior. Consider for example the genus-2 geon in $d=2$, which is prepared by the Euclidean path integral on half of a genus-2 surface \cite{Maxfield:2016mwh,Marolf:2017vsk}. Microscopically, the state can be described by
\be
\rsz \sim \sum_{i,j} C_{iij} e^{-E_i \beta_i/2-E_j \beta_j} \ket{E_j} \,,
\ee
where $\sim$ indicates that we have not been careful about the parametrization of the genus-2 surface, but $\beta_{i,j}$ are related to the moduli of the surface. The un-normalized overlap of this state corresponds to a genus-2 partition function in the dumbbell channel, where $\beta_j$ parametrizes the length of the two handles, and $\beta_i$ parametrizes the length of the neck between them. 

It is not straightforward to write down a metric that covers the entire space-time of such states. Outside the horizon whose size is controlled by $\beta_j$, they look exactly like the BTZ geometry. Inside the horizon, they have macroscopic time-dependence. A nice coordinate patch that covers the Wheeler-de Witt patch of the $t=0$ slice of the geometry can be written down in a very simple form
\be
ds^2= -dt^2 +\cos^2t \ d\Sigma_2^2 \,,
\ee
where $d\Sigma_2^2$ is the constant negative curvature metric on half of a genus-2 surface. This coordinate patch covers the entire $t=0$ slice of the geometry, which is precisely half of a genus-2 surface. The neck corresponds to the horizon, and there is topology (one handle) behind the horizon. From this metric, we explicitly see the time dependence of the geometry, even if a metric for the full spacetime is hard to write down. The interpretation of our operator is that the dressing is to the time-dependence of the geometry that sits inside the horizon. For end-of-the-world brane geometries, the situation is similar and the operator is dressed to the end-of-the-world brane.

\subsection{Typical states}

The question we would now like to ask is whether our prescription works in typical black hole microstates. Contrary to states with end-of-the-world branes or topology behind the horizon, it seems reasonable to expect that typical states should also look like the thermal state a finite distance inside the black hole (see for example \cite{deBoer:2018ibj,DeBoer:2019yoe}).

Whether or not our prescription works depends on the definition of a typical black hole microstate, and in particular on the energy spread we are choosing. One possibility is to define typical states using an ensemble of energy eigenstates with spread $\mathcal{O}(N^0)$ in energy (recall that there are still $e^S$ with $S\sim O(N^2)$ states in this energy band). In that situation, our prescription does not work, as the variance of energy is $\mathcal{O}(N^0)$ and the return probability will not decay fast enough. Another possibility is to consider typical states with an energy spread similar to that of the canonical ensemble, that is
\be
(\Delta E )^2 \sim \mathcal{O}(N^2) \,.
\ee
For such states, the return probability will decay following the behaviour \eqref{decay}. Therefore, we can follow our prescription and define the operators in the same way and they will satisfy the two properties of commuting with the Hamiltonian to all orders in $1/N$ and acting like HKLL operators to leading order at large $N$.

While these operators are certainly diff-invariant, since they are operators defined in the CFT, the bulk interpretation of their gravitational dressing on typical black hole microstates is not entirely clear. When the gravitational configurations are macroscopically time-dependent, our operators are dressed with respect to the features of the geometry. The typical states are still time-dependent, but only microscopically, as it seems plausible to assume that macroscopically they are  featureless. In some sense our operators are dressed to the microscopic time-dependence of the state (the phases of the $c_i$ in \eqref{statedef}), but it is unclear exactly what that means in the bulk.

Notice however, that if we start with a particular typical pure state $\rsz$, assume it has a smooth horizon without any perturbations behind it and then act with a unitary made out of the operator \eqref{defop}, associated to that state, then the predictions for what an infalling observer jumping into the black hole will see are unambiguous. For example, the operators \eqref{defop} will generally create an excitation in the bulk and the location in time relative to that of the infalling observer who jumps from the boundary at a particular boundary time, can be unambiguously computed for each state $\rsz$ and corresponding operators \eqref{defop}. We emphasize that for this interpretation it is important to remember that the operators \eqref{defop} are state-dependent and cannot generally be promoted to a single operator which acts in a specific way globally on most typical states.

We briefly comment on black hole interior reconstruction. Suppose we start with a typical black hole microstate with energy spread of order \eqref{vardef}. If we assume that the horizon is smooth, then the possibility of removing the dressing of the operators implies that we can deform the state behind the horizon by creating some particles there, in such a way that these excitations cannot be detected from the boundary CFT by the measurement of single-trace correlators, including the Hamiltonian, in the $1/N$ expansion. This was also discussed in \cite{Harlow:2014yoa,deBoer:2022zps}. We emphasize that this does not contradict the statements made in \cite{Papadodimas:2013jku,deBoer:2018ibj,DeBoer:2019yoe} that for typical states with {\it microcanonical} energy spread, it is impossible to add excitations without affecting single-trace correlators.

\subsection{Two entangled CFTs}\label{subsec:entangledcfts}

Similar considerations apply to geometries with two asymptotically AdS regions. Consider two non-interacting CFTs with total Hamiltonian $H=H_L+H_R$. We take the full system to be in a pure state $\rsz$ which may be entangled, but we will assume the pattern of entanglement is generic. In particular, we \textit{do not} consider states like the thermofield-double which have a very fine-tuned structure of entanglement. We can imagine the state $\rsz$ to be, for example, $U_L\ket{\textrm{TFD}}$, where $U_L$ is a complicated unitary acting on the left CFT. In this case we can consider the following generalization of our construction. Let us consider the 2-parameter family of time-shifted states
$$e^{-i (T_L H_L + T_R H_R)} \rsz.$$
We start with an HKLL operator $\Phi$ dressed with respect the to left system, which commutes with $H_R$ but not $H_L$. We now consider the following generalization of the operators \eqref{defop}
\be
\drop = c\int dT_L dT_R e^{-i ( T_L H_L + T_R H_R )} P_0 \hkll  P_0 e^{i (T_L H_L  + T_R H_R)}
\ee
using $P_0 = P_0^L \otimes P_0^R$ and  $[\hkll, P_0^R]=0$ then
\be
\drop = c\int dT_L  e^{-i T_L H_L} P_0^L \hkll P_0^L  e^{i T_L H_L}\otimes \int  dT_R P_{T_{R}}^R
\ee
The resulting operator commutes with both $H_L$ and $H_R$ on the relevant code subspaces. In this case, the operator is not dressed with respect to the overall time-dependence of the full system, but rather to the time dependence of the ``left'' subsystem. 

There are states with special entanglement pattern such as the TFD state, which was already discussed in section \ref{sec:tfd}. The generalized return amplitude $\braket{\Psi_0 | e^{-i (H_L T_L + H_r T_R)} | \Psi_0}$ which is a function of $T_L$ and $T_R$ does not decay in all directions for these special states. For example, in the TFD state it is constant along the line $T_L=-T_R$. In those cases we cannot set both commutators with $H_L,H_R$ to zero. So we can move the dressing from one side to another if we wish to, but there it is always dressed to one of the boundaries. This happens because the TFD state has a symmetry, it is annihilated by $H_L-H_R$.

\subsection{Island discussion}

Our prescription is also useful to resolve some paradoxes in the context of black hole evaporation and islands. Consider a setup where a holographic CFT is coupled to a bath such that the bulk description is given by an evaporating black hole. After the Page time, a non-trivial quantum extremal surface appears in the bulk delimiting an island, i.e. a part of the interior of the black hole that is encoded in the bath degrees of freedom rather than in those of the CFT \cite{penington2020entanglement,Almheiri:2019psf}.

There is an apparent tension in this context related to gravitational dressing already at the perturbative level \cite{Geng:2021hlu}. If we create an excitation in the island by acting with a local operator $\phi_{\textrm{island}}$, where does the gravitational dressing go? It appears that the only place for the dressing to go is the boundary CFT. This implies that the local operator will have the property
\be \label{nonzerocom}
[\phi_{\textrm{island}},H_{\textrm{CFT}}]\neq 0 \,.
\ee
where the RHS is $O(G_N)$. But this seems to be inconsistent, because since the operator is in the island, it should be reconstructable from the bath degrees of freedom, and commute with the CFT degrees of freedom.

Our operators provide a way out of this paradox in perturbation theory in $G_N$. We can apply our prescription above in terms of two entangled systems with a generic pattern of entanglement (there is a subtlety here since the bath and CFT are actually coupled rather than non-interacting, but we can treat this interaction as weak). In that case, even if we did start with an operator that had a non-trivial commutator \eqref{nonzerocom}, we would engineer a new operator which commutes with $H_{\rm CFT}$ up to exponentially small corrections. This new operator is now dressed with respect to the radiation, rather than the boundary CFT. Of course we expect that there is also a version of the operator in the island which exactly commutes with $H_{\rm CFT}$, even though we do not have an explicit expression for it. What we are pointing out is that the apparent paradox \eqref{nonzerocom} at the level of perturbative quantum gravity can be resolved.

The interpretation of the dressing is similar to that of the typical states. While it would be tempting to imagine dressing the operator to the quantum extremal surface, the bulk geometry only has extremely slow time-dependence so it is unclear if time-dependent features of the geometry are sharp enough to dress with respect to them. It appears that the dressing is towards the microscopic time-dependence of the radiation. The story becomes less subtle if we consider a doubly holographic model (see for example \cite{Almheiri:2019hni,Chen:2020hmv}). In that case, the dressing to the bath can be directly geometrized in the higher-dimensional geometry. Our operators can perhaps be thought as a counter-part of the operators in the doubly-holographic setup, but in cases where the dressing cannot be so easily geometrized.

Finally, we would like to clarify the distinction between reconstruction and dressing. To make things simple, let us consider the TFD state and consider an HKLL operator on the left $\phi_L$. This operator is dressed to the left CFT. Now we run our protocol, and as explained above, we can move the dressing to the right. The operator $\hat{\phi}_L$ now commutes with $H_L$ but no longer with $H_R$ \cite{Papadodimas:2015xma}. This \textit{does not} mean that it can be reconstructed from the right degrees of freedom, but that it can be detected from the right CFT via the Gauss law tail. It is still mostly built from the left CFT degrees of freedom, only its dressing has been pushed to the right.

\section{Discussion}\label{sec:discuss}

In this paper, we have investigated whether information can be localized in perturbative quantum gravity, in the context of the AdS/CFT correspondence. The challenge at hand is to construct local diff-invariant operators that are not dressed to the boundary where the CFT lives. We have presented evidence that such operators exist, at least around high energy states with a large energy variance. Such states include semi-classical geometries with features that break the symmetries of the dual CFT and for such states, local operators can be dressed to the features of the state. We have argued that there exist CFT operators that commute with all single-trace operators in a narrow time-band to all orders in the $1/N$ expansion, including the Hamiltonian and other charges that generate conformal transformations, while at the same time act like standard HKLL operators to leading order at large $N$. 

We have presented an explicit construction of such operators, and checked that they commute with the Hamiltonian to all orders in the $1/N$ expansion, and act like HKLL operators to leading order. Technically the construction of such operators is made possible due to the fact that different semi-classical states have exponentially small overlap. We have also discussed a generalization of our operators that would commute with all boundary charges of the conformal group. Moreover, we presented a definition of operators that commutes with all single-trace operators, not just conserved charges. The construction of these operators is slightly less explicit, and we define them by specifying their action on the code subspace around a semi-classical geometry. We argue that such operators commute with all single-trace operators in a narrow CFT time-band, while also acting like HKLL operators to leading order at large $N$. Acting with such operators creates excitations that are completely invisible to CFT correlation functions in a narrow time-band, even if they become accessible at later times when a lightray from the location where the bulk excitation was created reaches the boundary. This suggests that information can be lozalized in perturbative quantum gravity, to all orders in $G_N$ perturbation theory. We conclude with some open questions that we raised along the way.

\subsection{The variance of the energy from semi-classical gravity}

A quantity that played a primordial role throughout the paper is the variance of energy, which controls the early time decay of the return probability through \eqref{earlyr}. One question that would be interesting to understand better is how we can compute the variance $\braket{\Psi_0|\Delta H^2|\Psi_0}$ from semi-classical gravity. In appendix \ref{app:variance} we give an example that we can change the $O(N^2)$ coefficient of the variance of the Hamiltonian without changing the semi-classical geometry. This implies that the variance of the energy is not just a property of the geometry, but also of the quantum state of the fields on top of that geometry. Of course, if the metric changes as a function of time, this puts a bound on the variance through \eqref{deltaHbound}. This suggests that if we start with some time-dependent semi-classical geometry with a matter QFT state with large variance, it should not be possible to change the state in a way to make the variance decrease to $\mathcal{O}(1)$ without changing the metric towards a time-independent solution. The mechanism by which this would happen is unclear, and it would be interesting to pursue it further.

On a related note, we can ask how we can quantize the bulk mode associated to the Hamiltonian directly in gravity. The expectation value of the Hamiltonian is extracted through the fall-off of the metric near the AdS boundary, as is standard in AdS/CFT, but this does not capture its quantum 2-point function. If one computes the stress-tensor connected 2-point function on the geometry, takes the relevant components and performs the spatial integrals, one should obtain the variance. 
It would be desirable to have a more direct representation of the variance in terms of 
the bulk wavefunction of the non-propagating s-wave mode of the graviton and also understand from this point of view the lower bound on the variance for time-dependent geometries.

\subsection{Gravitational proof for the decay of the return probability }

A central part of this paper was played by the decay of the return probability. The physical interpretation of this decay for a semi-classical time-dependent geometry is that it computes (the square) of an overlap between two distinct geometries, namely the original one and the time-evolved one. The general expectation is that the overlap of two distinct coherent states should be given by
\be
\braket{\lambda_1 | \lambda_2} \sim e^{-N^2 f(\lambda_1,\lambda_2)} \,,
\ee
where $f$ is some $\mathcal{O}(1)$ function whose real part is positive (we have assumed that the states $\ket{\lambda_{1,2}}$ are normalized). The intuition is that $N^2$ plays the role of $1/\hbar$ which controls the overlap of coherent states, and from a gravitational stand-point, the on-shell action of any geometry will be proportional to $1/G_N$. However, this gravitational argument does not necessarily imply that the real part of $f$ is positive, which is required by reflection positivity of the CFT dual. As we have seen in \eqref{inequality}, interpreting geometries as quantum states implies constraints on various on-shell actions. 

It would be interesting to understand this problem directly in gravity. Can reflection positivity be proven directly at the level of the gravitational path integral? This requires proving \eqref{inequality} directly in gravity. A possible way to prove this is the following: we consider two states $\lambda_1$ and $\lambda_2$ with fixed sources, and their associated geometries contributing to the overlaps $\braket{\lambda_{1,2}|\lambda_{1,2}}$, with geometries $g_1$ and $g_2$ and on-shell actions $I_1$ and $I_2$. We start by considering a gravitational configuration which is half of $g_1$ (say the northern hemi-ball) and half of $g_2$ (the southern hemi-ball). This configuration has action 
\be
I_{\textrm{tot}}=\frac{I_1+I_2}{2} \,.
\ee
Note that the geometry is off-shell at the gluing surface between $g_1$ and $g_2$, and there could be another contribution $I_{\textrm{junction}}$ to the action coming from the gluing, which we will not include for now. To find the smooth saddle-point geometry, we need to let this geometry relax by modifying its configuration near the junction. One may be able to prove that this smoothing of the glued geometry comes with a definite sign in the action, therefore proving \eqref{inequality}. It would be interesting to pursue this idea.

\subsection{Microscopically time-dependent states}

We have seen that for any state with large energy and large energy variance, we can find bulk local operators who commute with the time-band algebra. The interpretation of these operators is that they are dressed with respect to features of the state (in particular the time-depdence of the state), rather than to the boundary CFT. This intuitive picture is clear when the state describes a semi-classical geometry that is macroscopically time-dependent, as the time-dependence can be seen directly from the background metric which has features with respect to which we can attach a gravitational dressing.

As we have discussed, our prescription also works for typical states with energy variance of $\mathcal{O}(N^2)$. In that context, the interpretation of the dressing is less clear. The dual geometry is not macroscopically time-dependent. We can declare that the operator is dressed with respect to the microscopic time-dependence, but it is unclear what that means. It would be interesting to have a better physical understanding of the dressing for such type of states. We hope to return to this question in the future.

It is also important to note that our operators are state-dependent, even outside the horizon. For a given typical state, we can use our construction to find the state-dressed local operator. However, if we now pick a different typical state then the operator will not act in the desired fashion. In this sense, our operators are similar to mirror operators \cite{Papadodimas:2012aq}, but they can live outside the horizon. Nevertheless, we wish to emphasize again that independently of questions surrounding the interpretation of these operators, an important message of this paper is that these operators exist and that {\it states} created by acting on the corresponding typical state with unitaries built from these state-dressed operators have identical correlators of single-trace operators in a narrow time-band in the $1/N$ expansion as the original state. Moreover, this can be done around any typical state once the state has been fixed.

\subsection{Microcanonical states and small energy variance}

There are also typical states with a small energy variance, of $\mathcal{O}(N^0)$. For example, when one refers to the microcanonical ensemble, one often has in mind picking a state with spread in energy which is $\mathcal{O}(N^0)$. For such states, the return probability does not decay to values which are exponentially small in $N^2$ after an order one time, which means we cannot use our construction to define state-dressed operators. The variance of the energy is a very coarse way to define how time-dependent a state is, and for states with energy variance of size $\mathcal{O}(N^0)$, the state is not time-dependent enough to dress operators to it. Of course, all these states look macroscopically time independent, and all the information is in the microscopic phases of the state. It would be interesting to study this further, and have a better physical picture of whether one can find state-dressed operators to these small variance states.

It is worthing mentioning that if the variance is $\mathcal{O}(N^c)$ for any $0<c<2$, our prescription does work. For typical states, this is some kind of intermediate regime between canonical states and microcanonical states. For coherent states that are macroscopically time-dependent, this situation would occur if the profile of the fields are not $\mathcal{O}(1)$, but rather scale with some positive power of $G_N$. In that case, backreaction is small, but the return probability still decays. It would be interesting to understand these regimes better, they interpolate between coherent states of the bulk quantum fields propagating on a frozen AdS background, and semi-classical geometries with a non-trivial metric.

\subsection{The AdS vacuum and low-energy states}

For low-energy states like the AdS vacuum or states with an $\mathcal{O}(N^0)$ energy above it, our construction does not work. Therefore, the results of this paper do not contradict the claims of \cite{Chowdhury:2021nxw}, that for perturbative excitations on top of the AdS vacuum one can reconstruct the state directly from the time-band. Technically, this happens because the return probability does not decay to exponentially small values for such states. Physically, states like the AdS vacuum have no features to which we can dress operators, so the only possible diff-invariant way to specify a point is with relation to the boundary. Even classically, there are no diff-invariant local observables in classical general relativity for the case of vacuum AdS. It thus appears that the failure of constructing approximately local diff-invariant operators around the AdS vacuum happens because of the special nature of the state, rather than a fundamental obstruction due to the non-locality of quantum gravity.

For excited states on top of the AdS vacuum, it is less obvious why local diff-invariant states cannot be constructed. One may imagine that if the VEV of a scalar field has a quantum lump in some region of space-time, we could dress an operator to the location of this lump. Technically, we see that at least our operators cannot achieve this goal. It would be interesting to have a more physical understanding of why it is not possible to dress operators to quantum profiles, rather than semi-classical ones. As we have seen in the previous subsection, it is not completely related to backreaction. If we consider a coherent state on top of vacuum AdS corresponding to a source which scales as $N^{1/4}$, the return probability would decay fast enough for our construction to work, even if backreaction can be neglected. Note however that such a state is not really part of the low-energy EFT on top of vacuum AdS, since it has energy that scales with some fractional power of the Planck scale. It would be interesting to understand this better.

\acknowledgments

We are happy to thank Micha Berkooz, Jan de Boer, Monica Guica, Elias Kiritsis, Shota Komatsu, Hong Liu, Olga Papadoulaki, Suvrat Raju, Erik Verlinde, Spenta Wadia, and Sasha Zhiboedov for stimulating discussions. EB and NV would like to thank CERN-TH for their hospitality during the preparation of this work and M. Bertolini for his invaluable support during this work. The work of EB and NV is partially supported by INFN Iniziativa Specifica - String Theory and Fundamental Interactions project.

\appendix

\section{Changing the variance of $H$ \label{app:variance}}

We would like to understand whether the variance of the energy is accessible within semi-classical gravity, simply from the geometry, or whether it requires more knowledge and in particular, the knowledge of the bulk quantum state for the fields propagating on the background. As we will see, knowledge of the quantum state seems to be required to extract the variance. 

The quantity we would like to compute is
\be
\braket{\Psi_0 | H^2 | \Psi_0}- \braket{\Psi_0 | H | \Psi_0}^2 \equiv \braket{\Psi_0 | H^2 | \Psi_0}_c \,.
\ee
This is a connected correlation function in holography, which usually would be compute from the 2-point function of the associated propagating fields on the relevant background. This 2-point function is sensitive both to the geometry and to the bulk quantum state of the propagating fields. However, here the situation is more subtle, because we are not studying the local correlation function of an operator, but rather the 2-point function of the spatial integral of a local operator. In this particular case, the situation is a lot more confusing because the dual bulk field would be the s-wave graviton, which is not a propagating degree of freedom in gravity.

So what computes this variance? We will not be able to answer this question, and we believe it to be an interesting open problem which we hope to return to in the future. Nevertheless, we will study some particular states that should be interpreted as adding an s-wave graviton in the bulk. Even though this mode doesn't propagate, we will see that adding it can affect the CFT variance. We will consider two type of deformations of the thermofield double (TFD) state, both of which are related to adding an integrated stress-tensor operator on the cylinder that prepares the TFD state. Let us start with some basics. We consider the TFD state
\be
\ket{\textrm{TFD}} = {1\over \sqrt{Z}}\sum_{i} e^{-\beta E_i/2} \ket{E_i} \ket{E_i} \,.
\ee
We assume that the partition function has the usual large $N$ behavior
\be
Z(\beta) = \exp\left[N^2\left(F_0(\beta) + {1\over N^2} F_1(\beta) + ...\right)\right] \,,
\ee
from which we can compute
\be
\langle H^n\rangle_\beta = (-1)^n {1\over Z}{d^n\over d\beta^n} Z \,.
\ee
where $H$ is  $H_L$ or $H_R$. We have
\bea
\bra{\textrm{TFD}} H \ket{\textrm{TFD}}&=& \braket{H}_\beta = - N^2 F_0'-F_1' \,, \\
\bra{\textrm{TFD}} H^2 \ket{\textrm{TFD}} -\bra{\textrm{TFD}} H \ket{\textrm{TFD}}^2&=&\braket{H^2}_{\beta,c} \equiv \braket{H^2}_{\beta} - \braket{H}_\beta^2 \,.
\eea
We have
\be
\braket{H^2}_{\beta,c}=N^2F_0''+F_1'' \,.
\ee
Now, consider the following state
\be
\ket{\psi}=H\ket{\textrm{TFD}} \,.
\ee
We now have
\be
\braket{\psi|\psi}= \braket{H^2}_\beta  
\ee
Let us now see how the energy and variance of the state have evolved. We have

\be
\frac{\bra{\psi}H\ket{\psi}}{\braket{\psi|\psi}}=  \frac{\bra{\textrm{TFD}}H^3\ket{\textrm{TFD}}}{\bra{\textrm{TFD}}H^2\ket{\textrm{TFD}}}=\frac{\braket{H}_\beta^3+3\braket{H^2}_{\beta,c}\braket{H}_\beta+\braket{H^3}_{\beta,c}}{\braket{H}_\beta^2+\braket{H^2}_{\beta,c}} \,,
\ee
where we defined
\be
\braket{H^3}_{\beta,c}\equiv \braket{H^3}_\beta -3\braket{H^2}_{\beta,c}\braket{H}_\beta-\braket{H}_\beta^3 \,.
\ee
Large $N$ factorization implies that we can expand this answer and we find
\bea
\frac{\bra{\psi}H\ket{\psi}}{\braket{\psi|\psi}}&=&  \braket{H}_\beta+2 \frac{\braket{H^2}_{\beta,c}}{\braket{H}_\beta} + \cdots \notag \\
&=&-N^2F_0' -F_1' - 2\frac{F_0''}{F_0'} +\cdots  \,.
\eea
We see that we obtain the TFD answer, up to a correction term, which is of size $N^0$. This means we have not changed the geometry classically, but only added a quantum particle on top of the TFD state. Similarly, one can compute
\bea
\frac{\bra{\psi}H^2\ket{\psi}}{\braket{\psi|\psi}}-\left(\frac{\bra{\psi}H\ket{\psi}}{\braket{\psi|\psi}}\right)^2&=&\frac{\braket{H}_\beta^4+6\braket{H^2}_{\beta,c}\braket{H}_\beta^2+\cdots}{\braket{H}_\beta^2+\braket{H^2}_{\beta,c}}-\left(\braket{H}_\beta^2+4 \braket{H^2}_{\beta,c}+\cdots \right) \notag \\
&=&\braket{H^2}_{\beta,c}+\cdots \notag \\
&=& N^2F_0''+ \cdots
\eea
We see that the energy has changed at $N^0$, but the variance has not changed at order $N^2$, only at order $N^0$. So this state modifies both the variance and the energy at subleading order compared to the TFD. We will now build a state that modifies the energy at subleading order, but the variance at leading order compared to the TFD.

Consider the state

\be
\ket{\phi}=(H-\braket{H}_\beta)\ket{\textrm{TFD}} \,.
\ee
We now have
\be
\braket{\phi|\phi}=\braket{H^2}_{\beta,c} \,,
\ee
and we can now compute the energy in this state:
\be
\frac{\bra{\phi}H\ket{\phi}}{\braket{\phi|\phi}}=  \frac{\braket{H^3}_\beta -2 \braket{H^2}_\beta\braket{H}_\beta + \braket{H}_\beta^3}{\braket{H^2}_{\beta,c}} = \braket{H}_\beta + \frac{\braket{H^3}_{\beta,c}}{\braket{H^2}_{\beta,c}}= -N^2F_0' -F_1' - 2\frac{F_0'''}{F_0''} +\cdots \,.
\ee
We see that this state modifies again the energy only at order $N^0$, and in a slightly different way than the previous state. In a similar way, we compute the variance and find
\bea
\frac{\bra{\phi}H^2\ket{\phi}}{\braket{\phi|\phi}}-\left(\frac{\bra{\phi}H\ket{\phi}}{\braket{\phi|\phi}}\right)^2&=&\braket{H}_\beta^2+ 3\braket{H}_{\beta,c}^2+\frac{2\braket{H^3}_{\beta,c}\braket{H}_\beta+\braket{H^4}_{\beta,c} }{\braket{H^2}_{\beta,c}}-\left(\braket{H}_\beta + \frac{\braket{H^3}_{\beta,c}}{\braket{H^2}_{\beta,c}} \right)^2 \notag \\
&=& 3\braket{H^2}_{\beta,c}  +\frac{\braket{H^4}_{\beta,c} }{\braket{H^2}_{\beta,c}} -\left( \frac{\braket{H^3}_{\beta,c}}{\braket{H^2}_{\beta,c}} \right)^2 \notag \\
&=&3 N^2 F_0'' + {3 (F_0'')^2 F_1'' - (F_0''')^2 + F_0''F_0'''' \over (F_0'')^2}+...
\eea
One can see that the change in the variance is order $N^2$ (it is three times the variance of the TFD state), so this is a modification of the variance at the order we were looking for.

From this, we can conclude that the semi-classical geometry is not enough to extract the variance of the energy. The quantum state of the bulk fields is equally important. For the state $\ket{\phi}$, we have the same leading large $N$ properties, but a different quantum state for the graviton. The fact that it is the s-wave of the graviton that enters is still puzzling, and it would be interesting how to propertly quantize this non-propagating degree of freedom. We leave this for the future.

\section{Boosts in global AdS \label{appendixboosts}}

As we have discussed in section \ref{sectionAdSCFT}, the conformal generators on the $d$-dimensional cylinder $\mathbb{R}\times S^{d-1}$ organize themselves as time-translations, rotations, and $2d$ remaining generators which correspond to boosts in the dual AdS geometry. The goal of this section is to discuss whether there exist states that can preserve the boost symmetry. As we have seen throughout the paper, symmetries that are broken by semi-classical states allow us to specify bulk points by dressing the location of a bulk point to the feature of the state that breaks the symmetry. It is important to understand which symmetries are broken, and which symmetries can be preserved by semi-classical states. For time translations and rotations, this is straightforward, but it is somewhat more subtle for boosts, which is the purpose of this section.

The $2d$ boost generators can be realized as $d$ non-independent copies of $SL(2,\mathbb{R})$ \cite{Freivogel:2011xc}. For simplicity, we will study the case of AdS$_3$, but the higher dimensional versions follow in a straight forward manner. In $d=2$, the two copies of $SL(2,\mathbb{R})$ are well-known and correspond to the left and right moving sectors of conformal transformation. The generators are given by $L_{-1},L_0,L_1$ and $\bar{L}_{-1},\bar{L}_{0},\bar{L}_{1}$. Time-translations and rotations are obtained by the combinations
\be
H=L_0+\bar{L}_0 \,, \qquad J=L_0-\bar{L}_0 \,.
\ee
The four residual generators correspond to boosts in AdS$_3$. For explicit expressions, see \cite{Maldacena:1998bw}. We would now like to analyze whether non-trivial states can be annihilated by these boosts. As a starting point, notice that there are obviously CFT states which are annilitated by $L_{-1}$ and $\bar{L}_{-1}$: primary states. However, we would like to consider generators that can be exponentiated to norm-preserving group elements. This means the generators should be Hermitian. The generators $L_{-1}$ and $\bar{L}_{-1}$ do not satisfy this property. However, we can assemble them into the combinations
\be
L_+= L_{-1} + L_{1} \qquad,\qquad L_- =i (L_{-1}- L_1)
\ee
Using that $L_{-1}^\dagger=L_1$, we see that $L_{\pm}$ are hermitian operators and can thus be exponentiated to form unitaries. 

The question we would like to ask is whether there are states in the Hilbert space that are eigenstates of $L_{\pm}$. We will see that the only finite energy eigenstates of these operators are those where the left-moving part of the CFT is in the vacuum. To see this, we consider the commutator
\be
[L_+, L_-]=4 i L_0
\ee
Suppose now that $|\psi\rangle$ is a normalizable eigenstate of ---say--- $L_+$. Computing the expectation value of this equation we find
\be 
\langle \psi | L_0 |\psi \rangle = 0
\ee
From the positivity of the energy spectrum this is possible only if $L_0 |\psi\rangle =0$. The only states with this property are states where the left moving sector of the CFT is in the vacuum.

Non-trivial states will thus break boost invariance, which can be use to specify the radial location of an operator. For the construction of operators presented in this paper, this would require considering the states obtained by acting with the unitary operators on semi-classical states $\ket{\psi_0}$ as
\be
e^{- i \gamma L_{\pm} } \ket{\psi_0} \,,
\ee
and studying the generalized return probability 
\be
R(\gamma)\equiv \left| \bra{\psi_0} e^{-i \gamma L_{\pm} } \ket{\psi_0} \right|^2 \,.
\ee
These return probabilities have not been studied but for semi-classical states, it is natural to expect them to be exponentially small for $\gamma\sim \mathcal{O}(1)$.

\section{Early time decay of the return probability}\label{appendixdecay}

We wish to estimate the early time decay of the return probability \eqref{radef}. We will see that at very early times, namely $t\sim \frac{1}{N}$, we can find the decay purely from large $N$ factorization. We will first recall a general property of coherent state overlaps which follows from large $N$ factorization, and then adapt the situation slightly to the return probability.

\subsection{Overlap of coherent states and large N factorization}

Coherent states of quantum gravity in AdS/CFT can be described by states prepared by a Euclidean path integral with sources turned on for single-trace operators. These states are thus given by
\be
\ket{\lambda}=e^{\int_{x_0<0}dx^d \lambda(x) {\cal O}(x)} \ket{0} \,,
\ee
where we have not written the appropriate time-ordering which is left implicit. We will now show that the overlap is given by
\be \label{overlapcoherent}
\braket{\lambda_1 | \lambda_2}=e^{\int_{\mathbb{R}^d} \lambda^*_1(y) \lambda_2(x) \braket{{\cal O}(y){\cal O}(x)}} +\mathcal{O}(1/N) \,,
\ee
where it should be understood that $y$ is integrated over the upper half plane while $x$ is integrated over the lower half plane.

We can explicitly expand out the integrals of the bra and the ket states, and use large $N$ factorization: this implies that the operators should be paired up and contracted using Wick's theorem, up to $1/N$ corrections. At a given power in the source, we will have a term of the form
\be
\left(\int dx dy\right)^k\frac{1}{(k!)^2}\lambda^*_1(y)^k \lambda_2(x)^k \bra{0}{\cal O}^k(y) {\cal O}^k(x)\ket{0} \,.
\ee
We can now apply Wick's theorem and find
\bea 
\left(\int dx dy\right)^k\frac{1}{(k!)^2}\lambda^*_1(y)^k \lambda_2(x)^k \bra{0} {\cal O}^k(y) {\cal O}^k(x)\ket{0} 
=\frac{1}{k!}\left(\int dx dy \lambda^*_1(y) \lambda_2(x) \bra{0}{\cal O}(y) {\cal O}(x)\ket{0} )\right)^k  \,, \notag
\eea
which we can re-exponentiate to find \eqref{overlapcoherent}. Note that we have not written the normalization of the states, which takes care of the Wick contraction between any two operators living both in the lower half plane, or upper half plane. Similarly, terms which have a different powers of upper and lower operators do not give contributions to leading order at large $N$ because we cannot pair the operators and use Wick's theorem.

For this to work, we have implicitly assumed that $\lambda\sim \mathcal{O}(N^0)$. To see this, note that the connected correlation functions of higher-point operators are suppressed by $1/N$, but also have more sources than lower-point functions. If we scale the sources as $\lambda\sim N^{1/2}$, which is the correct scaling to induce $\mathcal{O}(1)$ back-reaction on the dual spacetime\footnote{For operators that have unit 2-point function.}, we have to be more careful, as some of the terms we dropped involving connected correlators will be the same size as the Wick contractions. For example, we have
\bea
\lambda^*_1(y)\lambda_2(x) \braket{{\cal O}(y){\cal O}(x)} &\sim& N^2 \\
(\lambda^*_1(y)\lambda_2(x))^2 \braket{{\cal O}(y) {\cal O}(y){\cal O}(x){\cal O}(x)}_c &\sim& N^2 \,.
\eea
This means that we cannot truncate to the sector of Wick contraction, and we must resum the entire expansion. Note however that the contributions corresponding to loop diagrams in AdS are still suppressed by $1/N$, so we are resumming tree-level diagrams to build the backreacted geometry.

The upshot of this analysis is that we can use large-$N$ factorization to easily compute the overlap of coherent states, but only if the sources are $\mathcal{O}(1)$, in which case the exponent in the exponential is also $\mathcal{O}(1)$. If we try to make the sources scale with $N$, the exponent will be of order $N^2$ and then infinitely many contributions must be resummed. We will now apply this logic to the return probability.

\subsection{The return probability}

We can now apply the same logic as above, taking the operator $e^{-iHT}$ to be seen as an imaginary Euclidean source for the Hamiltonian (which is the integral of the stress-tensor). We want to compute
\be
R(T)=\lsz e^{-i H T} \rsz \lsz e^{i H T} \rsz \,.
\ee
Applying the logic above, we would find that to leading order we have
\be
R(T)= e^{-i T \lsz H_0 \rsz} e^{i T \lsz H_0 \rsz} = 1 +\mathcal{O}(1/N) \,.
\ee
So we see that the candidate leading term vanishes, and we must go to the next order. This is due to the nature of the return probability, which is a square of overlaps. A quick expansion of the exponentials shows that at order $T^2$, we have
\be
T^2 \Big(-\lsz H^2 \rsz +\big(\lsz H \rsz\big)^2\Big) = - T^2 \Delta H^2 \,.
\ee
For reasons similar to those explained above, this term can be exponentiated such that we find
\be
R(T)=e^{-T^2 \Delta H^2} +\mathcal{O}(1/N) \,.
\ee
As in the previous section, we can only trust this approximation if the exponent is $\mathcal{O}(1)$. Because we are considering states that have $\Delta H \sim N^2$, we see that we can trust this exponential decay of the return probability for time-scales up to $t\sim 1/N$. 

For larger time-scales, it may still hold, but it cannot be justified based solely on large $N$ factorization. It is instructive to consider the case of the thermofield double state and the spectral form factor, as we already discussed in section \ref{sec:tfd}. For simplicity, we set $d=2$ where we have
\be
Z(\beta)=e^{\frac{c}{12}\frac{4\pi^2}{\beta}} \,.
\ee
The spectral form factor then gives
\be
R(T)= e^{\frac{\pi^2 c}{3} \left(\frac{1}{\beta+IT} +\frac{1}{\beta-iT}\right)}= e^{\frac{2\pi^2 c}{3} \frac{\beta}{\beta^2+T^2}}\,.
\ee
We can expand this expression in $T$, as long as $T\ll \beta$, to find
\be
R(T)\approx Z(\beta)^2e^{-\frac{2\pi^2 c}{3} \frac{T^2}{\beta^3}} \,.
\ee
We find the exponential decay that goes like $T^2$. What is important is that even though $T$ must be much smaller than $\beta$, it is allowed to scale as $N^0$. This cannot be justified solely from large $N$ factorization, but still holds in this particular context. We expect the return probability to satisfy this property for holographic states more generally.

\section{LLM solutions in the bulk}\label{llm}

The LLM geometries correspond to solutions of type IIB supergravity with symmetry $SO(4)\cross SO(4) \cross R$. We assume the axion and dilaton are constant and the IIB three forms are vanishing. We introduce coordinates $x^\mu=(t,y,x_1,x_2)$ and $\Omega_3,\tilde{\Omega}_3$ for two 3-spheres corresponding to the $SO(4)$ isometries. We parametrize the five form as
\begin{equation}
    F_{5}=F_{\mu \nu}dx^{\mu}\wedge dx^{\nu} \wedge d\Omega_{3} + \tilde{F}_{\mu \nu}dx^{\mu}\wedge dx^{\nu} \wedge d\tilde{\Omega}_{3} \,,
\end{equation}
where the self duality of the five form implies that the two forms $F$ and $\widetilde{F}$ are dual to each other.

After demanding that the geometry preserves the Killing spinor in the presence of the five form, we arrive at the following solution for the ${1\over 2}$-BPS bulk states \cite{Lin:2004nb} 
\begin{equation}\label{13.5}
     ds^{2}=-\frac{(dt+V_{i}dx^{i})^{2}}{h^{2}}+h^{2}(dy^{2}+dx^{i}dx^{i})+ye^{G}d\Omega^{2}_{3}+\frac{y}{e^{G}}d\tilde{\Omega}^{2}_{3} \,,
\end{equation}
where every function in the metric is expressed in terms of a function $z(x_{1},x_{2},y)$ and we defined $z=\frac{1}{2}\text{tanh}\;G$, $h^{-2}=2y\;\text{cosh}\;G$, and 
\begin{align}\label{13.6}
    y\partial_{y}V_{i}=\epsilon_{ij}\partial_{j}z, && y(\partial_{i}V_{j}-\partial_{j}V_{i})=\epsilon_{ij}\partial_{y}z \,.
\end{align}
For the forms $F,\tilde{F}$ we have
\begin{align}\label{13.7}
    F=dB_{t}\wedge (dt+V)+ B_{t}dV + d\hat{B}, && \tilde{F}=d\tilde{B}_{t}\wedge (dt+V)+ \tilde{B}_{t}dV + d\hat{\tilde{B}} \,,
\end{align}
where $B_{t}=-\frac{1}{4}y^{2}e^{2G}$ and $\tilde{B}_{t}=-\frac{1}{4}y^{2}e^{-2G}$. On the other hand,
\begin{align}\label{13.8}
    d\hat{B}=-\frac{1}{4}y^{3}\star_{3}d(\frac{z+2}{y^{2}}), &&  d\hat{\tilde{B}}=-\frac{1}{4}y^{3}\star_{3}d(\frac{z-2}{y^{2}})  \,,
\end{align}
where $\star_{3}$ is the epislon symbol in the flat three dimensions.

The only free function, $z$, is constrained to solve the equation,
\begin{equation}\label{13.9}
    \partial_{i}\partial_{j}z+y\partial_{y}(\frac{\partial_{y}z}{y})=0 \,.
\end{equation}
We focus our attention on the plane $y=0$. Since the product of the radii of the two 3-spheres is $y$, there will be a conical singularity at $y=0$ unless the function $z$ has a special behaviour.

Let's consider the case where $R_{1}$ is kept finite, i.e, $e^{-G}\rightarrow 0 $ as $y\rightarrow 0$. Thus, one has, $z \sim 1/2 -e^{-2G}+ ...$ . If one assumes that $z=1/2$ at $y=0$, then one gets the expansion, $z \sim 1/2 - y^{2}f(x_{1},x_{2})+...$ for some positive function $f$, with our boundary conditions. Thus, $e^{-G} \sim y c(x_{1},x_{2})+...$ and $h^{2}\sim c(x_{1},x_{2})+...$ . Therefore, close to $y=0$, the part of the metric involving $R_{2}$ will look like,
\begin{equation}
    h^{2}dy^{2} + R_{2} d\tilde{\Omega}^{2}_{3}\approx c(dy^{2} +y^{2}d\tilde{\Omega}^{2}_{3}) \,.
\end{equation}
Thus the conical singularity is resolved. In the case where $R_{2}$ is kept fixed, the same argument goes through but now with the condition that $z=-1/2$ at $y=0$.

With these boundary values of $z$ at $y=0$ as a source, one can solve the Laplace equation\footnote{More precisely, it is a Laplace equation for $z/y^{2}$.} \eqref{13.9} and compute $z(x_{1},x_{2},y)$. In addition, $V_{i}$ can also be expressed in terms of an integral of $z(x_{1},x  _{2},0)$ over the two dimensional space.

\section{Notes on boundary states}

Some useful references for this section are \cite{Cardy:2004hm, Miyaji:2014mca,Guo:2017rji, Miyaji:2021ktr}.

\subsection{Boundary states in 2D CFT}

Boundary states in a 2d CFT need to satisfy \cite{Cardy:2004hm}
\begin{equation}\label{3}
    (L_n - \tilde{L}_n) \ket{B} =0.
\end{equation}
In any Verma module, one can find a simple solution to these conditions as 
\begin{equation}
    \ket{I_h} = \sum_{\Vec{k}} |\Vec{k},h\rangle_L \otimes |\Vec{k},h\rangle_R \,,
\end{equation}
where $|\Vec{k},h\rangle_L$ is a linear combination of Virasoro descendants of the primary state $|h\rangle$ characterized by an infinite dimensional vector $\Vec{k} = (k_1, k_2, ...)$  with  non-negative integer components. We identify these states by starting with descendants of the form
\begin{equation}\label{2d-basis}
    ... L_{-n}^{K_{n}}  ... L_{-1}^{K_{1}} \ket{h}_L.
\end{equation}
and forming an orthonormal basis selected such that $ _{L}\langle\Vec{k},h|\Vec{k'},h\rangle_L = \delta_{\Vec{k}, \Vec{k'}}$.

The state $ \ket{I_h}$ is called the Ishibashi state for the primary state $ \ket{h}_L$, where the states $|\Vec{k},h\rangle$ are the descendant on top of the primary labeled by $h$. It can be seen easily that 
\be
L_n |I_h\rangle = \tilde{L}_n |I_h \rangle \,.
\ee
It is clear that the Ishibashi states have maximal entanglement between the left-moving and right-moving sectors. Linear combinations of the Ishibashi states satisfy the constraint \eqref{3} as well. 

Physical boundary sates are given by special linear combinations of Ishibashi states  which are called Cardy states
\begin{equation}
    \ket{B_a} = \sum _h C_{a,h} \ket{I_h} \,.
\end{equation}
Physical boundary states should satisfy a consistency condition of the partition function on a finite cylinder related to open-closed duality \cite{Cardy:2004hm}.
   
The Cardy states are singular because
the norm of the Ishibashi states is divergent. One can define regularized boundary states by evolving in Euclidean time as 
\begin{equation}\label{2}
    \ket{B_{a,\beta}} = e ^{- \frac{\beta}{4} H_c} \ket{B_a},
\end{equation}
where $ \beta$ is a positive constant and $H_c = L_0 + \tilde{L}_0 - \frac{c}{12}$. Since $ [L_0-\tilde{L}_0,H_c] = 0 $, the state \eqref{2} is still space-translational invariant on the circle, but it is time-dependent.

Ishibashi states are orthogonal to each other. The amplitude of Euclidean time evolution by $ \beta /2$ between two such states is computed as
\begin{equation}
    \langle I_k | e^{- \beta H_c/2}| I_l \rangle = \delta_{kl} \chi_k (e^{- \beta /2}) \,.
\end{equation}
$ \chi_k$ is the character for the primary $k$.
On the other hand, the Cardy states are not orthogonal to each other but satisfy the
open-closed duality relation as follows
\begin{equation}
     \langle B_a | e^{- \frac{\beta}{2} H_c}| B_b \rangle = \sum _k N^{(k)}_{a,b} \Tr_k[e ^{-\frac{4 \pi ^2}{\beta}H_o}]
\end{equation}
where $ H_o = L_o - \frac{c}{24}$ denotes the Hamiltonian in the dual channel, characterized by the boundary conditions $a,b$.
On the right hand side, $\Tr_k[...]$ denotes a trace in the sector associated to a primary $k$ as well as its descendants. Moreover, $N^{(k)}_{a,b}$ counts the degeneracy of sectors which belong to the primary $k$ with boundary conditions $a$ and $b$.

In the high temperature limit $ \beta \rightarrow 0$, we find that 
\begin{equation}
     \langle B_a | e^{- \frac{\beta}{2} H_c}| B_b \rangle \simeq N^{(k_m)}_{a,b} e^{-\frac{4 \pi ^2}{\beta}(h_{a,b}^{(min)}- \frac{c}{24}) } \,,
\end{equation}
where $k_m$ is the lightest primary among those satisfy $ N^{(k_m)}_{a,b}\neq 0$, whose conformal dimension is denoted as $h_{a,b}^{(min)}$. 

We can estimate the inner products between two normalized boundary states in this limit as 
\begin{equation}
     \langle \psi_a | e^{- \frac{\beta}{2} H_c}| \psi_b \rangle
     = \frac{ \langle B_a | e^{- \frac{\beta}{2} H_c}| B_b \rangle }{\sqrt{ \langle B_a | e^{- \frac{\beta}{2} H_c}| B_a \rangle  \langle B_b | e^{- \frac{\beta}{2} H_c}| B_b \rangle }}
     \simeq \delta_{a,b} + N^{(k_m)}_{a,b} e^{-\frac{4 \pi ^2}{\beta} h_{a,b}^{(min)}} \,.
\end{equation}
Note that $ N_{a,a}^{(0)}=1$. In this way, a large gap in the open string channel leads to a large exponential suppression of off-diagonal elements of inner products.

In holographic BCFT, the inner product between two boundary states can be computed 
by evaluating the gravity action on the dual
background. When we consider the gravity dual of a cylinder, there are two candidates of classical gravity solutions depending on whether the end of the word brane is connected or disconnected which are called connected and disconnected solutions. 
When we consider the overlap for an identical boundary condition $a$, then both the connected and disconnected solution are allowed. In the limit $ \beta \rightarrow 0 $, the connected solution is favored and one can find that 
\begin{equation}\label{5}
     \langle B_a | e^{- \frac{\beta}{2} H_c}| B_a \rangle \simeq e^{\frac{\pi ^2 c}{6 \beta}}.
\end{equation}
We will use it later to calculate the return probability for boundary states. In addition to it, one can find the inner product between two boundary states with different boundary conditions. In this case, only the disconnected solutions are allowed and  
\begin{equation}
     \langle B_a | e^{- \frac{\beta}{2} H_c}| B_b \rangle \simeq e^{\frac{c \beta}{12} + S_{bdy}^{(a)} +S_{bdy}^{(b)} },
\end{equation}
where $S_{bdy}^{(i)}, i = a,b$
are the boundary entropies \cite{Miyaji:2021ktr}.

\subsection{Boundary states in higher dimensions}

One can generalize  to higher dimensions and define a boundary state $\ket{B_a}$ as a state associated to a $(d-1)$-dimensional boundary in $d$-dimensional CFT \cite{Fujita:2011fp,Miyaji:2021ktr}. Taking the boundary to be a torus ${\mathbb T}^{d-1}$, the inner product between two boundary states in a holographic BCFT can be computed as a partition function on a $d$-dimensional open manifold $ I_{\beta /2} \times {\mathbb T}^{d-1}$ where $ I_{\beta /2} $ is a length $ \beta /2$ interval. As in the 2d case, there are two bulk solutions, a connected and a disconnected one. In the $ \beta \rightarrow 0 $
limit the connected solution is dominant and one can find the inner product between two identical boundary states using the gravity solution as 
\begin{equation}\label{11}
 \langle B_a | e^{- \frac{\beta}{2} H_c}| B_a \rangle_{con} \simeq e^{\alpha_d/\beta ^{d-1}}   \,,
\end{equation}
where 
\begin{equation}\label{10}
  \alpha_d = (4 \zeta(T))^d \frac{R^{d-1}}{16 G_N} L^{d-1}  \,,
\end{equation}
where $R$ is the AdS radius, $ L$ is the length of the compactified spatial directions and 
$ \zeta(T)$ is a function of tension which is defined when $T<0$ as 
\begin{equation}
    \zeta(T) \equiv \frac{\Gamma(1/d)\Gamma(1/2)}{\Gamma(1/d + 1/2)} \frac{R|T|}{d(d-1)} (1- \frac{R^2 T^2}{(d-1)^2})^{1/d - 1/2} F(1, 1/d, 1/2 + 1/d;1- \frac{R^2 T^2}{(d-1)^2} ) \,,
\end{equation}
 and when $ T> 0$, $ \zeta (T) = \frac{2\pi}{d} - \zeta (-T)$. 
 The tension takes values in the range $ |T| < \frac{d-1}{R}$.
 For $ d> 2$, $ \zeta (T)$ non-trivially depends on $T$ and there is an upper bound of the tension $ T< T_*$ which $ T_* > 0$ and $ \zeta(T_*)=0$ \cite{Miyaji:2021ktr}.
 
\subsection{Correlation functions in BCFTs}

Let us first start with the simplest case where the CFT is defined on the upper half plane and the boundary state $|B\rangle$ is placed along the real axis.
We consider the 1-point function of a local operator placed at $z$ in the upper half plane. 
In the case of a CFT on the plane, the 1-point function of a primary operator in the vacuum is required to vanish by the symmetries. These are partly broken in a BCFT. The remaining symmetries constraint the 1-point function to have the form
\begin{equation}
    \langle {\cal O}(z)\rangle_{\rm UHP} = \frac{A_{\cal O}}{(2 \Im(z))^\Delta} \,,
\end{equation}
where $A_{\cal O}$ is determined by the details of the theory and the precise boundary state in
question. One could think of this as the boundary providing a source for the operator
${\cal O}$.

The 2-point function of a primary operator in a BCFT is more complicated than
the case with no boundaries where it is exactly fixed by the symmetries. Non-trivial
information about the operator content and OPE coefficients is necessary to compute
the 2-point function exactly in a BCFT. We assume that for large $N$ holographic CFTs the large $N$ 2-point function takes the form
\begin{equation}
     \langle {\cal O}(z_1) {\cal O}(z_2)\rangle_{UHP} =  \langle {\cal O}(z_1)\rangle_{UHP} \langle {\cal O}(z_2)\rangle_{UHP}
     +  \langle {\cal O}(z_1) {\cal O}(z_2)\rangle \pm  \langle {\cal O}(z_1) {\cal O}(z_2^*)\rangle \,,
\end{equation}
where 
\begin{equation}
     \langle {\cal O}(z_1) {\cal O}(z_2)\rangle= \frac{1}{|z_1 - z_2|^{2 \Delta}} \,,
\end{equation}
where the contribution from an image insertion placed at $z_2^*$. The sign of the last term is governed by the boundary conditions, being either Dirichlet ($-$)
or Neumann ($+$). 

Mapping the $z$ coordinate to a new coordinate $w$ by
\begin{equation}
    w \rightarrow z = \exp (2\pi w /\beta + i 2 \pi/4) \,,
\end{equation}
we can map the upper half plane to the a strip of width $ \beta/2$, where the positive (negative) real axis is mapped to the lower (upper) edge of the strip.

Since primary operators continue to transform in the usual way, the correlation functions now transform to
\begin{equation}\label{8}
    \begin{split}
       \langle {\cal O}(w)\rangle_{\text{strip}}& = \frac{A_{\cal O}}{(\frac{\beta}{\pi} \cos [\frac{2 \pi}{\beta} \tau])^\Delta} 
       \\ 
    \langle {\cal O}(w_1) {\cal O}(w_2)\rangle ^{\text{connected}}_{\text{strip}}& = 
    \frac{1}{|\frac{\beta}{\pi} \sinh[\frac{\pi}{\beta} (w_1 - w_2)]|^{2\Delta}} \pm
     \frac{1}{|\frac{\beta}{\pi} \cosh[\frac{\pi}{\beta} (w_1 - \Bar{w}_2)]|^{2\Delta}} \,,
    \end{split}
\end{equation}
where the second line is only the connected piece of the large $N$ 2-point function \cite{Almheiri:2018ijj}.
 Higher order correlation function can be found through large $N$ factorization.

Correlation functions on a state defined on a circle by
\begin{equation}
    \ket{B_\beta} = e^{- \beta H/4} \ket{B},
\end{equation}
can be thought of as correlation function on a cylinder of width $ \beta/2$ where the boundary state is placed on both sides. We can instead consider a strip of width $ \beta /2$, from $ \tau= - \beta /4$ to $ \tau= \beta /4$ with periodicity $ x \sim x + R$. We choose $ R= 2 \pi$ for simplicity from now on. In large $N$ holographic CFTs  correlation functions on the cylinder can be found from the correlation function on the strip using the method of images

 \begin{equation}
     \langle O(w_1) O(w_2)\rangle ^{\text{connected}}_{\text{cylinder}} = \sum _{n=0}^{\infty}  \langle O(w_1 + 2\pi n) O(w_2)\rangle ^{\text{connected}}_{\text{strip}} \,.
\end{equation}

\bibliography{references}

\bibliographystyle{JHEP}

\end{document}